\documentclass{article}
\usepackage{arxiv}

\usepackage[utf8]{inputenc}
\usepackage[T1]{fontenc}
\usepackage{lmodern}

\usepackage{microtype}
\usepackage{nicefrac}
\usepackage{bm}
\usepackage{makecell}

\usepackage{amsmath, amssymb, amsfonts}
\usepackage{amsthm}

\usepackage{graphicx}
\usepackage{float}
\usepackage{rotating}
\usepackage{multirow}
\usepackage{booktabs}
\usepackage{colortbl}
\usepackage[table,xcdraw]{xcolor}

\usepackage{tikz}
\usetikzlibrary{positioning, shapes, arrows.meta}

\usepackage{url}
\usepackage{hyperref}

\usepackage{natbib}
\usepackage{doi}

\usepackage{afterpage}
\usepackage{lscape}
\newtheorem*{definition*}{Definition}

\usepackage{titlesec}
\titleclass{\subsubsubsection}{straight}[\subsubsection]
\newcounter{subsubsubsection}[subsubsection]
\renewcommand\thesubsubsubsection{\thesubsubsection.\arabic{subsubsubsection}}
\titleformat{\subsubsubsection}
  {\normalfont\normalsize\bfseries}{\thesubsubsubsection}{1em}{}
\titlespacing*{\subsubsubsection}{0pt}{3pt}{3pt}

\definecolor{lightblue}{RGB}{173, 216, 230}
\definecolor{lightgreen}{RGB}{144, 238, 144}
\definecolor{lightorange}{RGB}{255, 228, 181}

\newtheorem{definition}{Definition}

\title{Opinion Dynamics: A Comprehensive Overview}
\renewcommand{\shorttitle}{\textit{Opinion Dynamics: A Comprehensive Overview}}

\author{
  Mohammad Shirzadi\thanks{Corresponding author} \\
  School of Computing \\
  Australian National University \\
  Canberra, Australia \\
  \texttt{mohammad.shirzadi@anu.edu.au}
  \And
  Emilio Cruciani \\
  European University of Rome \\
  Rome, Italy \\
  \texttt{emilio.cruciani@unier.it}
  \And
  Ahad N. Zehmakan \\
  School of Computing \\
  Australian National University \\
  Canberra, Australia \\
  \texttt{ahadn.zehmakan@anu.edu.au}
}

\hypersetup{
    pdftitle={Opinion Dynamics: A Comprehensive Overview},
    pdfsubject={Physics and Computing, Opinion Dynamics},
    pdfauthor={Mohammad Shirzadi, Emilio Cruciani, Serge Galam, Ahad N. Zehmakan},
    pdfkeywords={Opinion Dynamics, Polarization, Network Science, Social Influence},
    colorlinks=true,
    linkcolor=blue,
    citecolor=blue,
    urlcolor=blue
}

\begin{document}
\maketitle

\begin{abstract}
Opinion dynamics, the evolution of individuals through social interactions, is an important area of research with applications ranging from politics to marketing. Due to its interdisciplinary relevance, studies of opinion dynamics remain fragmented across computer science, mathematics, the social sciences, and physics, and often lack shared frameworks. This survey bridges these gaps by reviewing well-known models of opinion dynamics within a unified framework and categorizing them into distinct classes based on their properties. Furthermore, the key findings on these models are covered in three parts: convergence properties, viral marketing, and user characteristics. We first analyze the final configuration (consensus vs polarized) and convergence time for each model. We then review the main algorithmic, complexity, and combinatorial results in the context of viral marketing.  Finally, we explore how node characteristics, such as stubbornness, activeness, or neutrality, shape diffusion outcomes. By unifying terminology, methods, and challenges across disciplines, this paper aims to foster cross-disciplinary collaboration and accelerate progress in understanding and harnessing opinion dynamics.
\end{abstract}

\tableofcontents

\section{Introduction}

Humans continuously form and update their opinions on a wide range of topics, from everyday decisions such as choosing a new restaurant or a sunscreen brand, to consequential matters such as selecting political candidates or making investment choices. In shaping their opinions, individuals rely not only on their personal knowledge and judgment but also on the views of others, particularly close social contacts and influential public figures. As a result, opinions in a community evolve through ongoing interpersonal and societal interactions.

In recent decades, the rise of online platforms such as Facebook, TikTok, and WeChat has profoundly transformed how people communicate and influence one another. These platforms enable the rapid dissemination of information across geographical, cultural, and ideological boundaries. While phenomena such as disagreement, conformity, and polarization have long existed in human societies, their scale and dynamics have been reshaped, and in many cases intensified, by digital social engagement.

The way opinions form and spread through social interactions can have profound effects on various aspects of society, including politics, public health, economics, fashion, and culture. Accordingly, there has been increasing interdisciplinary interest, e.g.,~\cite{kempe2003maximizing}, in understanding the underlying mechanisms that drive opinion dynamics. Gaining such insights is crucial not only for advancing theoretical understanding but also for informing the design of more effective communication strategies, countering misinformation, mitigating polarization, and fostering safer, more constructive online environments.

One natural approach to studying opinion dynamics is to develop mathematical models that simulate how opinions evolve within a population. Although real-world opinion formation is highly complex and influenced by numerous contextual factors, abstract models can help uncover general principles and identify key parameters that govern opinion diffusion. A common modeling framework represents individuals as nodes in a graph (network) and encodes opinions as numerical values. Updating mechanisms then determine how individuals revise their opinions over time in response to their neighbors' opinions.

Due to the inherently interdisciplinary nature of opinion dynamics, the field has attracted researchers from diverse domains such as computer science~\cite{kempe2003maximizing}, applied mathematics~\cite{proskurnikov2017tutorial}, statistical physics~\cite{castellano2009statistical,starnini2025opinion,caldarelli2025physics}, sociology~\cite{degroot1974reaching}, and economics~\cite{jackson2011overview}. However, despite this shared interest, the field remains fragmented: researchers often develop parallel models and theories using domain-specific methodologies and terminology. For instance, physicists may draw on spin systems to model social influence, while computer scientists may frame it as information diffusion or learning dynamics. This disciplinary siloing has impeded the development of a unified framework for understanding opinion formation and evolution.

This survey provides a unified, systematic overview of the major classes of opinion diffusion models, presenting them with consistent notation and a comparative framework. Despite the growing body of research on opinion dynamics, no existing survey, to the best of our knowledge, comprehensively covers the full range of models across disciplines. By synthesizing perspectives from diverse research communities, this work aims to expose shared conceptual foundations, clarify underlying assumptions, and reveal connections that are often obscured by disciplinary boundaries.

\paragraph{Section~\ref{models}: Models.} We present a unified review of the main opinion dynamics models using consistent notation and terminology. Broadly, we categorize these models into two groups:
\begin{itemize}
    \item \textit{Discrete Models}: Ising, Sznajd, Majority, Voter, PUSH-PULL, Bootstrap Percolation, Linear Threshold, Independent Cascade, and epidemiological processes like SI, SIS, and SIR;
    \item \textit{Continuous Models:} French-DeGroot, Friedkin-Johnsen, Deffuant-Weisbuch, Hegselmann-Kraus, and Abelson models.
\end{itemize}
We establish the necessary mathematical notation, including graph definitions, that enable us to formulate each opinion dynamics model as a well-defined dynamic process over a graph structure with an opinion state and an updating mechanism.

\paragraph{Section~\ref{Convergence_Properties}: Convergence Properties.} Arguably, the most well-studied aspect of opinion dynamics is their convergence properties: how long does it take for the process to reach a ``final'' configuration? And what does such a final configuration look like? We review key findings on the convergence properties of the models studied. Leveraging mathematical techniques, such as Markov chain analysis and spectral graph theory, as well as experimental analysis and prior work (e.g.,~\cite{hassin2001distributed, elsasser2009runtime}), prior works have studied the convergence characteristics of different models. We review the key findings, including various bounds on the convergence time of the models in terms of graph and model parameters, and highlight some important existing gaps.

\paragraph{Section~\ref{Viral_Marketing}: Viral Marketing.} What subset of users should adopt a particular opinion to trigger its widespread adoption across a social network? This fundamental question, central to applications such as viral marketing and political campaigning, has been studied extensively within the framework of opinion dynamics. Several problem formulations have emerged in this context: (i) Optimization: given a budget $k$, which $k$ individuals should be selected to adopt the opinion to maximize its spread? (ii) Combinatorial: What are the theoretical bounds on the minimum number of initial adopters required for an opinion to become viral, expressed in terms of graph and model parameters? (iii) Probabilistic: What is the minimum probability $p$ such that, if each user independently adopts the opinion with probability $p$, a viral cascade is likely to occur?

To address these questions, researchers have drawn upon tools from network science, graph theory, optimization, machine learning, probability, and combinatorics. Prior work (e.g.,~\cite{kempe2003maximizing}) has proposed both algorithmic solutions and theoretical bounds for these formulations. These approaches are often evaluated empirically using real-world datasets from platforms such as Facebook and Twitter.

In this survey, we provide a structured categorization of the major problem formulations related to opinion virality. We summarize key theoretical results, highlight state-of-the-art algorithms across different models of opinion dynamics, and review the key empirical analyses.

\paragraph{Section~\ref{Users_Characteristics}: User Characteristics.} 
A common simplification in most opinion dynamics models is the assumption that all users behave identically. To address this limitation, a growing body of research (e.g.,~\cite{galam2016stubbornness}) has focused on understanding how individual user characteristics, such as stubbornness, agnosticism, or contrarianism, affect the diffusion of opinions. These studies reveal that heterogeneity in user behavior can significantly alter the dynamics and outcomes of opinion formation. We review key developments in this area, with particular emphasis on how user traits influence phenomena such as polarization, the influence of elite groups in social settings, and the design of effective vaccination strategies in epidemiological models.

\section{Models}\label{models}

In this section, we first define the graph and the notation for opinion dynamics. We then classify opinion diffusion models into two categories: \textit{discrete opinion dynamics}, where the opinion space is a finite set of values, and \textit{continuous models}, where opinions take values in (a subset of) the real numbers. After this classification, we briefly review each model.

\subsection{Basic Definitions}

\paragraph{Graph Definitions. }
\begin{definition*}[\textbf{Weighted Graph}]
A \textit{weighted graph} is a triple $G = (V, E, W)$ where $V = \{v_1, v_2, \dots, v_n\}$ is the set of nodes, $E \subseteq V \times V$ is the set of directed edges, and $W: E \to \mathbb{R}^+$ is a weight function that assigns a positive real-valued weight $w_{ij}$ to each edge $(v_i, v_j) \in E$.
\end{definition*}

\begin{definition*}[\textbf{Adjacency Matrix}]
Given a weighted graph $G = (V, E, W)$, the \textit{adjacency matrix} $\mathbf{A} \in \mathbb{R}^{n \times n}$ is defined as
\[
[\mathbf{A}]_{ij} =
\begin{cases}
w_{ij} & \text{if } (v_i, v_j) \in E, \\
0 & \text{otherwise}.
\end{cases}
\]
\end{definition*}

\begin{definition*}[\textbf{Undirected Graph}]
A graph is called \textit{undirected} if for every edge $(v_i, v_j) \in E$, the reverse edge $(v_j, v_i)$ is also in $E$, and the weights are symmetric: $w_{ij} = w_{ji}$ for all $(v_i, v_j) \in E$. In this case, the adjacency matrix $\mathbf{A}$ is symmetric.
\end{definition*}

\begin{definition*}[\textbf{In-Neighborhood and Out-Neighborhood}]
For a node $v_i$ in a directed graph, the \textit{in-neighborhood} is defined as
\[
N_{\text{in}}(v_i) = \{v_j \in V : (v_j, v_i) \in E\},
\]
and the \textit{out-neighborhood} is defined as
\[
N_{\text{out}}(v_i) = \{v_j \in V : (v_i, v_j) \in E\}.
\]
\end{definition*}

\begin{definition*}[\textbf{In-Degree and Out-Degree}]
For a node $v_i$ in a directed graph, the \textit{in-degree} is defined as
\[
\deg_{\text{in}}(v_i) = \sum_{v_j \in N_{\text{in}}(v_i)} w_{ji},
\]
and the \textit{out-degree} is defined as
\[
\deg_{\text{out}}(v_i) = \sum_{v_j \in N_{\text{out}}(v_i)} w_{ij}.
\]
\end{definition*}

\begin{definition*}[\textbf{Degree Matrix}]
The \textit{degree matrix} $\mathbf{D} \in \mathbb{R}^{n \times n}$ is a diagonal matrix defined by 
\[
[\mathbf{D}]_{ij} =
\begin{cases}
\deg_{\text{out}}(v_i) & \text{if } i=j, \\
0 & \text{otherwise}.
\end{cases}
\]
\end{definition*}
In the case of an undirected graph, the \textit{neighborhood} of a node $v_i$ in an undirected graph is defined as
\[
N(v_i) = \{v_j \in V : w_{ij} > 0\}, 
\]
and the in-degree and out-degree are equal, and we refer to the \textit{degree} of a node $v_i$ as
\[
\deg(v_i) = \sum_{v_j \in N(v_i)} w_{ij}.
\]
\begin{definition*}[\textbf{Unweighted or Simple Graph}]
A graph is called \textit{unweighted} or \textit{simple} if all weights are equal to $1$. In this case, the adjacency matrix becomes binary with $[\mathbf{A}]_{ij} = 1$ if $(v_i, v_j) \in E$, and $0$ otherwise. The degree of a node reduces to the number of its neighbors.
\end{definition*}
It should be mentioned that when edge weights are not explicitly considered, we assume the graph is simple.

\paragraph{Opinion Dynamics Notations.}

Consider a social network represented as a weighted graph $G = (V, E, W)$, where $V$ is the set of nodes (individuals), $E$ is the set of edges (potential connections), and $W$ is a weight function assigning weights to edges. Each node $v \in V$ holds an opinion at any given time, which evolves according to a predefined update rule.

\begin{definition}[\textbf{Opinion Space}]
The \textit{opinion space}, denoted by $\mathcal{O}$, defines the set of possible states or opinions a node can adopt. 
\end{definition}

\paragraph{Discrete vs Continuous.} The space can be discrete (e.g., $\{-1, 1\}$, $\{\text{blue, red}\}$, $\{\text{resident, mutant}\}$ or $\{\text{susceptible, infected, recovered}\}$) or continuous (e.g., $[-1, +1]$). The state of a node $v$ at time $t$ is denoted by $s_v(t) \in \mathcal{O}$, and the configuration of all node states at time $t$ is represented as state vector $\textbf{s}(t)=\big(s_{v_1}(t), s_{v_2}(t),\cdots, s_{v_n}(t)\big)^{\top}$. 

The evolution of opinions is governed by an \textit{update function} $\Phi$, which determines how nodes update their states based on the current state. Formally:
\begin{definition}[\textbf{Update Function}]
For each node $v$ at time $t$, the updated state is computed as
\[
s_v(t+1) = \Phi(G, s(t), v),
\]
where $\Phi$ depends on the graph $G$, the current state vector $\textbf{s}(t)$, and the node $v$. If node $v$ gets updated, then $s_v(t+1) = \Phi(G, \textbf{s}(t), v)$; otherwise, $s_v(t+1) = s_v(t)$.
\end{definition}
As we will see, an updating function can be deterministic or stochastic (random). 
\begin{definition}[\textbf{Synchronous/Asynchronous Update}]
In \textit{synchronous models}, the states of all nodes are updated simultaneously at each time step using the update function $\Phi$. In \textit{asynchronous models}, only a subset of nodes is updated at each time step.
\end{definition}

The choice of the updating mechanism (synchronous or asynchronous) can significantly influence the dynamics and outcomes of the opinion formation process. Unless otherwise mentioned, we always assume the update rule is applied synchronously.

To summarize, an opinion dynamics model is formally defined by an underlying graph $G=(V,E,W)$, an initial state (opinion) vector $\textbf{s}(0)$, and the update function $\Phi$ and its synchronous/asynchronous properties. This unified notation will be used throughout the paper to describe and analyze various opinion dynamics models.

\subsection{Discrete Models}

A discrete opinion space is typically used to model scenarios where a fixed number of opinions/states are possible, such as informed/uninformed, infected/uninfected, in favor/against, and political party A/B.

\subsubsection{Voter Model}
\begin{figure}[h]
\centering
\begin{tikzpicture}[
    node/.style={circle, draw, minimum size=9mm, inner sep=0pt, thick},
    red/.style={fill=red!30},
    blue/.style={fill=blue!30}
]

\node[node, blue] (A0) at (0,1) {$i$};
\node[node, red]  (B0) at (2,1) {$i+1$};
\node[node, red]  (C0) at (0,-1) {$i+2$};
\node[node, red]  (D0) at (2,-1) {$i+3$};
\draw[->] (A0) -- (D0);
\draw[->] (B0) -- (D0);
\draw[<->] (C0) -- (D0);
\draw[->] (A0) -- (B0);
\draw[->] (B0) -- (C0);
\node at (1,2.2) {\textbf{$t=0$}};

\node at (3.5,0) {\Huge$\rightarrow$};

\begin{scope}[xshift=5.5cm]
\node[node, blue] (A1) at (0,1) {$i$};
\node[node, blue] (B1) at (2,1) {$i+1$};
\node[node, red]  (C1) at (0,-1) {$i+2$};
\node[node, red]  (D1) at (2,-1) {$i+3$};
\draw[->] (A1) -- (D1);
\draw[->] (B1) -- (D1);
\draw[<->] (C1) -- (D1);
\draw[->] (A1) -- (B1);
\draw[->] (B1) -- (C1);
\node at (1,2.2) {\textbf{$t=1$}};
\end{scope}

\node at (9.3,0) {\textbf{or}};

\begin{scope}[xshift=10.75cm]
\node[node, blue] (A2) at (0,1)  {$i$};
\node[node, blue]  (B2) at (2,1) {$i+1$};
\node[node, red]  (C2) at (0,-1) {$i+2$};
\node[node, blue] (D2) at (2,-1) {$i+3$};
\draw[->] (A2) -- (D2);
\draw[->] (B2) -- (D2);
\draw[<->] (C2) -- (D2);
\draw[->] (A2) -- (B2);
\draw[->] (B2) -- (C2);
\node at (1,2.2) {\textbf{$t=1$}};
\end{scope}

\end{tikzpicture}
\caption{Voter Model: Initially, node $i$ is $+1$ (blue). Node $i+1$ will flip to $+1$ anyway as it adopts node $i$'s state. Node $i+3$ will be $+1$ with probability $w_{i,i+3}$.}
\label{Voter_model_outcomes}
\end{figure}
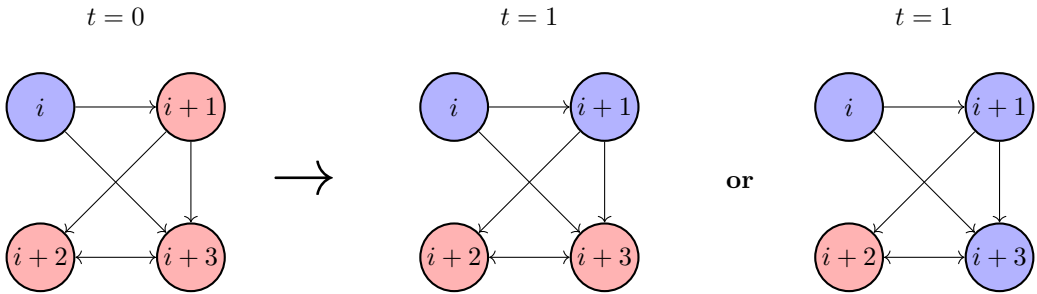

In the Voter model, the opinion space is $\mathcal{O} = \{-1,+1\}$. At each time step $t$, every node $v_i$ performs the following steps: node $v_i$ randomly adopts the state of one of its in-neighbors $v_j \in N_{\text{in}}(v_i)$ with probability proportional to $w_{ji}$, where $w_{ji}$ is the weight of the edge between $v_j$ and $v_i$. 
\[
s_{v_i}(t+1) = \Phi(G, \textbf{s}(t), v_i) = s_{v_j}(t),
\]
where $v_j$ is the in-neighbor selected by $v_i$ at time $t$. The process is synchronous, meaning all nodes update their states simultaneously at each time step $t$. However, the asynchronous version of this model is also studied extensively, mostly by the statistical physics community, see e.g.,~\cite{redner2019reality}. Figure \ref{Voter_model_outcomes} shows one round of updates in this model. In this graph, as well as in subsequent figures, blue nodes are $+1$, while red nodes are $-1$. This color scheme is used to improve clarity and avoid overloading figures with numbers. 

The Voter model gets its name from the intuitive way its rules mimic opinion dynamics. However, due to its simplicity, the model has been widely studied in areas far removed from social dynamics, such as population genetics (see e.g., \cite{castellano2009statistical}). Clifford and Sudbury~\cite{clifford1973model} first introduced Voter dynamics as a model for species competition, which was later named the ``Voter model'' in  \cite{holley1975ergodic}.

A model closely related to the Voter model is the \textbf{Moran process}~\cite{moran1958random}. In this process, while all nodes' states are $-1$, the \textit{resident state}, a new state $+1$, the \textit{mutant state}, is introduced. Then, nodes update their states by copying the state of a neighbor, with a preference determined by a key parameter called the \textit{mutant fitness advantage}, denoted by $\delta \geq 0$. This parameter controls how strongly state $+1$ propagates compared to $-1$ states: larger values of $\delta$ favor the spread of the $+1$, while $\delta = 0$ makes $+1$ and $-1$ equally likely to be copied. This process unfolds asynchronously.

\subsubsection{Majority Models}
From a sociophysics perspective, models based on repeated local-majority interactions within groups were first introduced by Galam~\cite{galam2002minority}.  Majority models, however, constitute a broad class that researchers from diverse communities across various frameworks and setups have extensively studied. The origins of such models can be traced back to graph-theoretical approaches documented as early as $1991$ in~\cite{agur1991fixed}, highlighting the interdisciplinary roots of this research area.

\paragraph{Galam Majority Model.} The Galam model describes how a population of individuals decides whether to accept or reject a reform proposal through repeated local majority-based interactions \footnote{However, the model is not limited to reforms but can be applied to any scenario involving a binary (or even multiple~\cite{galam2005heterogeneous,galam2013drastic}) decision-making process, such as predicting the result of an election~\cite{galam2021will} choosing between competing opinions, options, or products. 
}. The opinion space is binary, denoted by $\mathcal{O} = \{-1, +1\}$, where $-1$ represents opposition to the reform and $+1$ represents support. Initially, at time $t = 0$, each individual in the population of size $n$ is assigned an initial opinion, either $-1$ or $+1$. The proportions of supporters and opponents are given by $P^1(0) = \frac{n^1(0)}{n}$ and $P^{-1}(0) = \frac{n^{-1}(0)}{n}$, respectively, where $n^1(0)$ and $n^{-1}(0)$ are the number of supporters and opponents at $t = 0$. These probabilities satisfy the normalization condition $P^1(0) + P^{-1}(0) = 1$.

The update rule is as follows. At each time step, individuals interact in groups of size $k$, where $k$ can vary from $1$ to $L$ (the maximum group size). The probability of forming a group of size $k$ is $a_k$, with $\sum_{k=1}^L a_k = 1$ (one can think of this as all nodes being on a complete graph, but at each time, they interact with a fixed subset of their neighbors). Within each group, the local majority rule is applied: if the number of supporters ($j$) in a group of size $k$ satisfies $j > k/2$, the entire group adopts the reform ($+1$); if $j < k/2$, the group opposes the reform ($-1$); and if $j = k/2$, the tie results in opposition ($-1$) due to the status quo bias. The probabilities of supporting ($P^1$) and opposing ($P^{-1}$) the reform at time $t+1$ are updated as:
\[
P^1(t+1) = \sum_{k=1}^L a_k \sum_{j=\lceil k/2 + 1 \rceil}^k \binom{k}{j} (P^1(t))^j (P^{-1}(t))^{k-j}, \quad 
P^{-1}(t+1) = \sum_{k=1}^L a_k \sum_{j=0}^{\lfloor k/2 \rfloor} \binom{k}{j} (P^{-1}(t))^j (P^1(t))^{k-j},
\]
where $\lceil \cdot \rceil$ and $\lfloor \cdot \rfloor$ denote the ceiling and floor functions, respectively. The update rule is stochastic because the formation of groups and the distribution of opinions within groups are probabilistic, and the dynamics are synchronous because all individuals update their opinions simultaneously at each time step. The asynchronous version of the model has also been studied in~\cite{qian2015activeness}. 

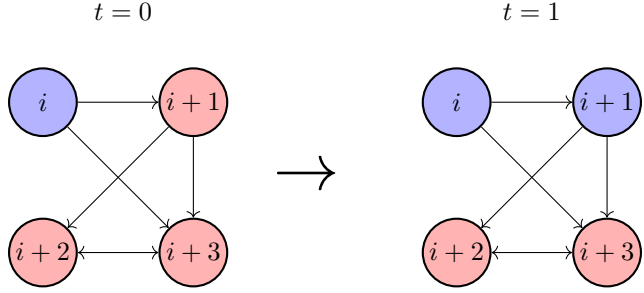
\begin{figure}[h]
\centering
\begin{tikzpicture}[
    node/.style={circle, draw, minimum size=9mm, inner sep=0pt, thick},
    red/.style={fill=red!30},
    blue/.style={fill=blue!30}
]

\node[node, blue] (A0) at (0,1) {$i$};
\node[node, red]  (B0) at (2,1) {$i+1$};
\node[node, red]  (C0) at (0,-1) {$i+2$};
\node[node, red]  (D0) at (2,-1) {$i+3$};
\draw[->] (A0) -- (D0);
\draw[->] (B0) -- (D0);
\draw[<->] (C0) -- (D0);
\draw[->] (A0) -- (B0);
\draw[->] (B0) -- (C0);
\node at (1,2.2) {\textbf{ $t=0$}};

\node at (3.5,0) {\Huge$\rightarrow$};

\begin{scope}[xshift=5.5cm]
\node[node, blue] (A1) at (0,1) {$i$}; 
\node[node, blue] (B1) at (2,1) {$i+1$}; 
\node[node, red]  (C1) at (0,-1) {$i+2$}; 
\node[node, red] (D1) at (2,-1) {$i+3$}; 
\draw[->] (A1) -- (D1);
\draw[->] (B1) -- (D1);
\draw[<->] (C1) -- (D1);
\draw[->] (A1) -- (B1);
\draw[->] (B1) -- (C1);
\node at (1,2.2) {\textbf{$t=1$}};
\end{scope}
\end{tikzpicture}
\caption{Majority Model: Each node updates its opinion based on the majority of its in-neighbors. Here, node $i+1$ changes from $-1$ (red) to $+1$ (blue).
}
\label{majority_model_update}
\end{figure}

\paragraph{Graph-based Majority Models. } In the graph-based majority models, the opinion space is $\mathcal{O} = \{ -1, +1 \}$ (sometimes, colors such as red/blue or black/white are used). Each node $v_i \in V$ has a state $s_{v_i}(t) \in \mathcal{O}$ at time $t$, representing its current opinion. At time $t = 0$, each node $v_i \in V$ is assigned an initial opinion $s_{v_i}(0) \in \mathcal{O}$.  At each time step $t$, the state of a node $v_i$ is updated based on the states of its in-neighbors. In the majority model, a node updates its state to match the majority of its in-neighbors' states. In case of a tie, the node retains its current state. Formally, the update function is defined as
\[
s_{v_i}(t+1)=\Phi(G, \mathbf{s}(t), v_i) =
\begin{cases} 
-1 & \text{if } | \{ v_j \in N_{\text{in}}(v_i) : s_{v_j}(t) = -1 \} | > | \{ v_j \in N_{\text{in}}(v_i) : s_{v_j}(t) = +1 \} |, \\
+1 & \text{if } | \{ {v_j} \in N_{\text{in}}(v_i) : s_{v_j}(t) = +1 \} | > | \{ {v_j} \in N_{\text{in}}(v_i) : s_{v_j}(t) = -1 \} |, \\
s_{v_i}(t) & \text{otherwise (in the case of a tie)}.
\end{cases}
\]
Figure \ref{majority_model_update} displays one round of updates in the majority model.

The update rule in the random majority model is the same as the majority model, except that in the case of a tie, the node randomly selects one of the two options ($-1$ or $+1$) with equal probability $\frac{1}{2}$. Formally, the update function for the random majority model is defined as
\[
\Phi(G, \mathbf{s}(t), v_i) =
\begin{cases} 
-1 & \text{if } | \{ {v_j} \in N_{\text{in}}(v_i) : s_{v_j}(t) = -1 \} | > | \{ {v_j} \in N_{\text{in}}(v_i) : s_{v_j}(t) = +1 \} |, \\
+1 & \text{if } | \{ {v_j} \in N_{\text{in}}(v_i) : s_{v_j}(t) = +1 \} | > | \{ {v_j} \in N_{\text{in}}(v_i) : s_{v_j}(t) = -1 \} |, \\
-1 \text{ or } +1 & \text{with probability } \frac{1}{2} \text{ each (in the case of a tie)}.
\end{cases}
\]

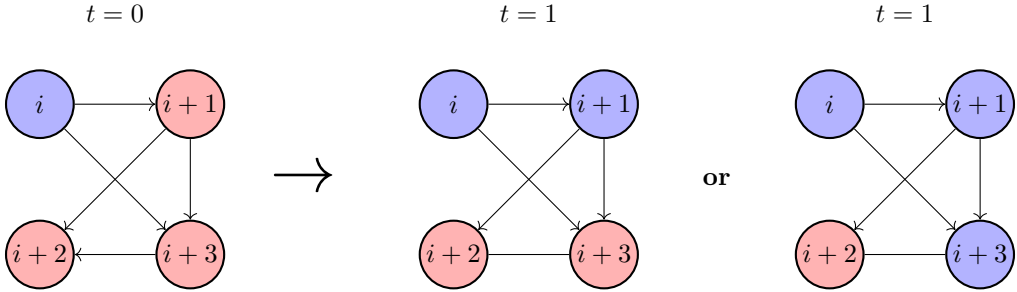
\begin{figure}[h]
\centering
\begin{tikzpicture}[
    node/.style={circle, draw, minimum size=9mm, inner sep=0pt, thick},
    red/.style={fill=red!30},
    blue/.style={fill=blue!30}
]

\node[node, blue] (A0) at (0,1) {$i$};
\node[node, red]  (B0) at (2,1) {$i+1$};
\node[node, red]  (C0) at (0,-1) {$i+2$};
\node[node, red]  (D0) at (2,-1) {$i+3$};
\draw[->] (A0) -- (D0);
\draw[->] (B0) -- (D0);
\draw[->] (D0) -- (C0); 
\draw[->] (A0) -- (B0);
\draw[->] (B0) -- (C0);
\node at (1,2.2) {\textbf{$t=0$}};

\node at (3.5,0) {\Huge$\rightarrow$};

\begin{scope}[xshift=5.5cm]
\node[node, blue] (A1) at (0,1) {$i$}; 
\node[node, blue] (B1) at (2,1) {$i+1$}; 
\node[node, red] (C1) at (0,-1) {$i+2$}; 
\node[node, red]  (D1) at (2,-1) {$i+3$}; 
\draw[->] (A1) -- (D1);
\draw[->] (B1) -- (D1);
\draw[-] (D1) -- (C1); 
\draw[->] (A1) -- (B1);
\draw[->] (B1) -- (C1);
\node at (1,2.2) {\textbf{$t=1$}};
\end{scope}

\node at (9,0) {\textbf{or}};

\begin{scope}[xshift=10.5cm]
\node[node, blue] (A2) at (0,1) {$i$}; 
\node[node, blue] (B2) at (2,1) {$i+1$}; 
\node[node, red]  (C2) at (0,-1) {$i+2$}; 
\node[node, blue]  (D2) at (2,-1) {$i+3$}; 
\draw[->] (A2) -- (D2);
\draw[->] (B2) -- (D2);
\draw[-] (D2) -- (C2); 
\draw[->] (A2) -- (B2);
\draw[->] (B2) -- (C2);
\node at (1,2.2) {\textbf{$t=1$}};
\end{scope}

\end{tikzpicture}
\caption{Random Majority Model: Each node updates based on the majority of its in-neighbors, with ties resolved randomly. Here, $i+1$ updates to blue ($+1$) from $i$ regardless. 
Two possible outcomes (with equal probability) are shown: $i+3$ remains red (\(-1\)) or changes to blue (\(+1\) ).
}
\label{random_majority_model}
\end{figure}

Figure \ref{random_majority_model} shows one round of updates in the random majority model.

While most studies consider the synchronous variant of the (random) majority model, e.g.,~\cite{zehmakan2023random}, the asynchronous version of majority models has also been studied, e.g.,~\cite{schoenebeck2018consensus, bahrani2019asynchronous}. In the asynchronous case, at each time step, an individual is selected uniformly at random to update their opinion, while the opinions of other nodes remain unchanged.

\subsubsection{Ising Model}

The Ising model~\cite{ising1925beitrag}, often considered the earliest model of opinion dynamics, was originally developed to explain ferromagnetism (see, e.g.,~\cite{brush1967history}), but has since become a fundamental framework for studying binary opinion formation. Given a weighted, undirected graph $G = (V, E, W)$, the opinion space is defined as $\mathcal{O} = \{ -1, +1 \}$, and the system evolves toward configurations that minimize a predefined energy function, typically the Hamiltonian, given by
\begin{equation*}
H(s) = -\beta \sum_{(v_i, v_j) \in E} w_{ij} s_{v_i} s_{v_j} - B \sum_{v_i \in V} s_{v_i},    
\end{equation*}
where $w_{ij}$ denotes the weight of the edge between $v_i$ and $v_j$, $\beta \geq 0$ controls the strength of alignment between connected nodes, and $B \in \mathbb{R}$ represents the influence of a uniform external field. The case $\beta > 0$ corresponds to the \textit{ferromagnetic} Ising model, encouraging agreement among neighbors, while $\beta < 0$ defines the \textit{antiferromagnetic} Ising model, encouraging disagreement. The system's equilibrium distribution is given by the Gibbs measure (which arises from maximizing entropy under fixed average energy, derived via the Boltzmann distribution in statistical mechanics, see e.g.,~\cite{bovier2006statistical} for a comprehensive reference)
\begin{equation*}
\mu\big(s\big) = \frac{1}{Z} \exp\big(-\beta H(s)\big), \quad 
Z = \sum_{s \in \{-1, +1\}^n} \exp\big(-\beta H(s)\big).
\end{equation*}
As $Z$ is computationally intractable for large systems, equilibrium properties are typically estimated via Markov Chain Monte Carlo (MCMC) methods, generally using the Gibbs sampler.  
Hence, the process is asynchronous and stochastic. 

The Ising model was created to explain how magnets work at the atomic level. In $1920$, physicist Wilhelm Lenz proposed that small atomic magnets (now known as ``spins'') can only point in two directions: up and down. This hypothesis contradicts earlier theories that allowed free rotation. His student, Ernst Ising, developed this idea further in $1925$. He added two essential rules: first, only neighboring spins affect each other; and second, spins prefer to align in the same direction. Ising solved this model for atoms arranged in a straight line, i.e., the one-dimensional case. While he did not find the expected magnetic behavior in this simple case, his model became fundamental for understanding more complex systems in higher dimensions, see~\cite {duminil2022100,mullick2025sociophysics} for historical accounts and standard references on the model. The original model was defined on regular lattices~\cite{duminil2022100}, but later has been extended to diverse network topologies, including scale-free networks~\cite{herrero2004ising}, random graphs~\cite{dommers2010ising}, locally tree-like graphs~\cite{dembo2010ising}, and general graphs~\cite{mossel2013exact}.

\subsubsection{Sznajd Model}

\begin{figure}[H]
\centering
\begin{tikzpicture}[
    node/.style={circle, draw, minimum size=9mm, inner sep=0pt, thick},
    opinion1/.style={fill=red!30},
    opinion2/.style={fill=blue!30},
    selected/.style={ultra thick, draw=green!80!black},
    arrow/.style={->, >=stealth, thick}
]

\begin{scope}[shift={(0,0)}]

\node[node, opinion1] (v1) at (0,2) {$i-1$};
\node[node, opinion1, selected] (v2) at (1.5,2) {$i$};
\node[node, opinion2] (v3) at (3,2) {$i+1$};
\node[node, opinion1] (v4) at (3,0) {$i+2$};
\node[node, opinion1] (v5) at (1.5,0) {};
\node[node, opinion2] (v6) at (0,0) {};

\draw[thick] (v1) -- (v2) -- (v3) -- (v4) -- (v5) -- (v6) -- (v1);

\node[above=2mm of v2] {Selected node};
\node at (1.5,3.5) {t=0};

\begin{scope}[xshift=5cm]
    \node[node, opinion2] (v1) at (0,2) {$i-1$};
    \node[node, opinion1, selected] (v2) at (1.5,2) {$i$};
    \node[node, opinion2] (v3) at (3,2) {$i+1$};
    \node[node, opinion1] (v4) at (3,0) {$i+2$};
    \node[node, opinion1] (v5) at (1.5,0) {};
    \node[node, opinion2] (v6) at (0,0) {};

    \draw[thick] (v1) -- (v2) -- (v3) -- (v4) -- (v5) -- (v6) -- (v1);

    \node at (1.5,3.5) {t=1};
\end{scope}

\draw[arrow] (3.5,1) -- (4.5,1) node[midway, above] {};

\end{scope}
\end{tikzpicture}
\caption{Sznajd Model: Node $i$ disagrees with neighbor $i+1$, causing $i-1$ to flip its opinion. Node $i+2$ already disagrees with $i+1$ and remains unchanged. 
}
\label{sznajd}
\end{figure}
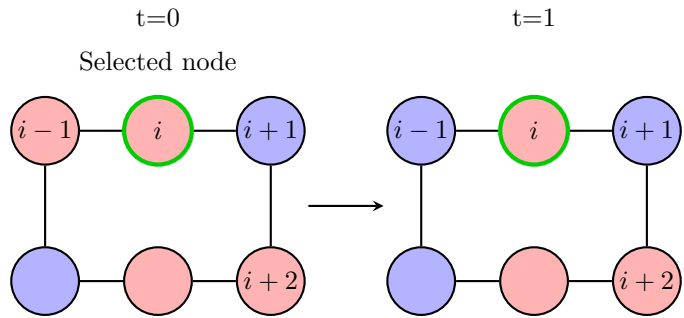

This model, introduced in 2000 by Sznajd-Weron and Sznajd~\cite{sznajd2000opinion}, is inspired by the Ising model. In this model, individuals are arranged on a one-dimensional loop, which can be visualized as a cycle graph $C_n=v_1,v_2,\cdots,v_n,v_1$ (that is, $G$ is always set to $C_n$). The opinion space is $\mathcal{O} = \{-1, +1\}$ representing two opposing states. The process unfolds as follows. A node $v_i$ is selected uniformly at random from the network. The opinion of $v_i$, denoted as $s_{v_i}$, and its neighbor $v_{i+1}$ is examined. In the case of agreement, i.e., $s_{v_i}(t) = s_{v_{i+1}}(t)$, they influence their neighbors $v_{i-1}$ and $v_{i+2}$. Specifically
\begin{equation}
s_{v_{i-1}}(t+1) = s_{v_i}(t), \quad s_{v_{i+2}}(t+1) = s_{v_{i+1}}(t).
\end{equation}
This means that the shared opinion is imposed on the neighbors, reinforcing the local consensus. In the case of disagreement, i.e., $s_{v_i}(t) \neq s_{v_{i+1}}(t)$, their neighbors $v_{i-1}$ and $v_{i+2}$ adopt the opposite opinions of their neighbors. Specifically
\begin{equation}
s_{v_{i-1}}(t+1) =-s_{v_i}(t), \quad s_{v_{i+2}}(t+1) =-s_{v_{i+1}}(t).
\end{equation}
This reflects a situation in which disagreement leads to divergent opinions among neighbors. A simple visualization of the model is shown in Figure \ref{sznajd}. The model updates asynchronously, where only one node and its neighbors are updated at each step. (However, a synchronous version has also been studied by Stauffer~\cite{stauffer2004difficulty}.) The updated node is chosen randomly, making the process stochastic.

Initially, the Sznajd model was introduced on a one-dimensional lattice (a cycle $C_n$). However, as such a structure is not a realistic representation of social systems, the model has been extended to various topologies, including the square lattice~\cite{stauffer2000generalization}, the randomly diluted square lattice~\cite{moreira2001sznajd}, the triangular lattice~\cite{chang2001sznajd}, and Barab\'{a}si-Albert networks~\cite{bonnekoh2003monte}.

\subsubsection{PUSH-PULL Protocols}
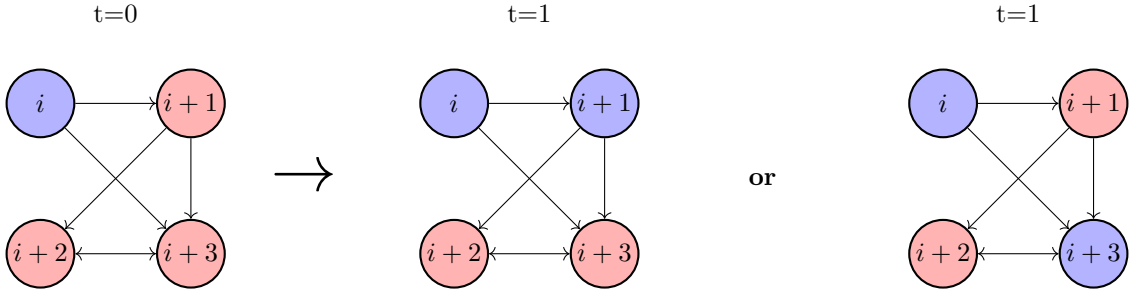
\begin{figure}[h]
\centering
\begin{tikzpicture}[
    node/.style={circle, draw, minimum size=9mm, inner sep=0pt, thick},
    red/.style={fill=red!30},
    blue/.style={fill=blue!30}
]
\node[node, blue] (A0) at (0,1) {$i$};
\node[node, red]  (B0) at (2,1) {$i+1$};
\node[node, red]  (C0) at (0,-1) {$i+2$};
\node[node, red]  (D0) at (2,-1) {$i+3$};
\draw[->] (A0) -- (D0);
\draw[->] (B0) -- (D0);
\draw[<->] (C0) -- (D0);
\draw[->] (A0) -- (B0);
\draw[->] (B0) -- (C0);
\node at (1,2.2) {t=0};

\node at (3.5,0) {\Huge$\rightarrow$};

\begin{scope}[xshift=5.5cm]
\node[node, blue] (A1) at (0,1) {$i$};
\node[node, blue] (B1) at (2,1) {$i+1$};
\node[node, red]  (C1) at (0,-1) {$i+2$};
\node[node, red]  (D1) at (2,-1) {$i+3$};
\draw[->] (A1) -- (D1);
\draw[->] (B1) -- (D1);
\draw[<->] (C1) -- (D1);
\draw[->] (A1) -- (B1);
\draw[->] (B1) -- (C1);
\node at (1,2.2) {t=1};
\end{scope}

\node at (9.6,0) {\textbf{or}};

\begin{scope}[xshift=12cm]
\node[node, blue] (A2) at (0,1)  {$i$};
\node[node, red]  (B2) at (2,1) {$i+1$};
\node[node, red]  (C2) at (0,-1) {$i+2$};
\node[node, blue] (D2) at (2,-1) {$i+3$};
\draw[->] (A2) -- (D2);
\draw[->] (B2) -- (D2);
\draw[<->] (C2) -- (D2);
\draw[->] (A2) -- (B2);
\draw[->] (B2) -- (C2);
\node at (1,2.2) {t=1};
\end{scope}
\end{tikzpicture}
\caption{PUSH Model: Node $i$ is initially blue ($+1$) and randomly pushes its opinion to one of its out-neighbors, node $i+1$ and $i+3$.  Two possible outcomes (with equal probability) are shown after one round.}
\label{push_model}
\end{figure}

In these models, the opinion space is binary given as $\mathcal{O} = \{-1,+1\}$, where $-1$ represents an \textit{uninformed} node (does not know the rumor) and $+1$ represents an \textit{informed} node (knows the rumor). The state of a node $v_i$ at time $t$ is denoted by $s_{v_i}(t) \in \mathcal{O}$. At time $t=0$, an arbitrary node $v_0 \in V$ holds the rumor, i.e., 
\[ s_{v_0}(0) = +1, \quad \text{and} \quad s_v(0) = -1 \text{ for all } v \neq v_0. \]
The update function $\Phi$ depends on the protocol used (PUSH, PULL, or PUSH-PULL). Considering the PUSH protocol, in each round, every informed node $v_i$ (i.e., $s_{v_i}(t) = +1$) sends the rumor to a randomly chosen out-neighbor $v_j$, and each uninformed node switches from $-1$ to $+1$ if it receives the rumor from at least one in-neighbor. The update function can be written as
\[
\Phi_{\text{PUSH}}(G, s(t), v_j) =
\begin{cases}
    +1, & \text{if } s_{v_j}(t) = -1 \text{ and } \exists v_i \in N_{\text{in}}(v_j) \text{ such that } \\
         & \quad s_{v_i}(t) = +1,\ \text{ and } \text{RandOutNeighbor}(v_i) = v_j, \\
    s_{v_j}(t), & \text{otherwise},
\end{cases}
\]
where $\text{RandOutNeighbor}(v_i)$ refers to the random out-neighbor selected by node $v_i$ in round $t$. Figure \ref{push_model} shows one round of updates with the PUSH protocol.

\begin{figure}[h]
\centering
\begin{tikzpicture}[
    node/.style={circle, draw, minimum size=9mm, inner sep=0pt, thick},
    red/.style={fill=red!30},
    blue/.style={fill=blue!30}
]

\node[node, blue] (A0) at (0,1) {$i$};
\node[node, red]  (B0) at (2,1) {$i+1$};
\node[node, red]  (C0) at (0,-1) {$i+2$};
\node[node, red]  (D0) at (2,-1) {$i+3$};
\draw[->] (A0) -- (D0);
\draw[->] (B0) -- (D0);
\draw[<->] (C0) -- (D0);
\draw[->] (A0) -- (B0);
\draw[->] (B0) -- (C0);
\node at (1,2.2) {t=0};

\node at (3.5,0) {\Huge$\rightarrow$};

\begin{scope}[xshift=5.5cm]
\node[node, blue] (A1) at (0,1) {$i$};
\node[node, blue] (B1) at (2,1) {$i+1$};
\node[node, red]  (C1) at (0,-1) {$i+2$};
\node[node, red]  (D1) at (2,-1) {$i+3$};
\draw[->] (A1) -- (D1);
\draw[->] (B1) -- (D1);
\draw[<->] (C1) -- (D1);
\draw[->] (A1) -- (B1);
\draw[->] (B1) -- (C1);
\node at (1,2.2) {t=1};
\end{scope}

\node at (9.5,0) {\textbf{or}};

\begin{scope}[xshift=11.5cm]
\node[node, blue] (A2) at (0,1) {$i$};
\node[node, blue] (B2) at (2,1) {$i+1$};
\node[node, red]  (C2) at (0,-1) {$i+2$};
\node[node, blue]  (D2) at (2,-1) {$i+3$};
\draw[->] (A2) -- (D2);
\draw[->] (B2) -- (D2);
\draw[<->] (C2) -- (D2);
\draw[->] (A2) -- (B2);
\draw[->] (B2) -- (C2);
\node at (1,2.2) {t=1};
\end{scope}

\end{tikzpicture}
\caption{PULL Model: Node $i$ is initially blue ($+1$). Every $-1$ node requests the rumor from a randomly chosen in-neighbor. Node $i+1$ will be $+1$ and node $i+3$ will be informed with probability $1/3$.
}
\label{pull_protocal}
\end{figure}
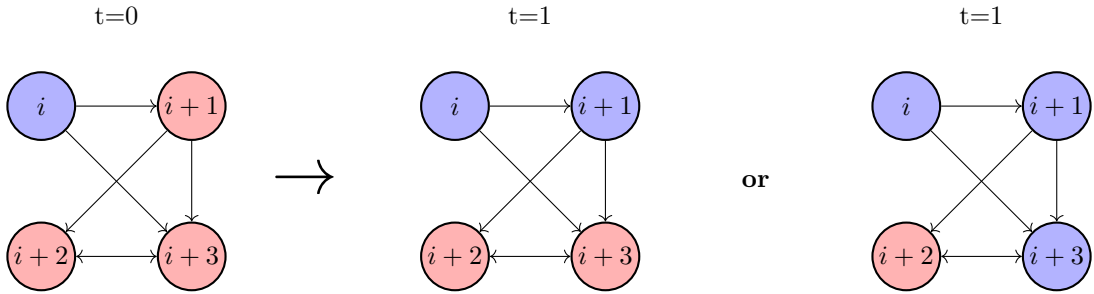

Considering the PULL protocol, in each round, every uninformed node $v_j$ (i.e., $s_{v_j}(t) = -1$) requests the rumor from a randomly chosen in-neighbor $v_i$. If $v_i$ knows the rumor (i.e., $s_{v_i}(t) = 1$), then $v_j$ becomes informed, i.e., $s_{v_j}(t+1) = 1$. \
The update function can be written as
\[
\Phi_{\text{PULL}}(G, s(t), v_j) =
\begin{cases}
    +1, & \text{if } s_{v_j}(t) = -1 \text{ and } \exists v_i \in N_{\text{in}}(v_j) \text{ such that } \\
    & \quad s_{v_i}(t) = +1 \text{ and } \text{RandInNeighbor}(v_j) = v_i,, \\
    s_{v_j}(t), & \text{otherwise}.
\end{cases}
\]
Figure \ref{pull_protocal} shows one round of updates with the PULL protocol.

Finally, in the PUSH-PULL protocol, in each round every informed node $v_i$ (i.e., $s_{v_i}(t) =+ 1$) sends the rumor to a randomly chosen out-neighbor (PUSH) \textbf{and} also every uninformed node $v_j$ (i.e., $s_{v_j}(t) = -1$) requests the rumor from a randomly chosen in-neighbor (PULL), i.e., PULL and PUSH protocols run simultaneously. 
Figure~\ref{pull_protocal} also shows one round of updates with the PUSH-PULL protocol.

These protocols have been primarily studied under the assumption of \textit{synchronous} updates. However, the impact of asynchronous updates has also been investigated by  Giakkoupis et al.~\cite{giakkoupis2016asynchrony} and Panagiotou and Speidel~\cite{panagiotou2017asynchronous}. These protocols were first introduced by Frieze
and Grimmett~\cite{frieze1985shortest} for the solving of the shortest-path problem, and then later were used by Demers et al.~\cite{demers1987epidemic} in the context of replicated databases as a solution to the problem of distributing updates and ensuring replica consistency. Since then, it has been applied in various domains, including failure detection in distributed systems~\cite{van1998gossip}, peer sampling~\cite{jelasity2007gossip}, adaptive machine discovery~\cite{harchol1999resource}, and distributed averaging in sensor networks~\cite{boyd2005gossip}.

\subsubsection{Bootstrap Percolation Model}

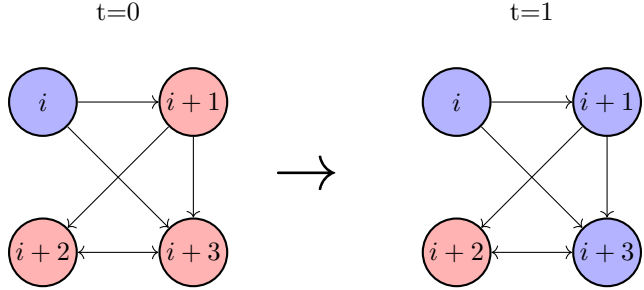
\begin{figure}[h]
\centering
\begin{tikzpicture}[
    node/.style={circle, draw, inner sep=0pt, minimum size=9mm, thick},
    red/.style={fill=red!30},
    blue/.style={fill=blue!30}
]

\node[node, blue] (A0) at (0,1) {$i$};
\node[node, red]  (B0) at (2,1) {$i+1$};
\node[node, red]  (C0) at (0,-1) {$i+2$};
\node[node, red]  (D0) at (2,-1) {$i+3$};
\draw[->] (A0) -- (D0);
\draw[->] (B0) -- (D0);
\draw[<->] (C0) -- (D0);
\draw[->] (A0) -- (B0);
\draw[->] (B0) -- (C0);
\node at (1,2.2) {t=0};

\node at (3.5,0) {\Huge$\rightarrow$};

\begin{scope}[xshift=5.5cm]
\node[node, blue] (A3) at (0,1) {$i$};
\node[node, blue] (B3) at (2,1) {$i+1$};
\node[node, red] (C3) at (0,-1) {$i+2$}; 
\node[node, blue] (D3) at (2,-1){$i+3$}; 
\draw[->] (A3) -- (D3);
\draw[->] (B3) -- (D3);
\draw[<->] (C3) -- (D3);
\draw[->] (A3) -- (B3);
\draw[->] (B3) -- (C3);
\node at (1,2.2) {t=1};
\end{scope}

\end{tikzpicture}
\caption{Bootstrap Percolation Model with $r=1$. If the process is two-way, then node $i$ will switch to $-1$. 
}
\label{r_bp_two}
\end{figure}

In the $r$-bootstrap percolation ($r$-BP) model, the opinion space is binary, denoted by $\mathcal{O} = \{-1,+1\}$, where $s_{v_i}(t) = +1$ indicates that node $v_i$ is active at time $t$, and $s_{v_i}(t) = -1$ demonstrates that it is inactive. Here, $r$ represents the activation threshold. A commonly studied setup involves assigning initial opinions randomly: each node $v_i \in V$ is independently set to $s_{v_i}(0) = 1$ with probability $p$ and $s_{v_i}(0) = -1$ with probability $1-p$. At each step, a node whose state is $-1$ and has at least $r$ in-neighbors with state $+1$ will switch to $+1$. Depending on whether a node remains active ($+1$) once activated or can switch back to inactive ($-1$), the model has two variants. In the $r$-BP, once a node becomes $+1$, it remains $+1$ forever. The update rule is given by
\[
\Phi(G, \textbf{s}(t), v_i) =
\begin{cases}
+1, & \text{if } s_{v_i}(t) =+ 1 \text{ or } \sum_{v_j \in N_{\text{in}}(v_i)} s_{v_j}^+(t) \geq r, \\
-1, & \text{otherwise}.
\end{cases}
\]
where $s_{v_j}^+(t):=\max(0,s_{v_j}(t))$. This model simulates processes where activation is \textit{irreversible}. The second variant of $r$-BP is \textit{two-way $ r$-BP}, in which nodes can switch between $+1$ and $-1$ states based on the states of their in-neighbors. The update rule is given as
\[
\Phi(G, \textbf{s}(t), v_i) =
\begin{cases}
+1, & \text{if } \sum_{v_j \in N_{\text{in}}(v_i)} s_{v_j}^+(t) \geq r, \\
-1, & \text{otherwise}.
\end{cases}
\]
Figure \ref{r_bp_two} shows one round of updates with $1$-BP. 

Bootstrap percolation\footnote{Roughly speaking, percolation is the process where connections grow in a system until a path forms that allows something to flow through it.} was introduced by Chalupa et al.~\cite{chalupa1979bootstrap} in 1979 in the context of magnetic disordered systems and has been rediscovered since then by several authors, mainly due to its connections with various physical models. Bootstrap percolation helps explain how systems transition from disconnected to connected configurations, with applications in magnetic materials, brain activity, network dynamics, and natural landscapes. We refer the interested reader to~\cite{saberi2015recent}, which provides a comprehensive review of the (physical) applications of this theory. 

\subsubsection{Linear Threshold Model}

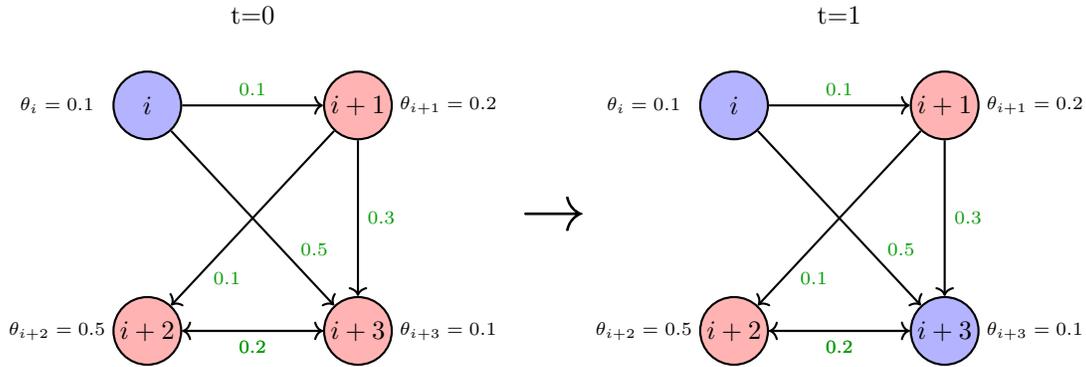
\begin{figure}[h]
\centering
\begin{tikzpicture}[
    node/.style={circle, draw, minimum size=9mm, inner sep=0pt, thick},
    red/.style={fill=red!30},
    blue/.style={fill=blue!30},
    every edge/.style={draw,->,thick},
    weight/.style={font=\scriptsize\color{green!60!black}}
]
\node[node, blue] (A0) at (0,1.5) {$i$};
\node[node, red]  (B0) at (2.8,1.5) {$i+1$};
\node[node, red]  (C0) at (0,-1.5) {$i+2$};
\node[node, red]  (D0) at (2.8,-1.5) {$i+3$};
\path (A0) edge node[weight, above] {0.1} (B0);
\path (A0) edge node[weight, pos=0.8, above right, xshift=-4pt, yshift=2pt] {0.5} (D0);
\path (B0) edge node[weight, right] {0.3} (D0);
\path (B0) edge node[weight, pos=0.8, right, yshift=-3pt] {0.1} (C0);
\path (C0) edge node[weight, below] {0.2} (D0);
\path (D0) edge node[weight, below] {0.2} (C0);
\node at (1.4,2.7) {t=0};
\node at (-1.2,1.5) {\scriptsize $\theta_{i} = 0.1$};
\node at (4.0,1.5) {\scriptsize $\theta_{i+1} = 0.2$};
\node at (-1.2,-1.5) {\scriptsize $\theta_{i+2} = 0.5$};
\node at (4.0,-1.5) {\scriptsize $\theta_{i+3} = 0.1$};

\node at (5.4,0) {\Huge$\rightarrow$};

\begin{scope}[xshift=7.8cm]
\node[node, blue] (A1) at (0,1.5) {$i$};
\node[node, red] (B1) at (2.8,1.5) {$i+1$};
\node[node, red]  (C1) at (0,-1.5) {$i+2$};
\node[node, blue] (D1) at (2.8,-1.5) {$i+3$};
\path (A1) edge node[weight, above] {0.1} (B1);
\path (A1) edge node[weight, pos=0.8, above right, xshift=-4pt, yshift=2pt] {0.5} (D1);
\path (B1) edge node[weight, right] {0.3} (D1);
\path (B1) edge node[weight, pos=0.8, right, yshift=-3pt] {0.1} (C1);
\path (C1) edge node[weight, below] {0.2} (D1);
\path (D1) edge node[weight, below] {0.2} (C1);
\node at (1.4,2.7) {t=1};
\node at (-1.2,1.5) {\scriptsize $\theta_{i} = 0.1$};
\node at (4.0,1.5) {\scriptsize $\theta_{i+1} = 0.2$};
\node at (-1.2,-1.5) {\scriptsize $\theta_{i+2} = 0.5$};
\node at (4.0,-1.5) {\scriptsize $\theta_{i+3} = 0.1$};
\end{scope}
\end{tikzpicture}
\caption{Linear Threshold Model: One update of the model given thresholds and edge weights labeled in green. 
}
\label{lt_model}
\end{figure}

The Linear Threshold (LT) model \cite{kempe2003maximizing}  describes how individuals in a social network transition from a $-1$ to a $+1$ state based on the influence of their in-neighbors. The opinion space is binary: $\mathcal{O} = \{-1,+1\}$, where $-1$ stands for inactive (has not adopted the innovation) and $+1$ denotes active (has adopted the innovation). The model is inspired by the idea that a user becomes $+1$ if a sufficient number of its in-neighbors are $+1$ (somewhat similar to the Bootstrap Percolation model). 
At $t=0$, an initial set of nodes $A_0$ is in state $+1$, and all other nodes are $-1$. Each node $v_i$ is influenced by its in-neighbors according to weights $w_{ji}$, where $v_j \in N_{\text{in}}(v_i)$ and 
\[
\sum_{v_j \in N_{\text{in}}(v_i)} w_{ji} \leq 1.
\]
Each node $v_i$ selects a threshold $\theta_{v_i}$ uniformly at random from the interval $[0,1]$. 
At each time step $t$, the state of node $v_i$ is updated as
\[
\Phi(G, \textbf{s}(t), v_i) = 
\begin{cases} 
+1, & \text{if } s_{v_i}(t) = + 1 \text{ or } \sum_{v_j \in N_{\text{in}}(v_i)} s_{v_j}^+(t) w_{ji} \geq \theta_{v_i}, \\ 
-1, & \text{otherwise}.
\end{cases}
\]
where $s_{v_j}^+(t):=\max(0,s_{v_j}(t))$. Figure \ref{lt_model} shows one round of updates in this model. The update rule itself is deterministic given the fixed thresholds, but the overall process is stochastic as thresholds are chosen uniformly at random from $[0,1]$. The dynamics are synchronous because all nodes update their states simultaneously at each time step. However, an asynchronous variant, where only one agent updates at a time, can also be considered.

It is worth noting that there are also anti-threshold models~\cite{aghaeeyan2023discrete,ramazi2017asynchronous} (also known as anti-coordination models or nonconformist models), which operate in the opposite direction to the linear threshold model. In these models, a node adopts state $-1$ or $+1$ only if fewer than a given threshold of its in-neighbors have already adopted that state. These models are particularly relevant in scenarios such as voluntary participation in collective actions, traffic flow dynamics, and situations where diversity of choices is encouraged.

\subsubsection{Independent Cascade Model}
The independent cascade (IC) model~\cite{kempe2003maximizing} is a simple and widely used framework for studying diffusion processes, particularly in marketing contexts. It describes how influence or information propagates through a network, starting from an initial set of active nodes. The opinion space is binary: given as $\mathcal{O} = \{-1, +1\}$, where $-1$ and $+1$ correspond to active and inactive. An initial set of nodes $A_0$ is in state $+1$, and all other nodes are $-1$.  The process unfolds in discrete steps according to the following rules. At each time step $t$, every node $u$ that was \emph{newly} activated at time $t$ (i.e., $u \in A_t \setminus A_{t-1}$) gets \emph{one chance} to activate each of its currently inactive out-neighbors $v$. Each activation attempt succeeds with probability $w_{uv}$, independently of other attempts; if successful, $v$ becomes $+1$ (or active) at time $t+1$; if unsuccessful, $u$ cannot attempt to activate $v$ again in subsequent rounds. Formally, for each node $v_i \in V$, the state update rule is
\[
\Phi(G, \textbf{s}(t), v_i) = \begin{cases}
+1 & \text{if } s_{v_i}(t) = +1 \text{ or } \exists u \in N_{\text{in}}(v_i) \cap (A_t \setminus A_{t-1}) \\
   & \text{that successfully activates } v_i \text{ (with probability } w_{uv_i}), \\
-1 & \text{otherwise},
\end{cases}
\]

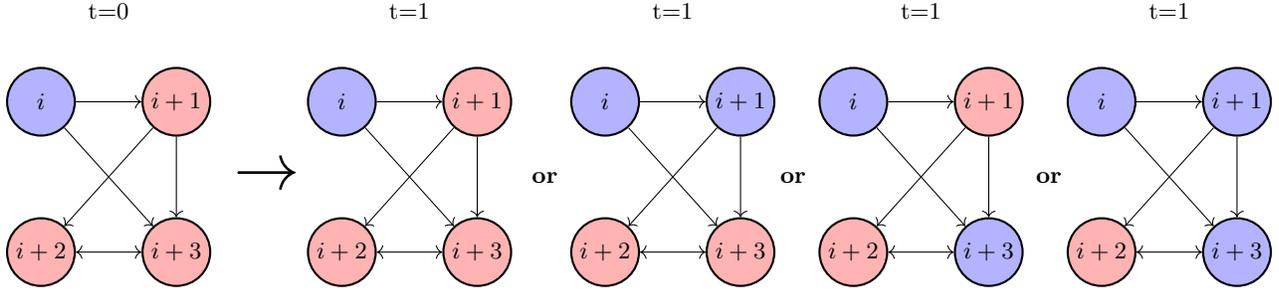
\begin{figure}[h]\small
\centering
\begin{tikzpicture}[
    node/.style={circle, draw, minimum size=9mm, inner sep=0pt, thick},
    red/.style={fill=red!30},
    blue/.style={fill=blue!30}
]
\node[node, blue] (A0) at (0,1) {$i$};
\node[node, red]  (B0) at (1.8,1) {$i+1$};
\node[node, red]  (C0) at (0,-1) {$i+2$};
\node[node, red]  (D0) at (1.8,-1) {$i+3$};
\draw[->] (A0) -- (D0);
\draw[->] (B0) -- (D0);
\draw[<->] (C0) -- (D0);
\draw[->] (A0) -- (B0);
\draw[->] (B0) -- (C0);
\node at (0.9,2.2) {t=0};

\node at (3,0) {\Huge$\rightarrow$};

\node[node, blue] (A0b) at (4.0,1) {$i$};
\node[node, red]  (B0b) at (5.8,1) {$i+1$};
\node[node, red]  (C0b) at (4.0,-1) {$i+2$};
\node[node, red]  (D0b) at (5.8,-1) {$i+3$};
\draw[->] (A0b) -- (D0b);
\draw[->] (B0b) -- (D0b);
\draw[<->] (C0b) -- (D0b);
\draw[->] (A0b) -- (B0b);
\draw[->] (B0b) -- (C0b);
\node at (4.9,2.2) {t=1};

\node at (6.7,0) {\textbf{or}};

\begin{scope}[xshift=7.5cm]
\node[node, blue] (A1) at (0,1) {$i$};
\node[node, blue] (B1) at (1.8,1) {$i+1$}; 
\node[node, red]  (C1) at (0,-1) {$i+2$};
\node[node, red]  (D1) at (1.8,-1) {$i+3$};
\draw[->] (A1) -- (D1);
\draw[->] (B1) -- (D1);
\draw[<->] (C1) -- (D1);
\draw[->] (A1) -- (B1);
\draw[->] (B1) -- (C1);
\node at (0.9,2.2) {t=1};
\end{scope}

\node at (10,0) {\textbf{or}};

\begin{scope}[xshift=10.8cm]
\node[node, blue] (A2) at (0,1) {$i$};
\node[node, red]  (B2) at (1.8,1) {$i+1$};
\node[node, red]  (C2) at (0,-1) {$i+2$};
\node[node, blue] (D2) at (1.8,-1) {$i+3$}; 
\draw[->] (A2) -- (D2);
\draw[->] (B2) -- (D2);
\draw[<->] (C2) -- (D2);
\draw[->] (A2) -- (B2);
\draw[->] (B2) -- (C2);
\node at (0.9,2.2) {t=1};
\end{scope}

\node at (13.4,0) {\textbf{or}};
\begin{scope}[xshift=14.1cm] 
\node[node, blue] (A2) at (0,1) {$i$};
\node[node, blue]  (B2) at (1.8,1) {$i+1$};
\node[node, red]  (C2) at (0,-1) {$i+2$};
\node[node, blue] (D2) at (1.8,-1) {$i+3$}; 
\draw[->] (A2) -- (D2);
\draw[->] (B2) -- (D2);
\draw[<->] (C2) -- (D2);
\draw[->] (A2) -- (B2);
\draw[->] (B2) -- (C2);
\node at (0.9,2.2) {t=1};
\end{scope}

\end{tikzpicture}
\caption{Independent Cascade Model: Initially, only node $i$ is $+1$ which attempts to activate nodes $i+1$ and $i+3$, with probability proportional to the edges, independently.
}
\label{independent_cascade}
\end{figure}

Figure \ref{independent_cascade} shows one round of updates in this model.
The update rule is stochastic because activation attempts are probabilistic, with success probability proportional to the edge weights. The dynamics are synchronous because all nodes update their states simultaneously at each time step.

\subsubsection{Epidemiological Processes}

Research on epidemic dynamics can be traced back to $1927$~\cite{kermack1927contribution}, aiming to model the spread of diseases within a fixed population using differential equations, focusing on the evolution of infection rates over time. It is worth mentioning that these processes also have been studied under the name of Reed-Frost (which itself is a modification of Soper's model~\cite{soper1929interpretation}), e.g.,~\cite{abbey1952examination}. While these models initially didn't consider a network structure, they can be adapted to networked settings, where individuals (nodes) interact through predefined connections (edges). Here, we review some well-known epidemic-based models.  While these models were initially designed to simulate epidemic dynamics, they are also regularly used to study the spread of information and the diffusion of opinions in social networks.

\paragraph{The Susceptible-Infected (SI) Model.}
The SI model describes the spread of a rumor or infection in a population where each individual can be in one of two states: susceptible ($S$) or infected ($I$). The opinion space is defined as $\mathcal{O} = \{-1, +1\}$, where $-1$ represents the susceptible state and $+1$ represents the infected state. Initially, one node $v_0 \in V$ is infected (patient zero), and all other nodes are susceptible. At each time step, a node $v_j$ is selected uniformly at random from the infected population. This node then chooses one of its out-neighbors $v_i \in N_{\text{out}}(v_j)$ uniformly at random. If $v_i$ is susceptible, it becomes infected. Once a node becomes infected, it remains infected permanently. 

The update rule is \textit{stochastic} because the selection of the infected node $v_i$ and its out-neighbor $v_j$ is random. The process is \textit{asynchronous}, meaning that in every round, one node is updated. \textit{It is worth mentioning that the process is equivalent to the asynchronous PUSH protocol}. An example of the update visualization is similar to the PUSH protocol in Figure \ref{push_model}.

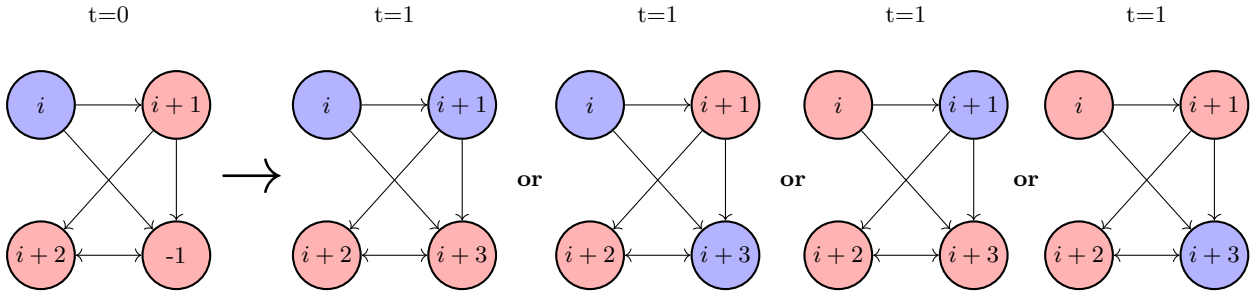
\begin{figure}[h]\small
\centering
\begin{tikzpicture}[
    node/.style={circle, draw, minimum size=9mm, inner sep=0pt, thick},
    red/.style={fill=red!30},
    blue/.style={fill=blue!30}
]
\node[node, blue] (A0) at (0,1) {$i$};
\node[node, red]  (B0) at (1.8,1) {$i+1$};
\node[node, red]  (C0) at (0,-1) {$i+2$};
\node[node, red]  (D0) at (1.8,-1) {-1};
\draw[->] (A0) -- (D0);
\draw[->] (B0) -- (D0);
\draw[<->] (C0) -- (D0);
\draw[->] (A0) -- (B0);
\draw[->] (B0) -- (C0);
\node at (0.9,2.2) {t=0};

\node at (2.8,0) {\Huge$\rightarrow$};
\begin{scope}[xshift=3.8cm]
\node[node, blue] (A1) at (0,1) {$i$};
\node[node, blue] (B1) at (1.8,1) {$i+1$};
\node[node, red]  (C1) at (0,-1) {$i+2$};
\node[node, red]  (D1) at (1.8,-1) {$i+3$};
\draw[->] (A1) -- (D1);
\draw[->] (B1) -- (D1);
\draw[<->] (C1) -- (D1);
\draw[->] (A1) -- (B1);
\draw[->] (B1) -- (C1);
\node at (0.9,2.2) {t=1};
\end{scope}
\node at (6.5,0) {\textbf{or}};
\begin{scope}[xshift=7.3cm]
\node[node, blue] (A2) at (0,1) {$i$};
\node[node, red]  (B2) at (1.8,1) {$i+1$};
\node[node, red]  (C2) at (0,-1) {$i+2$};
\node[node, blue] (D2) at (1.8,-1) {$i+3$};
\draw[->] (A2) -- (D2);
\draw[->] (B2) -- (D2);
\draw[<->] (C2) -- (D2);
\draw[->] (A2) -- (B2);
\draw[->] (B2) -- (C2);
\node at (0.9,2.2) {t=1};
\end{scope}
\node at (10,0) {\textbf{or}};
\begin{scope}[xshift=10.6cm]
\node[node, red] (A2) at (0,1) {$i$};
\node[node, blue]  (B2) at (1.8,1) {$i+1$};
\node[node, red]  (C2) at (0,-1) {$i+2$};
\node[node, red] (D2) at (1.8,-1) {$i+3$};
\draw[->] (A2) -- (D2);
\draw[->] (B2) -- (D2);
\draw[<->] (C2) -- (D2);
\draw[->] (A2) -- (B2);
\draw[->] (B2) -- (C2);
\node at (0.9,2.2) {t=1};
\end{scope}
\node at (13.1,0) {\textbf{or}};
\begin{scope}[xshift=13.8cm]
\node[node, red] (A2) at (0,1) {$i$};
\node[node, red]  (B2) at (1.8,1) {$i+1$};
\node[node, red]  (C2) at (0,-1) {$i+2$};
\node[node, blue] (D2) at (1.8,-1) {$i+3$};
\draw[->] (A2) -- (D2);
\draw[->] (B2) -- (D2);
\draw[<->] (C2) -- (D2);
\draw[->] (A2) -- (B2);
\draw[->] (B2) -- (C2);
\node at (0.9,2.2) {t=1};
\end{scope}
\end{tikzpicture}
\caption{Susceptible-Infected-Susceptible (SIS) Model: Four possible outcomes after one time step. The last two outputs happen with probability $\beta$.
}
\label{SIS}
\end{figure}

\paragraph{The Susceptible-Infected-Susceptible (SIS) Model.} The SIS model extends the SI model by allowing infected nodes to recover and become susceptible again. Hence, the opinion space is $\mathcal{O} = \{-1, +1\}$, the same as the SI model, and the process is similar to the SI model except that every infected node transitions back to the susceptible state in each round with probability $\beta$. 
Figure \ref{SIS} displays one round of updates in this model.

\begin{figure}[h]\small
\centering
\begin{tikzpicture}[
    node/.style={circle, draw, minimum size=9mm, inner sep=0pt, thick},
    red/.style={fill=red!30},
    blue/.style={fill=blue!30},
    green/.style={fill=green!30}
]
\node[node, blue] (A0) at (0,1) {$i$};
\node[node, red]  (B0) at (1.8,1) {$i+1$};
\node[node, red]  (C0) at (0,-1) {$i+2$};
\node[node, red]  (D0) at (1.8,-1) {-1};
\draw[->] (A0) -- (D0);
\draw[->] (B0) -- (D0);
\draw[<->] (C0) -- (D0);
\draw[->] (A0) -- (B0);
\draw[->] (B0) -- (C0);
\node at (0.9,2.2) {t=0};

\node at (2.8,0) {\Huge$\rightarrow$};
\begin{scope}[xshift=3.8cm]
\node[node, blue] (A1) at (0,1) {$i$};
\node[node, blue] (B1) at (1.8,1) {$i+1$};
\node[node, red]  (C1) at (0,-1) {$i+2$};
\node[node, red]  (D1) at (1.8,-1) {$i+3$};
\draw[->] (A1) -- (D1);
\draw[->] (B1) -- (D1);
\draw[<->] (C1) -- (D1);
\draw[->] (A1) -- (B1);
\draw[->] (B1) -- (C1);
\node at (0.9,2.2) {t=1};
\end{scope}
\node at (6.5,0) {\textbf{or}};
\begin{scope}[xshift=7.3cm]
\node[node, blue] (A2) at (0,1) {$i$};
\node[node, red]  (B2) at (1.8,1) {$i+1$};
\node[node, red]  (C2) at (0,-1) {$i+2$};
\node[node, blue] (D2) at (1.8,-1) {$i+3$};
\draw[->] (A2) -- (D2);
\draw[->] (B2) -- (D2);
\draw[<->] (C2) -- (D2);
\draw[->] (A2) -- (B2);
\draw[->] (B2) -- (C2);
\node at (0.9,2.2) {t=1};
\end{scope}
\node at (10,0) {\textbf{or}};
\begin{scope}[xshift=10.6cm]
\node[node, green] (A2) at (0,1) {$i$};
\node[node, blue]  (B2) at (1.8,1) {$i+1$};
\node[node, red]  (C2) at (0,-1) {$i+2$};
\node[node, red] (D2) at (1.8,-1) {$i+3$};
\draw[->] (A2) -- (D2);
\draw[->] (B2) -- (D2);
\draw[<->] (C2) -- (D2);
\draw[->] (A2) -- (B2);
\draw[->] (B2) -- (C2);
\node at (0.9,2.2) {t=1};
\end{scope}
\node at (13.1,0) {\textbf{or}};
\begin{scope}[xshift=13.8cm]
\node[node, green] (A2) at (0,1) {$i$};
\node[node, red]  (B2) at (1.8,1) {$i+1$};
\node[node, red]  (C2) at (0,-1) {$i+2$};
\node[node, blue] (D2) at (1.8,-1) {$i+3$};
\draw[->] (A2) -- (D2);
\draw[->] (B2) -- (D2);
\draw[<->] (C2) -- (D2);
\draw[->] (A2) -- (B2);
\draw[->] (B2) -- (C2);
\node at (0.9,2.2) {t=1};
\end{scope}
\end{tikzpicture}
\caption{Susceptible-Infected-Recovered (SIR) Model: Four possible outcomes after one time step. The green circle indicates state $0$, which occurs with probability $\gamma$.} 
\label{sir_model_four_outcomes}
\end{figure}
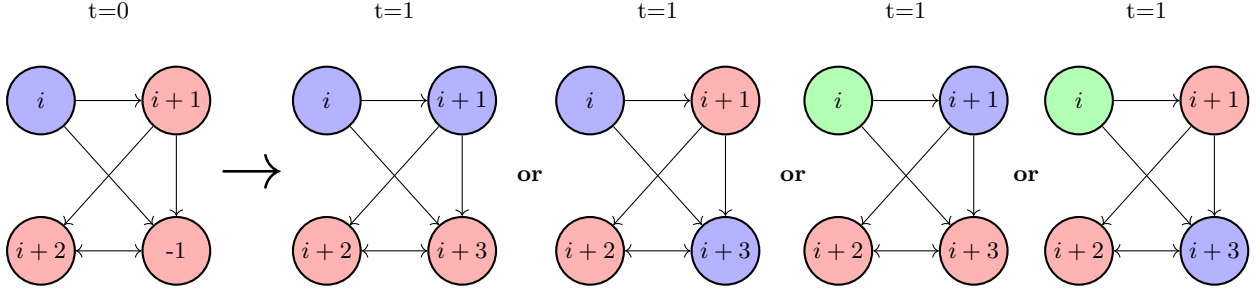

\paragraph{Susceptible-Infected-Recovered (SIR) Model.}
The SIR model introduces a third state, recovered ($R$), denoted by $0$, in which nodes no longer spread the rumor or disease. Hence, the opinion space is $\mathcal{O}=\{-1, 0, +1\}$ and the process is the same as the SI model, except every infected node transitions to the recovered state with probability $\gamma$. Once a node enters this state, it cannot be reinfected or infect other nodes. This is a key distinction from the SI model, where nodes can transition back to the susceptible state. Figure \ref{sir_model_four_outcomes} shows one round of updates in this model.

It is worth emphasizing that many prior works, e.g.,~\cite{kermack1927contribution}, assume a homogeneous mixing condition, i.e., that nodes are connected via a complete graph. This then allows the process to be formulated as a set of differential equations, which are easier to analyze. However, this condition is quite unrealistic, as real-world networks are far from being complete graphs. As a result, recent work has highlighted the importance of underlying networks, e.g.,~\cite{chatterjee2023effective}.

\subsection{Continuous Models}

In the continuous setting, opinions are represented by real values within an interval such as $[-1, +1]$, capturing varying degrees of sentiment. This framework is well-suited for modeling scenarios such as the level of support for a policy reform or the perceived quality of a product.

\subsubsection{French-DeGroot Model}\label{sec_DeGroot}

The French-DeGroot model~\cite{degroot1974reaching} is the first continuous model of opinion formation, providing a framework for understanding how rational nodes can achieve consensus through iterative opinion pooling. The opinion space can be either scalar opinions where \( s_i \in \mathcal{O}=\mathbb{R} \) for each node \( v_i \) or vector opinions where \( \textbf{s}_i = (s_{i1}, \dots, s_{im}) \in \mathcal{O}=\mathbb{R}^m \) for each node \( v_i \) (which represents the opinions of node $v_i$ on $m$ different topics). At \( t = 0 \), the initial opinions of the nodes are given as \( \mathbf{s}(0) = (s_{1}(0), \dots, s_{n}(0))^\top \in \mathbb{R}^n \) for scalar opinions and as \( \mathbf{S}(0) = (s_{il}(0)) \in \mathbb{R}^{n \times m} \) for vector opinions (where each row represents a node's opinion vector). The opinions evolve over time according to an influence matrix \( \mathbf{P} = [p_{ij}] \), where \( p_{ij} \geq 0 \) represents the influence weight of node \( v_j \) on node \( v_i \). The matrix \( \mathbf{P} \) satisfies \( \sum_{j=1}^n p_{ij} = 1 \) for all \( i \), ensuring that the weights are normalized, i.e., $\textbf{P}$ can be defined as $\textbf{P}=\textbf{D}^{-1} \textbf{A}$. For scalar opinions, the update rule of node $v_i$ is the weighted average of the opinions of the in-neighbors given as 
\begin{equation}\label{DeGroot}
\Phi(G, \mathbf{s}(t), v_i) = \sum_{v_j \in N_{\text{in}}(v_i)} p_{ij} s_{j}(t), \quad t = 0, 1, \dots,
\end{equation} 
which can be written as \( \Phi(G, \mathbf{s}(t)) = \mathbf{P} \mathbf{s}(t) \) for all nodes in \( V \). For vector opinions, the update rule generalizes to
\begin{equation}\label{DeGroot_matrix}
\Phi(G, \mathbf{S}(t)) = \mathbf{P} \mathbf{S}(t), \quad t = 0, 1, \dots,
\end{equation}
where each column of \( \mathbf{S}(t) \) evolves independently according to the scalar update rule. The update rule is deterministic and synchronous because the evolution of opinions is fully determined by the influence matrix \( \mathbf{P} \) and the current opinions, and all nodes update simultaneously. The asynchronous version of the model has also been analyzed by Elboim et al.~\cite{elboim2024asynchronous} and Aldous~\cite{aldous2013interacting}. An example of the opinion dynamics over time in this model on a Barab\'{a}si--Albert graph is shown in Figure~\ref{DeGroot_FJ}.

\subsubsection{Friedkin-Johnsen Model}\label{sec_FJ}

The FJ model~\cite{friedkin1990social} extends the DeGroot model by incorporating nodes’ \textit{innate opinions} or \textit{prejudices}, which remain fixed over time, while still accounting for the opinions of their in-neighbors. More precisely, each node has an innate opinion that remains fixed and private. It also has a public opinion, known as the \textit{expressed opinion}, that evolves as a function of its innate opinion and the expressed opinion of its neighbors. In this model, a susceptibility parameter $\lambda_i$ denotes the susceptibility of node $v_i$ to its neighbors' expressed opinion (vs its own innate opinion). A value of $\lambda_i = 1$ indicates that the node is fully susceptible to social influence, whereas $\lambda_i = 0$ implies that the node is stubborn and retains only its innate opinion indefinitely. The opinion space is scalar, $\mathcal{O} = \mathbb{R}$, and each node $v_i$ holds an opinion $s_{i} \in \mathbb{R}$ (usually referred to as expressed opinion). Nodes also have innate opinions that are represented by the column vector $\mathbf{u} = (u_1, \dots, u_n)^\top$. The opinions evolve according to the update rule
\[
\Phi(G, \mathbf{s}(t)) = \bm\Lambda \mathbf{P} \mathbf{s}(t) + (\mathbf{I} - \bm\Lambda) \mathbf{u}, \quad \mathbf{s}(0) = \mathbf{u},
\]
where $\bm\Lambda = \mathrm{diag}(\lambda_1, \dots, \lambda_n)$ is the diagonal susceptibility matrix, and $\mathbf{P} = \mathbf{D}^{-1} \mathbf{A} \in \mathbb{R}^{n \times n}$ is the same row-stochastic matrix as in the DeGroot model, with $p_{ij} \geq 0$ and $\sum_{j=1}^n p_{ij} = 1$ for all $i$.

\begin{figure}[H]
\begin{center}
\includegraphics[scale=0.5]{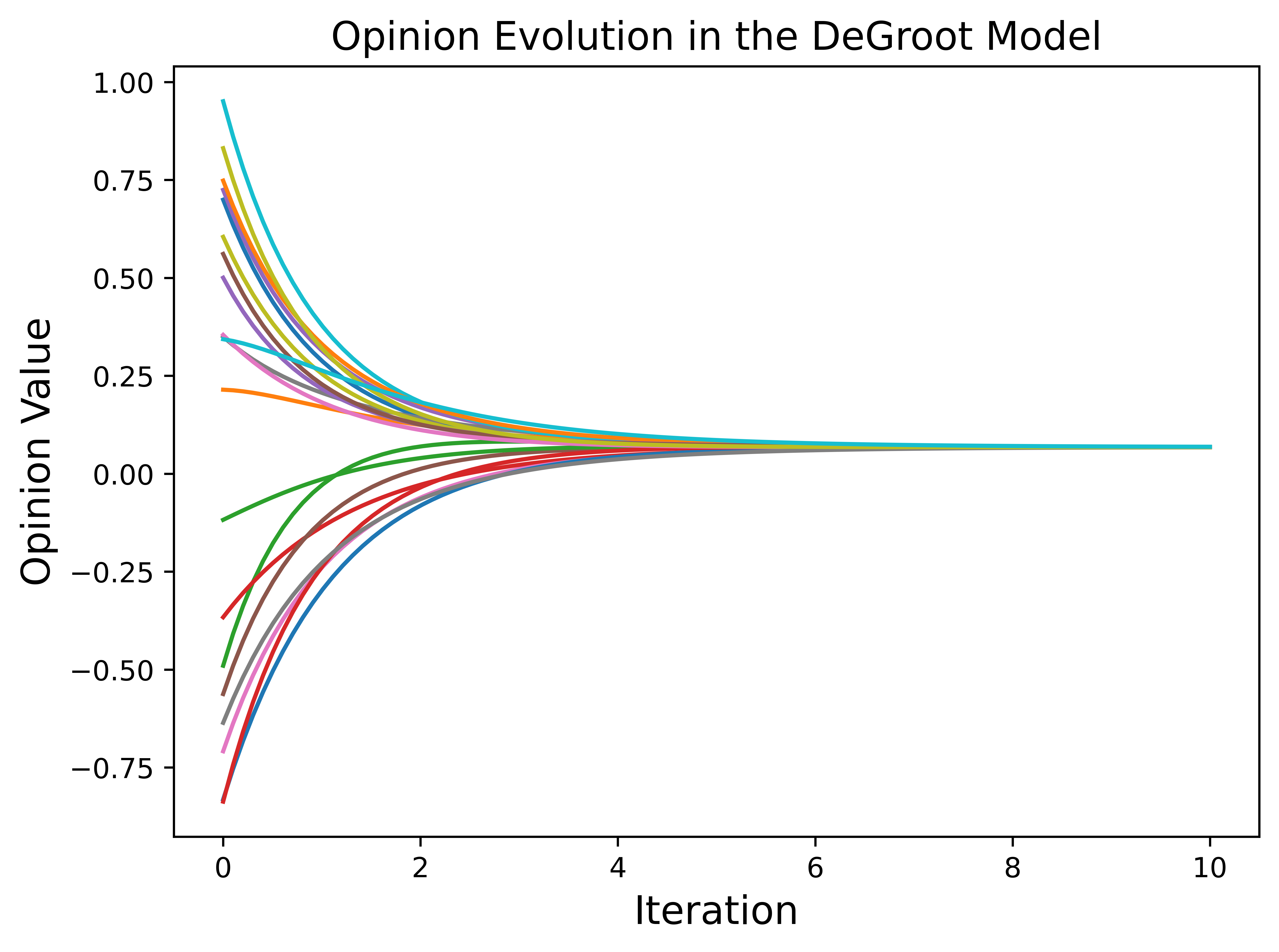}
\includegraphics[scale=0.5]{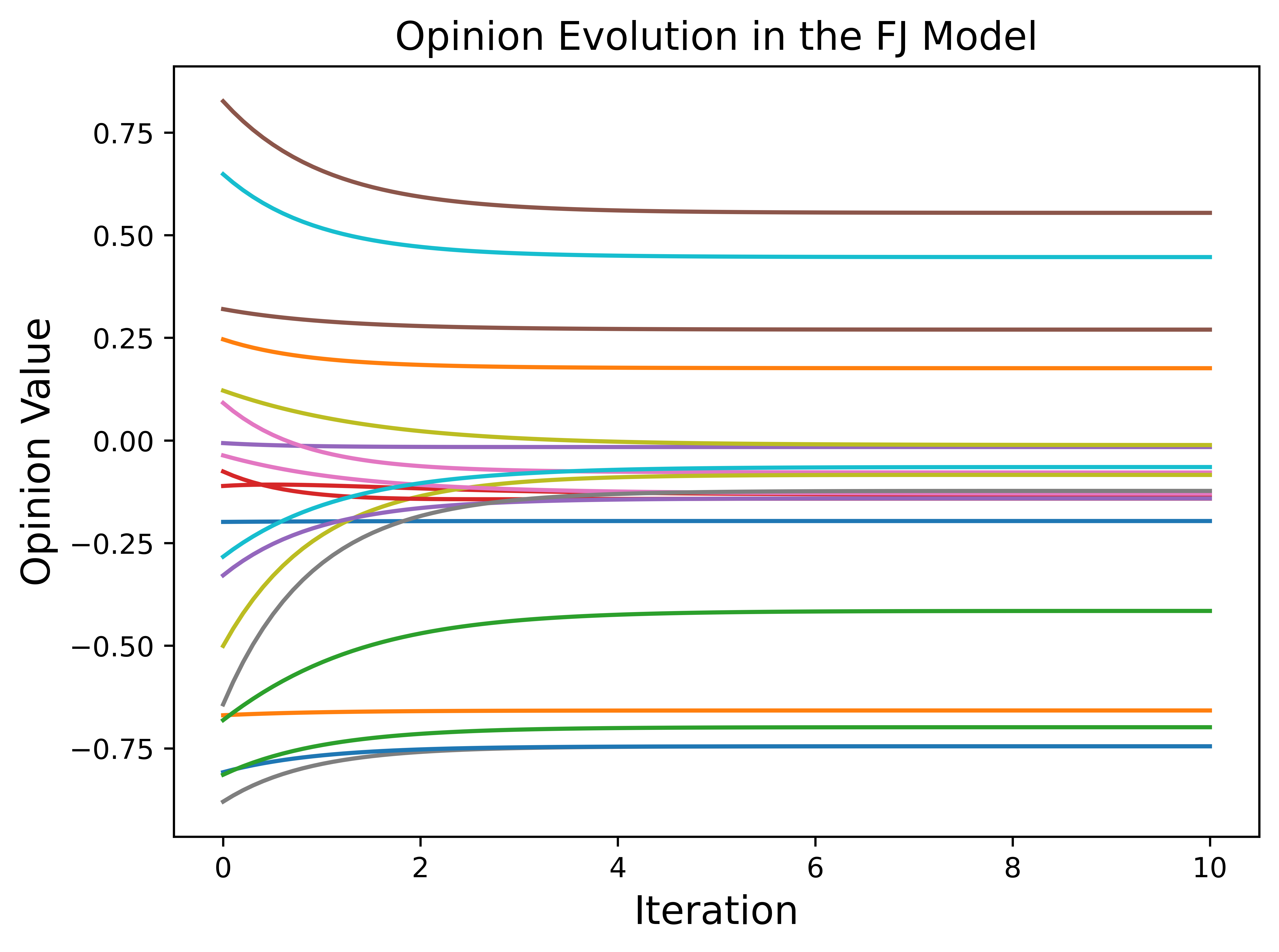}
\end{center}
\caption{Evolution of opinion values in the French-DeGroot and FJ models on a BA graph with $n = 100$ and $m = 7$. In the FJ model, both innate opinions and susceptibility parameters are drawn uniformly at random. For better visualization, the evolution is shown for a subset of 20 nodes.}
\label{DeGroot_FJ}
\end{figure}

The opinion dynamics in this model are \textit{deterministic}, as they are entirely determined by the matrices $\mathbf{P}$ and $\bm\Lambda$, along with the innate vector $\mathbf{u}$. They are also \textit{synchronous}, since all nodes update their opinions simultaneously at each time step. An example of the evolution of opinions in this model, using the same setup as in the French-DeGroot model, is shown in Figure~\ref{DeGroot_FJ}.

\subsubsection{Deffuant-Weisbuch Model}\label{sec_DW}

The DW model~\cite{deffuant2000mixing} is a continuous opinion dynamics model in which nodes interact pairwise and update their opinions based on a bounded confidence threshold. It describes how opinions evolve when nodes only interact with others whose opinions are sufficiently close to their own. The opinion space is continuous: $\mathcal{O} = \mathbb{R}$, where each node $v_i \in V$ has an opinion $s_{i} \in \mathbb{R}$. At $t = 0$, the initial opinions of the nodes are given by the vector $\mathbf{s}(0) = (s_{1}(0), \dots, s_{n}(0))^\top \in \mathbb{R}^n$. At each time step, a pair of nodes $\{v_i, v_j\}$ is selected randomly, where $v_i$ is the out-neighbor of $v_j$ or vice versa. The update function $\Phi(G, \mathbf{s}(t), v_j)$ and $\Phi(G, \mathbf{s}(t), v_i)$ can be written as
\begin{equation} \label{DW}
\Phi(G, \mathbf{s}(t), v_j) = \begin{cases}
s_{j}(t) + \alpha (s_{i}(t) - s_{j}(t)) & \text{if } |s_{j}(t) - s_{i}(t)| \leq r, \\
s_{j}(t) & \text{otherwise},
\end{cases}    
\end{equation}
\begin{equation} \label{DW2}
\Phi(G, \mathbf{s}(t), v_i) = \begin{cases}
s_{i}(t) + \alpha (s_{j}(t) - s_{i}(t)) & \text{if } |s_{i}(t) - s_{j}(t)| \leq r, \\
s_{i}(t) & \text{otherwise},
\end{cases}
\end{equation}
where $\alpha \in (0,1)$ is known as the \textit{weighting factor} and $r$ is a threshold parameter in $(0,1)$ which can be either a constant or variable for different nodes, e.g.,~\cite{chen2020convergence}. The evolution of the opinion values over time is shown in Figure \ref{abelson_time} for this model. The update rule is \textit{stochastic} because the selection of node pairs is random. The process is \textit{asynchronous} because only one pair of nodes updates their opinions at each time step.

\subsubsection{Hegselmann-Krause Model}\label{sec_HK}
The HK model~\cite{rainer2002opinion} is a continuous opinion dynamics model in which nodes update their opinions by averaging the opinions of others within their confidence bounds. It describes how opinions evolve when nodes are influenced only by those with sufficiently similar opinions. This model is essentially similar to the French-DeGroot model, but nodes consider only the opinion of a subset of their neighbors. The opinion space is continuous: $\mathcal{O} = \mathbb{R}$, where each node $v_i$ has an opinion $s_{i} \in \mathbb{R}$. At $t = 0$, the initial opinions of the nodes are given by the vector $\mathbf{s}(0) = (s_{1}(0), \dots, s_{n}(0))^\top \in \mathbb{R}^n$. At each time step, every node $v_i$ updates its opinion by averaging the opinions of its in-neighbors whose opinions lie within its confidence range $r$. The update function $\Phi(G, \mathbf{s}(t), v_i)$ can be written as
\[\label{HK}
\Phi(G, \mathbf{s}(t), v_i) = \frac{1}{|N_{v_i}(t)|} \sum_{v_j \in N_{v_i}(t)} s_{j}(t),
\]
where $N_{v_i}(t) = \{v_j \in N_{\text{in}}(v_i) \mid |s_{j}(t) - s_{i}(t)| \leq r\}$ is the set of in-neighbors within the confidence bound $r$ of node $v_i$, and the number of such in-neighbors is given by $|N_{v_i}(t)|$. The evolution of the opinion values over time for this model is shown in Figure~\ref{abelson_time}. This model is deterministic and synchronous. The HK model with asynchronous update has also been studied by Berenbrink et al.~\cite{berenbrink2024asynchronous}, in which nodes' opinions are updated one after another at random.  

\subsubsection{Abelson Model} \label{abelson_sec}

Considering the opinion space as $\mathcal{O} = \mathbb{R}$, where each node $v_i$ has an opinion $s_{i} \in \mathbb{R}$ and the the initial vector opinions of the nodes $\mathbf{s}(0) = (s_{1}(0), \dots, s_{n}(0))^\top \in \mathbb{R}^n$, Abelson's model~\cite{abelson1964mathematical} is a time-varying version of the French-DeGroot model where the opinion vector in any time follows the \textit{Laplacian flow dynamics} given as 
\begin{equation}\label{abel}
    \Phi(G, \mathbf{s}(t)) = \dot{\mathbf{s}}(t) = -\mathbf{L} [\mathbf{A}] \mathbf{s}(t), \quad t \geq 0,
\end{equation}
where $\mathbf{L}[\mathbf{A}]:=\textbf{D}-\textbf{A}$ denotes the Laplacian matrix(see Proskurnikov and Tempo~\cite{proskurnikov2017tutorial,proskurnikov2018tutorial} for the detailed calculation and also a comprehensive tutorial for this model). An example of the evolution of the opinion values over time is shown in Figure \ref{abelson_time}. Hendrickx and Tsitsiklis~\cite{hendrickx2012convergence} give a time-dependent generalization of the Abelson model 
\begin{equation}\label{abelson_time}
    \Phi(G, \mathbf{s}(t)) = \dot{\mathbf{s}}(t) = -\mathbf{L}[\mathbf{A}(t)]\mathbf{s}(t), \quad \mathbf{A}(t) \geq 0,
\end{equation}
where $\mathbf{A}(t) = (a_{ij}(t))$ is a non-negative matrix with locally $L^1$-summable entries \cite{proskurnikov2018tutorial}. This means that for any finite interval $[a,b]$, the integral $\int_a^b a_{ij}(t) \, dt$ exists and is finite for all $i, j$. Remarkably, this time-varying opinion dynamic was later extended to \textit{Altafini's model} for signed graphs by Altafini~\cite{altafini2012consensus}.

\begin{figure}[H]
\begin{center}
\includegraphics[scale=0.33]{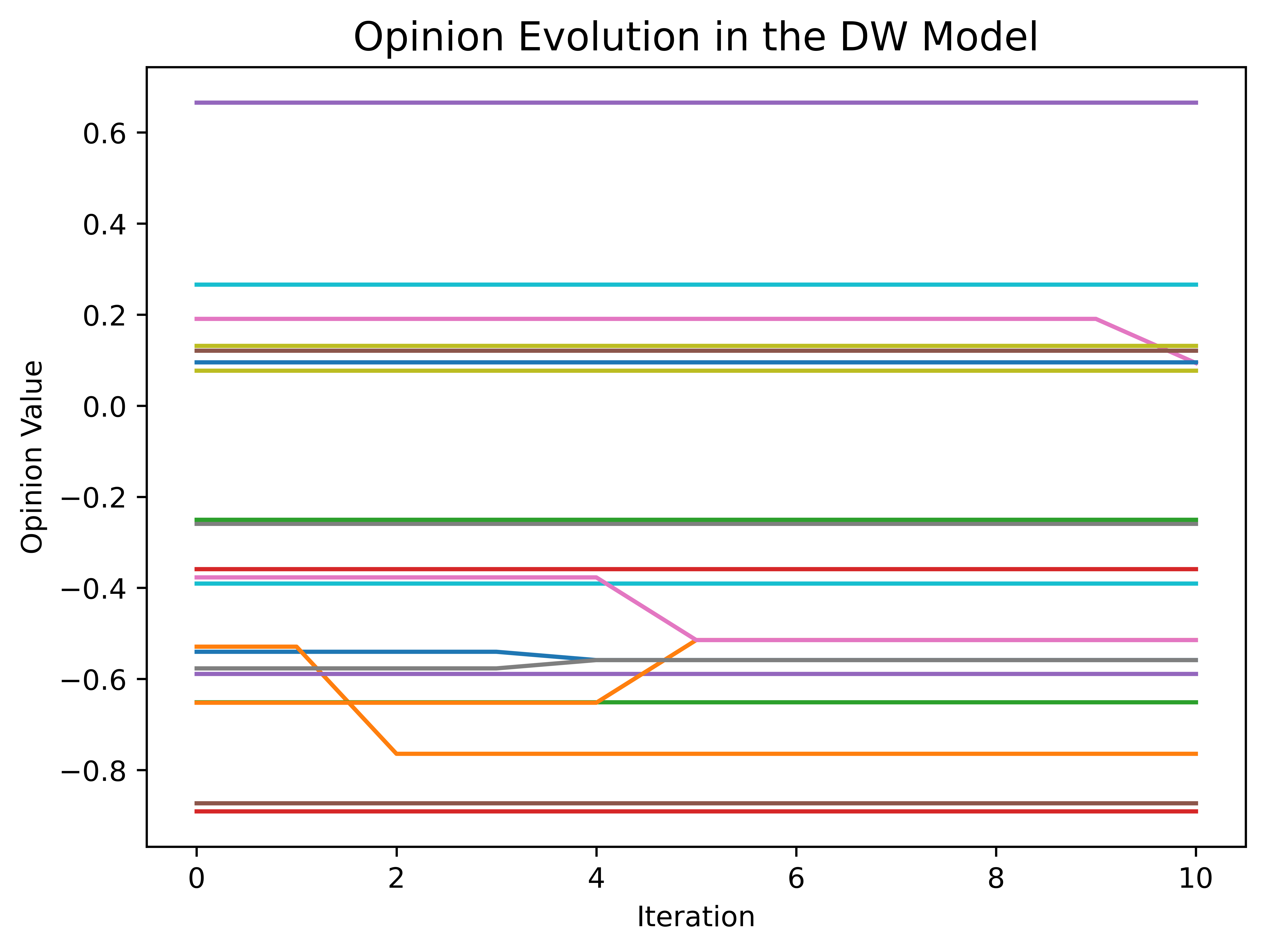}
\includegraphics[scale=0.33]{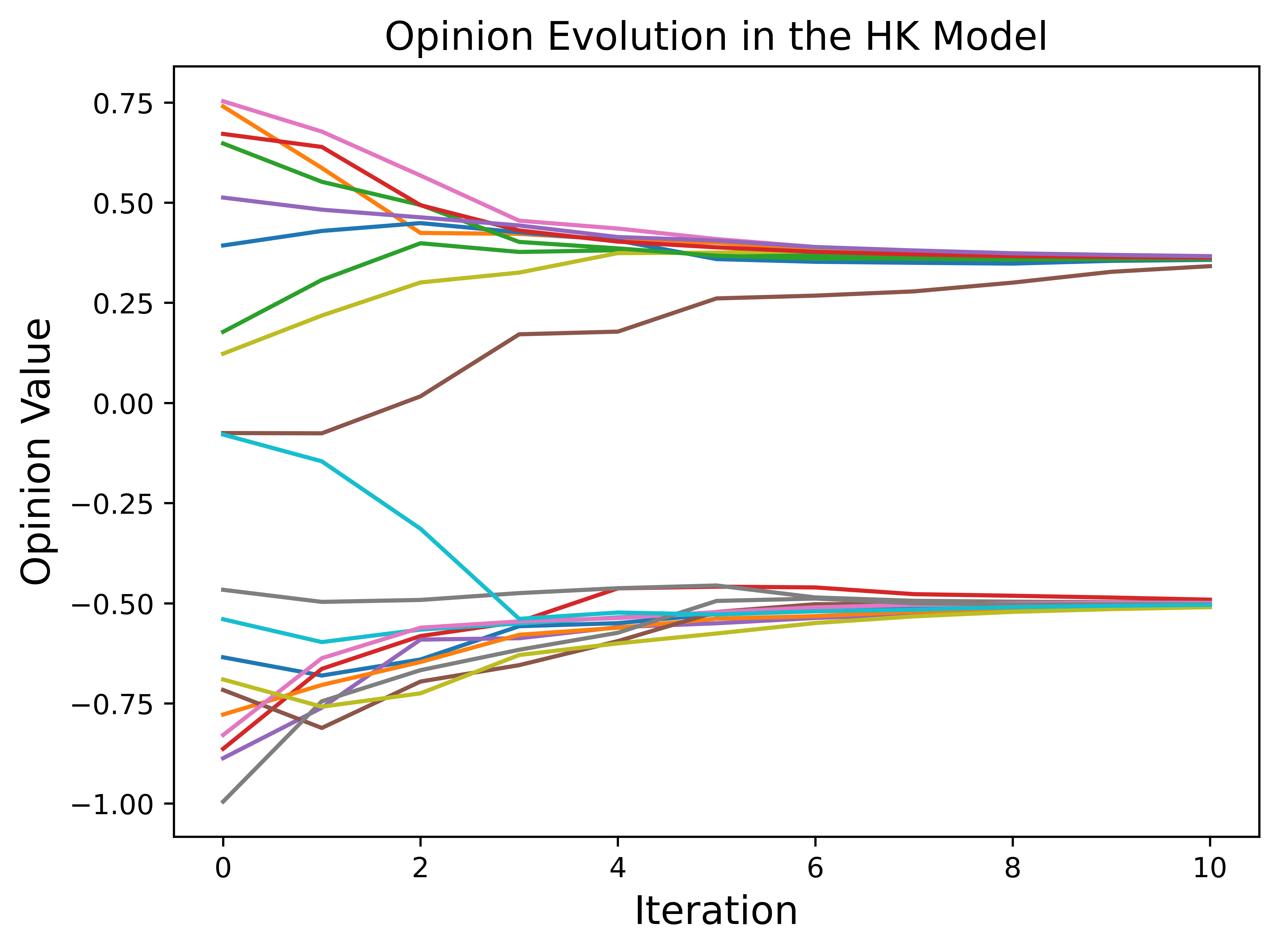}
\includegraphics[scale=0.33]{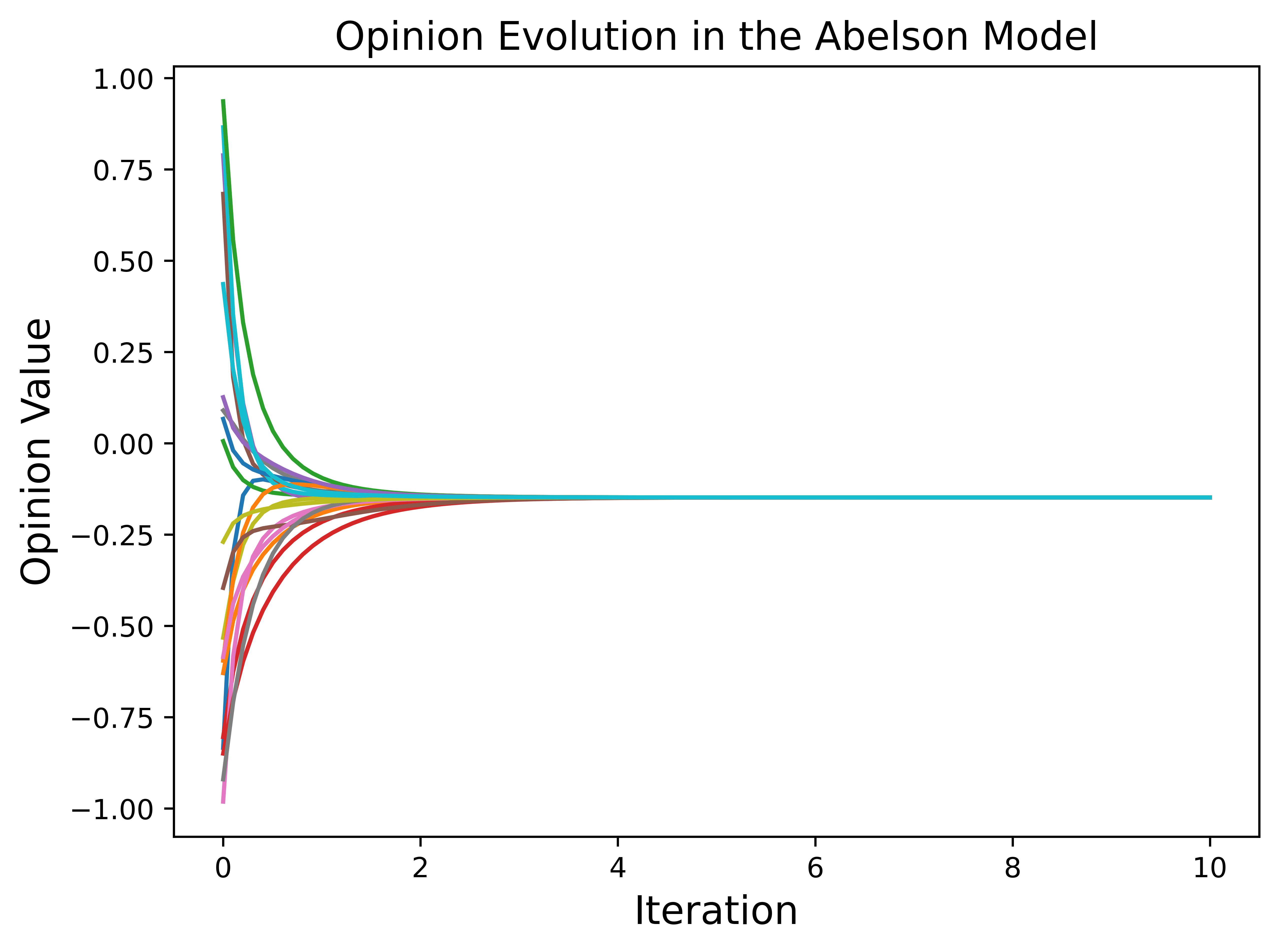}
\end{center}
\caption{
Evolution of opinion values in the DW model (with parameters $\alpha = r = 0.5$), the HK model (with parameter $r = 0.5$), and the Abelson model on a BA graph with \(n=100\), \(k=7\) (showing the evolution for 20 nodes for clarity). Initial opinions are drawn uniformly at random.}\label{abelson_time}
\end{figure}  

The summary of all models in this section is summarized in Table~\ref{tab_model_comparison}.

\begin{table}[htbp!]\scriptsize
    \centering
    \renewcommand{\arraystretch}{1.3}
    \begin{tabular}{|>{\centering\arraybackslash}p{2.3cm}|>{\centering\arraybackslash}p{0.6cm}|>{\centering\arraybackslash}p{0.6cm}|
    >{\centering\arraybackslash}p{0.6cm}|>{\centering\arraybackslash}p{0.6cm}|
    >{\centering\arraybackslash}p{0.6cm}|>{\centering\arraybackslash}p{0.6cm}|
    p{6.5cm}|}
        \hline
        \textbf{Model} & \multicolumn{2}{c|}{\textbf{Disc. / Cont.}} & \multicolumn{2}{c|}{\textbf{Sync. / Async.}} & \multicolumn{2}{c|}{\textbf{Det. / Stoch.}} & \textbf{Update Rule} \\
        \cline{2-7}
         & Disc. & Cont. & Sync. & Async. & Det. &  Stoch. & \\
        \hline
        Voter Model & \checkmark &   & \checkmark & \checkmark &  & \checkmark & Each node adopts the state of a randomly selected in-neighbor (with probability proportional to edge weights). \\
        \hline
        Galam Majority Model & \checkmark &  & \checkmark & \checkmark  &  & \checkmark & Opinions are updated by repeatedly forming random groups and applying a majority rule, where each group adopts the majority opinion, and in case of a tie, defaults to opposition. \\
        \hline
        Majority Models & \checkmark &  & \checkmark & \checkmark  & \checkmark & \checkmark  & Each node adopts the majority opinion of its in-neighbors; ties result in no change (\textbf{majority model}) or a random choice between opinions (\textbf{random majority model}).\\
        \hline
        Ising Model & \checkmark &  &  & \checkmark  &  & \checkmark  & Each node updates its binary opinion to reduce an energy function that encourages agreement with neighboring nodes. \\
        \hline 
        Sznajd Model & \checkmark &  &  \checkmark & \checkmark  &  & \checkmark  & A node is selected randomly; if two adjacent agents agree, their opinions are imposed on their neighbors; otherwise, the neighbors adopt the opposite opinions. \\
        \hline
        PUSH Protocol & \checkmark &  & \checkmark & \checkmark  &  & \checkmark & In each round, every $+1$ (informed) node, send the rumor to a randomly chosen out-neighbor. \\
        \hline
        PULL Protocol & \checkmark &  & \checkmark & \checkmark  &  & \checkmark & In each round, every $-1$ (uninformed) node requests the rumor from a randomly chosen in-neighbor.  \\
        \hline
        PUSH-PULL Protocol & \checkmark &  & \checkmark &  \checkmark &  & \checkmark & PUSH and PULL run simultaneously. \\
        \hline
        Bootstrap Percolation Model & \checkmark &  & \checkmark & \checkmark & \checkmark &  & Each node becomes and stays active if at least $r$ (a fixed given threshold) of its in-neighbors are active. \\
        \hline
        Linear Threshold Model & \checkmark &  & \checkmark & \checkmark &  & \checkmark  & Every $-1$ (inactive) node switches to $+1$ (active) when the combined weighted influence of their active in-neighbors surpasses a personal, randomly chosen threshold. \\
        \hline
        Independent Cascade Model & \checkmark &  & \checkmark & \checkmark  &  & \checkmark & Newly $+1$ (activated) nodes get one chance to activate each $-1$ (inactive) out-neighbor with probability proportional to the weight. \\
        \hline
        Epidemiological Processes & \checkmark &  &  & \checkmark &  & \checkmark & In the \textbf{SI model}, every $+1$ (infected) node randomly infects $-1$ (susceptible) neighbors and remains infected forever. The \textbf{SIS model} extends SI by allowing $+1$ nodes to become $-1$ again with a certain probability. The \textbf{SIR model} adds a $0$ (recovered) state where $+1$ nodes become $0$ with some probability and no longer spread or contract the infection. 
        \\
        \hline
        French-DeGroot Model &  & \checkmark & \checkmark & \checkmark & \checkmark &  & All nodes iteratively update their opinions by averaging the opinions of their neighbors weighted by an influence matrix. \\
        \hline
        Friedkin-Johnsen Model &  & \checkmark & \checkmark &  & \checkmark &  & Extends the French-DeGroot model by allowing nodes to retain a weighted portion of their innate opinions based on individual susceptibilities to influence. \\
        \hline
        Deffuant-Weisbuch Model &  & \checkmark &  & \checkmark &  & \checkmark & At each step, a random pair of connected nodes updates their opinions if they are within a confidence bound $r$, moving closer by a factor $\alpha$. \\
        \hline
        Hegselmann-Krause Model &  & \checkmark & \checkmark & \checkmark  & \checkmark &  & At each step, every node updates its opinion by averaging those of its neighbors within a confidence bound $r$.   \\
        \hline
        Abelson Model &  & \checkmark & \checkmark &  & \checkmark &  & The \textbf{Abelson model} captures time-varying opinion dynamics as Laplacian flow: $\dot{\mathbf{s}}(t) = -\mathbf{L}[\mathbf{A}] \mathbf{s}(t)$, where $\mathbf{L(\textbf{A})}:=\textbf{D}-\textbf{A}$ is the Laplacian matrix. \\
        \hline
    \end{tabular}
    \vspace{+0.1cm}
    \caption{Summary of various opinion dynamics models at a glance. 
    }\label{tab_model_comparison}
\end{table}

\section{Convergence Properties}\label{Convergence_Properties}

Arguably, the most well-studied aspect of opinion dynamics is
their convergence properties.
In this section, we first review the definition of periodicity and stabilization time in (discrete/continuous) opinion dynamics, under both deterministic and stochastic settings. We then present known convergence results for the models introduced in Section~\ref{models}.

\subsection{Basic Definitions}

\begin{definition*}[\textbf{Convergences in Deterministic and Discrete Opinion Dynamics}]
Consider a discrete deterministic opinion dynamics with $k$ opinions (states) on an $n$-node graph. There are $k^n$ possible configurations. Since the process is deterministic, it will eventually reach a cycle of configurations and remain there forever, starting from any initial configuration. The number of rounds needed to get this cycle is called the \textbf{stabilization time}, and the length of the cycle is referred to as the \textbf{periodicity} of the process. If this length is one, we call the process \textit{aperiodic}.     
\end{definition*}

\begin{definition*}[\textbf{Strongly Connected Graph}]
A graph is termed \textit{strongly connected} (or \textbf{strong}) if there exists a walk between any two nodes.
\end{definition*}

A node that is connected to all other nodes in a graph by walks is called a \textit{root node}. If a graph has at least one root node but is not necessarily strongly connected, it is called \textit{quasi-strongly connected} or \textit{rooted}.

\begin{definition*}[\textbf{Strongly Connected Component}]
A \textit{strongly connected component} (or \textit{strong component}) of a graph $G$ is a maximal subgraph $G'$ that is strongly connected and is not contained within any larger strongly connected subgraph.
\end{definition*}

\begin{definition*}[\textbf{Closed Strong Component}]
A strong component is called \textit{closed} if it has no incoming arcs from other components.
\end{definition*}

If the graph is not strongly connected, it contains two or more strong components, at least one of which is closed. A graph is quasi-strongly connected if and only if this closed strong component is unique. In this case, any node of this strong component is a root node.

When the update rule is stochastic, a discrete opinion dynamics can be modeled as a Markov chain. For $k$ opinions, the $k^n$ possible configurations form the state space of the Markov chain. The stochastic update rule of the process defines the transition probabilities between the configurations. More precisely, the transition probability from $i$ to $j$ in the Markov chain is the probability that the process moves to configuration $j$ in the next step, given that it is currently in configuration $i$. This is well-defined for both a synchronous and asynchronous setup.

Since the underlying structure of a Markov chain is a directed graph, it can be partitioned into maximal strongly connected components. One says that a component is \textit{absorbing} if it has no outgoing edges (transitions). Contracting each maximal strongly connected component to a single node yields a directed acyclic graph. In this graph, each absorbing component corresponds to a sink (a node with no outgoing edge). Thus, starting from any initial configuration, the opinion process will eventually reach and remain in one of these absorbing components.

\begin{definition*}[\textbf{Convergences in Stochastic and Discrete Opinion Dynamics}]
A stochastic opinion dynamic is a discrete-time Markov chain over the $k^n$ possible configurations, where transitions occur based on a random update rule applied to node opinions. The \textbf{stabilization time} is defined as the expected number of rounds needed to reach an absorbing component, and the \textbf{periodicity} is defined as the number of configurations within that absorbing component.
\end{definition*}

In simpler terms, the underlying stochastic process eventually settles into a subset of configurations and continues to transition between them. The stabilization time measures how long it takes to reach this subset, and the periodicity measures the size of the subset. 
 
\begin{definition*}[\textbf{With High Probability}]
 The term \textit{with high probability} (w.h.p.) means that as $n$ grows, the probability of an event (e.g., reaching an absorbing component within a certain time) approaches one at a rate of at least $1 - o(1)$.  
\end{definition*}

Consider a graph with \( n \) nodes, where each node holds a real-valued opinion that evolves continuously over time. That is, the opinion dynamic is continuous, and the system state is represented by a vector in \( \mathbb{R}^n \).

\begin{definition*}[\textbf{Convergence in Deterministic and Continuous Opinion Dynamics}]
Starting from an initial configuration, if the opinions converge to a fixed point as time approaches infinity, the process is said to converge. Formally, the system converges if there exists a vector \( \mathbf{s}^* \in \mathbb{R}^n \) such that \( \lim_{t \to \infty} \mathbf{s}(t) = \mathbf{s}^* \). The \textit{stabilization time} refers to the time required for the opinions to become arbitrarily close to their limiting values, within a specified tolerance $\epsilon>0$.
\end{definition*}

\paragraph{Consensus vs Polarization.} In both discrete and continuous settings, convergence can result in \textbf{consensus}, where all nodes eventually share the same opinion (i.e., $s_v(t) = s_u(t)$ for all $u,v \in V$ as $t \to \infty$). Note, however, that consensus may not necessarily require $t \to \infty$; it might be reached in finite time. If consensus is not achieved, the system converges to a \textbf{polarization} configuration, in which nodes stabilize into multiple distinct opinion groups.

\begin{definition*}[\textbf{Irreducible Matrix}]
A nonnegative adjacency matrix $\mathbf{A}$ is called \textit{irreducible} if its associated graph $G$ is strongly connected.
\end{definition*}

\begin{definition*}[\textbf{Periodicity/Aperiodicity}]
A graph is said to be \textit{periodic} if it contains at least one cycle and the length of every cycle in the graph is divisible by some integer $h > 1$. The largest such integer $h$ is called the \textit{period} of the graph. If no such integer exists, the graph is called \textit{aperiodic}.
\end{definition*}

A cycle graph $C_n$ is typically considered periodic because its cycle lengths are multiples of $n$. In contrast, a classic example of an aperiodic graph is one with self-loops. A self-loop introduces a cycle of length $1$, which ensures that not all cycles in the graph share a common divisor greater than $1$. Thus, the graph cannot be periodic (as periodicity requires all cycles to be multiples of some $h > 1$).

\begin{definition*}[\textbf{Primitive Matrix}]
An irreducible matrix $\mathbf{A}$ is \textit{primitive} if and only if its associated graph $G$ is aperiodic.
\end{definition*}

\subsection{Discrete Models}

\subsubsection{Voter Model}
The Voter model (on a connected non-bipartite graph) is a finite Markov chain with absorbing configurations (all $+1$ or all $-1$), so the system will eventually reach consensus, as shown by Hassin and Peleg~\cite{hassin2001distributed}.

The Voter model is the dual of the coalescing random walk model \cite{hassin2001distributed,cooper2013coalescing}, which can be described as follows. Initially, there is a pebble on every node of the graph. In each round, every pebble selects an out-neighboring node uniformly at random and moves to that node. Whenever two or more pebbles meet at the same node, they merge into a single pebble, which then continues its random walk. The process terminates when only one pebble remains. The time taken until only one pebble remains is called the \textit{coalescing time}. It is known that the coalescing time for a graph $G$ equals the stabilization time of the Voter model on $G$ when each node initially holds a distinct opinion~\cite {hassin2001distributed}. For weighted undirected graphs, Hassin and Peleg~\cite{hassin2001distributed} bound the expected coalescing time (and thus the expected stabilization time of the Voter model) in terms of the expected meeting time $t_{\text{meet}}$ of two random walks and proved a bound of $O(t_{\text{meet}} \log n) = O(n^3 \log n)$, which was later improved to $O(n^3)$ by Kanade et al.~\cite{kanade2023coalescence}. Cooper et al.~\cite{cooper2013coalescing} provide another upper bound of
\[
O\left(\frac{1}{1 - \lambda_2} \Big( \log^4 n + \frac{n}{\rho}\Big) \right),
\]
on the expected coalescing time for any graph $G$, where $\lambda_2$ is the second eigenvalue of the transition matrix of a random walk on $G$, and 
\[
\rho= \frac{(\sum_{u \in V} d(u))^2}{\sum_{u \in V} d(u)^2}. 
\]
Berenbrink et al.~\cite{berenbrink2016bounds} analyze the Voter model on dynamic graphs, where the edges are rewired in every round. Their work provides bounds on the stabilization time for graphs with rewired edges and conductance \footnote{Here, the \textit{conductance} of a connected graph is a value $\phi$ in the range $0 < \phi \leq 1$. This value is significant for graphs that are well connected (e.g., the complete graph) and small for graphs that exhibit poor connectivity (e.g., graphs with communication bottlenecks).} \( \phi \), showing a result of \( O\left(\frac{|E|}{d_{\text{min}} \cdot \phi}\right) \) for graphs with minimum degree \( d_{\text{min}} \).

The expected stabilization time for the Voter model under \textit{asynchronous} update, where at each step a node is selected uniformly at random and adopts the state of a random in-neighbor, is studied extensively by the statistical physics community, e.g.,~\cite{redner2019reality}. For graphs with arbitrary degree distributions, the expected stabilization time has been proved to follow~\cite{sood2008Voter}
\[
T(n) = n_{\text{eff}} \left( (1-n) \ln\Big(\frac{1}{1-n}\Big)+n\ln\Big(\frac{1}{n}\Big)\right)
\]
where $n_{\text{eff}}$ is the effective population size given by \( n  \mu_1^2  / \ \mu_2  \), where \( \mu_i :=\frac{1}{n} \sum_{j=1}^n d_j^i \) is the moments of the degree distribution. For a scale-free network with a degree distribution $n_k \sim k^{-\alpha}$, the mean stabilization time $T(n)$ depends on $n$ as follows~\cite{sood2008Voter}
\[\label{consensus_Voter_scale_free}
T(n) = 
\begin{cases}
\Theta(n) & \alpha > 3, \\
\Theta(n / \ln n) & \alpha = 3, \\
\Theta(n^{2(\alpha - 2)/(\alpha - 1)}) & 2 < \alpha < 3, \\
\Theta(\ln(n)^n) & \alpha=2 \\
\Theta(1) & \alpha < 2.
\end{cases}
\]
The main feature is that consensus is achieved quickly in the Voter model, i.e., $T_n \ll n$ for all $\alpha < 3$. This consensus time is significantly faster than the corresponding behavior on regular lattices in spatial dimensions $d$, as shown by Liggett and Liggett~ \cite{liggett1985interacting} and   Krapivsky~\cite{krapivsky1992kinetics}, where $T(n) =\Theta(n^2$) for $d = 1$, $T_n=\Theta( n \ln n)$ for $d = 2$, and $T(n) =\Theta( n)$ for $d \geq 3$.

\subsubsection{Majority Models}
\paragraph{Galam Majority Model. }
The iterative update process drives the population towards consensus, i.e., the possible outcomes are either unanimous support ($P^{+1} = 1$), where everyone supports the reform, or unanimous opposition ($P^{-1} = 1$), where everyone opposes the reform. Gärtner and Zehmakan~\cite{gartner2020threshold} show that the stabilization time to reach consensus is $O(\log(\log(n)))$ if all rooms are of size greater than $3$, and this bound is tight. Additionally, they demonstrated that if all groups are of size $2$, the consensus time increases to $O(\log(n))$.

\paragraph{Graph-based Majority Models.} The majority model can achieve two types of configurations, as shown by Goles and Olivos~\cite{goles1980periodic}. First, it may reach a stable configuration in which no node changes its state; this occurs when every node aligns with the majority state of its in-neighbors. In this case, the model is aperiodic, indicating that the system stabilizes with all nodes fixed in their states. Alternatively, the model can exhibit polarization. This might happen, for example, when a fixed portion of the population remains unchanged and only a subset of nodes switches. As another example, in bipartite graphs where nodes in one partition take the opposite state of nodes in the other partition at each step, polarization emerges (nodes constantly change their states).

The majority model on an $n$-node undirected and unweighted graph stabilizes in $O(n^2)$ time steps, as shown in early works such as~\cite{fogelman1983transient} and~\cite{winkler2008puzzled}. However, Frischknecht et al.~\cite{frischknecht2013convergence} establish almost tight bounds on the stabilization time by constructing a family of $n$-node unweighted graphs that require $\Omega\left(\frac{n^2}{\log^2 n}\right)$ rounds to stabilize. It is worth noting that for certain classes of graphs, such as $d$-regular graphs with strong conductance, the stabilization time improves to $O(\log_d n)$, as shown by Zehmakan~\cite{zehmakan2020opinion}. In the undirected weighed case, an $n$-node network $G$ stabilizes in $\min\{2\omega(G), 2^n\}$, where $\omega(G)=\sum_{\{u,v\}\in E} w_{uv}$~\cite{keller2014even}. For the case of the weighted directed graphs, however, this bound is proven to be lower-bounded by $2^{\Omega(n)}$ in the same paper.

For the random majority model, unlike the majority model, there are undirected unweighted graphs for which both stabilization time and periodicity are exponential~\cite{zehmakan2023random}. A similar exponential stabilization time behaviour has also been proven for the biased majority model (where nodes adopt $+1$ with probability $\alpha$ and update based on majority update with probability $1-\alpha$)~\cite{lesfari2022biased}.

From an algorithmic perspective, determining whether, for a given configuration of a (directed) graph, opinions form consensus (or polarization emerges) in the majority model is a PSPACE-complete problem~\cite{chistikov2020convergence}. This means that while the problem can be solved using a reasonable amount of memory (polynomial space), it may require a significant amount of time to compute. 

Considering the majority-based preference diffusion model (where nodes have a preference ranking over $\alpha$ candidates and update the order of a randomly selected pair of candidates following a majority rule), Brill et al.~\cite{brill2016pairwise} prove that asynchronous update converges on complete graphs while synchronous update converges on directed acyclic graphs. In undirected graphs, the process always converges~\cite{zehmakan2024majority}. For the asynchronous set-up, Zehmakan~\cite{zehmakan2024majority} proves that this happens in $O(n|E|\alpha^4)$ for any $n$-node graph.

\subsubsection{Ising Model}

Mossel and Sly~\cite{mossel2013exact} show that the stabilization time of Gibbs samplers for the ferromagnetic Ising model (although the same stabilization time is proven for the antiferromagnetic Ising model by Chen et al.~\cite{chen2021optimal}) on any graph of $n$ nodes and maximal degree $d_{\max}$, with bounded parameter $\beta$ and arbitrary external fields $B$ is $O(n\log n)$ if $(d_{\max}-1)\tanh(\beta)<1$.  Mossel and Sly~\cite{mossel2013exact} also show that when \( d_{\max} \tanh\beta < 1 \), with high probability over the Erdős–Rényi random graph with $p=\frac{d_{\max}}{n}$ the stabilization time of Gibbs samplers is \( n^{1 + \Theta(1/\log (\log n))} \). Both results are tight as it is known that the stabilization time is, with high probability, exponential in $n$ for random $d$-regular graphs when $(d - 1)\tanh\beta > 1$~\cite{gerschenfeld2007reconstruction}, and for Erdős–Rényi graphs when $d_{\max} \tanh\beta >1$~\cite{dembo2010ising}. The same exponential stabilization time has been shown for random d-regular graphs by Can et al.~\cite{can2019glauber} and Dommers~\cite{dommers2017metastability}.

\subsubsection{Sznajd Model}

Through standard Monte Carlo simulations, Sznajd-Weron and Sznajd~\cite{sznajd2000opinion} observe that the Sznajd update rule can lead to either consensus or polarization, with the stabilization time distribution exhibiting a power-law with exponent $3/2$, meaning that the time it takes for the group to make a decision follows a pattern where most decisions are made quickly but there are a few cases where decisions take a very long time.

Most studies on this model are numerical (see, e.g.,~\cite{sznajd2021review} for a comprehensive review of these methods), but some analytical results have also been obtained. Notably, González et al.~\cite{gonzalez2006renormalizing} and Araújo et al.~\cite{araujo2015mean} demonstrate, through renormalization and mean-field approximation approaches, respectively, that their analytical estimations align with the aforementioned Monte Carlo results.

\subsubsection{PUSH-PULL Protocol}

\paragraph{PUSH. }
Frieze and Grimmett~\cite{frieze1985shortest} demonstrate that for a complete unweighted, undirected graph with $n$ nodes, the stabilization time is $(1 + o(1))(\log_2 n + \log n)$ with high probability, where $\log n$ denotes the natural logarithm. For hypercubes and Erd\H{o}s--R\'enyi graphs with $p = \frac{c\log n}{n}$ (where $c > 1$), a stabilization time of $O(\log n)$ was established by Feige et al.~\cite{feige1990randomized}. This bound was later refined by Panagiotou et al.~\cite{panagiotou2015randomized} to $(1 + o(1))(\log_2 n + \gamma(c) \log n)$, where $\gamma(c) = c \log\left(\frac{c}{c-1}\right)$. They also prove that the stabilization time of the PUSH protocol on expander graphs is asymptotically the same as on the complete graph. Additionally, lsässer and Sauerwald~\cite{elsasser2009runtime} prove a lower bound of $\log_2 n + \log n - o(\log n)$ for any regular graph. 

\paragraph{PULL. } 
Considering an undirected and unweighted graph where a subset of nodes $S$ forms a dominating set (every node is either in $S$ or adjacent to $S$), where all nodes in $S$ initially know a rumor, the stabilization time for the PULL protocol is $O(\log(n))$ with very high probability~\cite{giakkoupis2014tight}. Given that $S \subseteq V$ is an (arbitrary) initial set of $+1$ vertices, and  $\beta > 0$ is a fixed parameter, the stabilization time of the PULL protocol is 
$50(\beta+2)\, \log n ({\phi}^{-1} +  \frac{d_{\max}}{\lceil \phi\, \mathrm{vol}(S) \rceil})$
with probability at least $1 - O(n^{-\beta})$ where $\mathrm{vol}(S)$ denotes the sum of degrees of vertices in $S$~\cite{giakkoupis2011tight}. For (an undirected and unweighted) $d$-regular graph with vertex expansion \footnote{The vertex expansion \( \alpha \in (0, 1] \) of a graph is, roughly speaking, the minimum ratio of the number of out-neighbors a set of nodes has (which are not in the set) to the cardinality volume of the set of size at most $n/2$.} at least $\alpha > 0$, the stabilization time of the PULL protocol is $O\left( \alpha^{-1} \log n \log d \right)$ with high probability~\cite{giakkoupis2012rumor}. This result also applies to the PUSH protocol. 

\paragraph{PUSH-PULL Protocol. }
Several studies have advanced this line of research by deriving bounds on the protocol's stabilization time in terms of the graph's expansion parameters, namely, conductance and vertex expansion. Specifically, for a undirected and  unweighted network with conductance $\phi$, the stabilization time for the PUSH–PULL protocol is $O\left(\frac{\log^2 \phi^{-1}}{\phi} \log n\right)$ with high probability~\cite{chierichetti2010almost}. This bound was later improved to $O\left(\frac{\log n}{\phi}\right)$ by  Giakkoupis~\cite{giakkoupis2011tight}. The bound is proven to be tight by Chierichetti et al.~\cite{chierichetti2018rumor}, who show that there exist graphs on $n$ nodes with diameter \( \Omega\left(\frac{\log n}{\phi}\right) \).

Furthermore, if the degrees of the two endpoints of each edge in the network differ by at most a constant factor, then both the PUSH and PULL protocols, when used individually, achieve a stabilization time of \( O\left(\frac{\log n}{\phi}\right) \) with high probability~\cite{giakkoupis2011tight}. For graphs with vertex expansion \( \alpha \), Giakkoupis and Sauerwal~\cite{giakkoupis2012rumor} show that the stabilization time for the PUSH–PULL protocol is \( O\left( \alpha^{-1} \log^2 n \sqrt{\log n} \right) \) with high probability. This bound was later improved to \( O\left( \alpha^{-1} \log n \log \Delta \right) \), where \( \Delta \) is the maximum degree~\cite{giakkoupis2014tight}.

In the case of the preferential attachment model, Chierichetti et al.~\cite{chierichetti2011rumor} show that the stabilization time of the PUSH–PULL strategy is $O(\log^2 n)$ with high probability.

In the PUSH–PULL rumor spreading model, nodes communicate in synchronized rounds; however, this model has also been extended to the asynchronous setting by  Giakkoupis et al.~\cite{giakkoupis2016asynchrony}, where, for example, each node has an independent Poisson clock with rate $+1$ and contacts a random out-neighbor whenever its clock ticks. In this case, if the stabilization time of the synchronous PUSH–PULL protocol is $T(G)$, the stabilization time for the asynchronous protocol is $O(T(G) + \log n)$.

\subsubsection{Bootstrap Percolation Model}
The periodicity of the \(r\)-BP is one, and the stabilization time is bounded by \(n - 1 \). This bound is tight for sequential activation, such as on a path graph with $1$-BP, where all nodes are initially $-1$ except for one leaf. In two-way \(r\)-BP, \(2^n\) is a trivial upper bound on both the periodicity and stabilization time. However, Goles and Olivos~\cite{goles1980periodic} prove that the periodicity is always either one or two. Fogelman et al.~\cite{fogelman1983transient} show that the stabilization time is bounded by \(O(|E|)\), which is asymptotically tight (consider a cycle \(C_n\) in which all nodes have the opinion $-1$ except for two adjacent $+1$ nodes for \(r = 1\).)

Considering the 2-BP on an $L \times L$ square grid, where $L$ denotes the side length of the grid (so the grid contains $L^2$ vertices in total), Benevides and Przykucki~\cite{benevides2015maximum} prove that the stabilization time is $13L^2/18 + O(L)$. In a similar setup, i.e., $2$-BP on the hypercube\footnote{
The $L$-dimensional hypercube $Q_L$ is the graph with vertex set $\{0,1\}^L$ and edge set $\{\{x,y\} : x,y \in \{0,1\}^L,\; |\{i : x_i \neq y_i\}| = 1\}$.}, the stabilization time was shown to be $\lfloor L^2/3 \rfloor$~\cite{przykucki2012maximal}. 

\subsubsection{Linear Threshold Model}
It is evident that the process eventually reaches a fixed configuration, so the model's periodicity is always 1. A trivial upper bound on the stabilization time is $n-1$. This is because initially there is at least one node with state $+1$ (otherwise the process is already at convergence), and at each step at least one node adopts the $+1$ state, and this continues until no node can become active ($+1$), i.e., the process has reached convergence.

\subsubsection{Independent Cascade Model}

Similar to the LT model, the IC process reaches a configuration in which no nodes change their states; hence, it converges. Therefore, the model's periodicity is 1. To analyze the stabilization time, we use the live-edge concept introduced by Kempe et al.~\cite{kempe2003maximizing}. When all coin flips are predetermined, the process becomes easier to analyze. Each edge in the graph \( G \) is labeled as either \textit{live} (if its coin flip results in a successful activation) or \textit{blocked} (if not). Once we address these outcomes and initiate with an initial active set \( A_0 \), we can track how the activation spreads through the network and determine which nodes will be active by the end of the process. A node $v \in V$ becomes $+1$ if and only if there exists a path from some node in $A$ to $v$ consisting entirely of live edges, referred to as a \emph{live-edge path}. The stabilization time can thus be upper-bounded by the diameter of the live-edge subgraph, which, in the case of the simple graph, is itself bounded above by $O(n)$. This bound is tight for sequential activation, such as on a path graph where the initial node is $+1$ and all other nodes are initially $-1$.

\subsubsection{Epidemiological Processes}

The convergence properties of epidemiological processes have been extensively studied within the framework of partial differential equations, under the homogeneous mixing condition (i.e., complete graph). For example, Chalub and Souza~\cite{chalub2011sir} prove that, over time, the SIR model's limiting distribution converges to a Dirac measure concentrated at the isolated equilibria, i.e., the system settles into a single steady state (e.g., a disease becoming endemic at a fixed level). For the SIRS model, the limit is a Radon measure supported on a segment of nonisolated equilibria; i.e., instead of a single steady state, there's a whole range of possible final configurations (because once the epidemic ends, the remaining susceptible population can vary). These processes can also be modeled and analyzed using stochastic differential equations; see, e.g.,~\cite{nguyen2019stochastic}.

However, from a network-based perspective, little is known about the stabilization time. Most existing work has focused on determining whether a virus or disease will spread throughout the population or eventually fade away, an issue we will explore in the next section. For the SI model, since the process is equivalent to the asynchronous PUSH protocol, the stabilization time can be inferred from that context.

\subsection{Continuous Models}

\subsubsection{French-DeGroot Model}

The French-DeGroot model is said to be convergent if, for any initial opinion vector \( \mathbf{s}(0) \), the opinions stabilize over time, meaning the $\lim_{k \to \infty} \mathbf{s}(k) = \lim_{k \to \infty} \mathbf{P}^k \mathbf{s}(0)$ exists. In this case, $\textbf{P}$ is called a regular matrix~\cite{proskurnikov2017tutorial}. A convergent model reaches consensus if all nodes eventually share the same opinion, i.e., $\mathbf{s}_1(\infty) = \dots = \mathbf{s}_n(\infty)$. In this case, the matrix $\textbf{P}$ is called a fully regular matrix~\cite{proskurnikov2017tutorial}. 

The convergence and consensus properties of the model depend on the structure of the influence matrix \( \mathbf{P} \). Specifically, the model is convergent, i.e., \( \mathbf{P} \) is regular, if and only if \( \lambda = 1 \) is the only eigenvalue of \( \mathbf{P} \) on the unit circle \( \{ \lambda \in \mathbb{C} \mid |\lambda| = 1 \} \)~\cite{macduffee2012theory}. The model reaches consensus if \( \mathbf{P} \) is fully regular, meaning \( \lambda = 1 \) is a simple eigenvalue (i.e., its eigenspace is spanned by the all-ones vector \( \mathbf{1} \))~\cite{macduffee2012theory}.  

For an irreducible stochastic matrix $\mathbf{P}$, the model is convergent if and only if $\mathbf{P}$ is primitive, meaning that for some power $k$, $\mathbf{P}^k$ has all positive entries. In this case, the model also reaches consensus~\cite{proskurnikov2017tutorial}. In general, if the graph $G$ is strongly connected, then the French-DeGroot model reaches consensus if and only if $G$ is aperiodic~\cite{proskurnikov2017tutorial}; otherwise, the model does not converge, and opinions exhibit polarization for almost all initial conditions.

In the general case, where $G$ has more than one strongly connected component, it is evident that the evolution of opinions in any closed strong component is independent of the rest of the network. Two different closed components cannot, in general, reach consensus for an arbitrary initial condition. This implies that, for convergence of opinions, all closed strong components must be aperiodic~\cite{demarzo2003persuasion}. For consensus to be reached, the graph should have a single closed, strongly connected component (i.e., it should be quasi-strongly connected), and that component must be aperiodic. Both of these conditions are, in fact, sufficient~\cite{demarzo2003persuasion}.

Considering the synchronous update of the model, the stabilization time of the DeGroot model is given as $O(\frac{1}{\log (\frac{1}{|\lambda_2|})})$ by Ding et al.~\cite{ding2019consensus}, where $\lambda_2$ is the second-largest eigenvalue of $\textbf{P}$ and satisfies $|\lambda_2| \leq 1$. This bound is also given as $O(\lambda_2^{t} \sqrt{\frac{d_{\max}}{d_{\min}}})$ by Becchetti et al.~\cite{becchetti2020step}. The first bound gives the total stabilization time, while the second bound gives the per-step ($t$) convergence rate. In the case of asynchronous updates, the expected $\varepsilon$-stabilization time is given by $O(4d_{\max}n^2 \lceil \log(1/\varepsilon) \rceil)$~\cite{elboim2024asynchronous}, where the $\varepsilon$-stabilization time of the dynamics is defined as the stopping time
\[
\tau_\varepsilon := \min\left\{ t : \max_{v \in V} s_t(v) - \min_{v \in V} s_t(v) \leq \varepsilon \right\}.
\]

\subsubsection{Friedkin-Johnsen Model}
The FJ model is convergent if, for any vector $ \bm{u} \in \mathbb{R}^{n} $, the sequence $ \bm{s}{(t)} $ has a limit $\bm{s}' = \lim_{t \to \infty} \bm{s}{(k)}$, which gives $\bm{s}' = \bm\Lambda \textbf{P} \bm{s}' + (\textbf{I} - \bm\Lambda) \bm{u}$. Notably, the limit value $ \bm{s}' = \bm{s}'(\bm{u}) $ generally depends on the innate condition $\bm{u}$. A special case arises when all solutions converge to the same equilibrium, which corresponds to the asymptotic stability of the system (\ref{sec_FJ}). This condition holds if $ \bm\Lambda \textbf{P} $ is Schur stable~\cite{parsegov2016novel}, meaning $\rho(\bm\Lambda \bm{P}) < 1$, where $\rho$ is the symbole of the largest eigenvalue. 

We define a node $ v_i $ as \textit{stubborn} if $ \lambda_{i} < 1 $ and \textit{totally stubborn} if $ \lambda_{i} = 0 $. A node that is neither stubborn nor influenced by a stubborn node (i.e., not connected to any stubborn node by a walk in the directed graph $G$). is called \textit{oblivious}. After renumbering the nodes, we assume that stubborn nodes and those influenced by them are indexed from $1$ to $n'$, where $n' \leq n$, while oblivious nodes (if they exist) are indexed from $n' + 1$ to $n$. For an oblivious node $v_i$, we have $\lambda_{i} = 1$ and $p_{ij} = 0$ for all $j \leq n'$. Indeed, if $p_{ij} > 0$ for some $j \leq n'$, then node $v_i$ would be connected by a walk to a stubborn node via node $v_j$ and thus would not be oblivious. The matrices $\textbf{P}$ and $\bm\Lambda$, as well as the vectors $\bm{s}{(t)}$, can therefore be decomposed as follows~\cite{parsegov2016novel}
\[
\textbf{P} =
\begin{bmatrix}
\textbf{P}_{11} & \textbf{P}_{12} \\
\textbf{0} & \textbf{P}_{22}
\end{bmatrix},
\quad
\bm\Lambda =
\begin{bmatrix}
\bm\Lambda_{11} & \textbf{0} \\
\textbf{0} & \textbf{I}
\end{bmatrix},
\quad
\bm{s}{(t)} =
\begin{bmatrix}
\bm{s}_1{(t)} \\
\bm{s}_2{(t)}
\end{bmatrix},
\]
where $\bm{s}_1 \in \mathbb{R}^{n'}$, and $\textbf{P}_{11}$ and $\bm\Lambda_{11}$ have dimensions $n' \times n'$. If $n' = n$, then $\bm{s}_2{(k)}$, $\textbf{P}_{12}$, and $\textbf{P}_{22}$ are absent. Otherwise, the oblivious nodes follow the conventional French-DeGroot dynamics 
\[
\bm{s}_2{(t+1)} = \textbf{P}_{22} \bm{s}_2{(t)},
\]
meaning their evolution is independent of the remaining nodes. The matrix $\bm\Lambda_{11} \textbf{P}_{11}$ is Schur stable~\cite{parsegov2016novel}. The system (\ref{sec_FJ}) is stable if and only if there are no oblivious nodes, meaning $\bm\Lambda\textbf {P} = \bm\Lambda_{11} \textbf{P}_{11}$~\cite{parsegov2016novel}. The FJ model with oblivious nodes is convergent if and only if $\textbf{P}_{22}$ is regular, i.e., the limit $\textbf{P}_{22}^* = \lim_{k \to \infty} \textbf{P}_{22}^k$ 
exists. In this case, the limiting opinion $\bm{s}' = \lim_{t \to \infty} \bm{s}{(t)}$ is given by
\[
\bm{s}' =
\begin{bmatrix}
(\textbf{I} - \bm\Lambda_{11} \textbf{P}_{11})^{-1} & \textbf{0} \\
\textbf{0} & \textbf{I}
\end{bmatrix}
\begin{bmatrix}
\textbf{I} - \bm\Lambda_{11} \textbf{P}_{12} \textbf{P}_{22}^* & \textbf{0} \\
\textbf{0} & \textbf{P}_{22}^*
\end{bmatrix}
\bm{u}.
\]
Moreover, if $G$ is strongly connected and $\bm\Lambda \neq \textbf{I}$ (i.e., at least one stubborn node exists), then the FJ model (\ref{sec_FJ}) is stable.

\subsubsection{Deffuant-Weisbuch Model}
By representing the update rule in (\ref{sec_DW}) in matrix form as \( \textbf{s}(t+1) = \mathbf{A}(\textbf{s}(t), t)\textbf{s}(t) \), it has been shown that the opinions always converge with high probability for all \( \alpha \in (0,1) \)~\cite{lorenz2005stabilization}. This convergence is guaranteed because the matrix \( [\mathbf{A}]_{ij}=a_{ij} \) satisfies three key properties: (i) for every node \( v_i \), \( a_{ii} > 0 \); (ii) for any pair of nodes \( v_i \) and \( v_j \), \( a_{ij} > 0 \) if and only if \( a_{ji} > 0 \); and (iii) there exists a constant \( \delta > 0 \) such that \( \min_{i,j} a_{ij} > \delta \). Zhang and Chen~\cite{zhang2015convergence} analyze the convergence rate of the DW opinion dynamics model, proving that opinions converge exponentially fast to consensus but only polynomially fast when forming polarization.

Considering the heterogeneous setup (where nodes have different threshold parameters $r$), Chen et al.~\cite{chen2020convergence} proved that, for a weighting factor of at least $1/2$, opinions converge to a limit with an exponential convergence rate in mean square with high probability. The same convergence result was later extended to the case for any weighting factors $\alpha \in (0,1)$~\cite{chen2024convergence}.

\subsubsection{Hegselmann-Krause Model}
The HK dynamics~(\ref{sec_HK}) satisfies the same three structural properties as the DW model and thus also guarantees convergence~\cite{dittmer2001consensus,lorenz2005stabilization}. The stabilization time for a graph with $n$ nodes and scalar opinions is at least $\Omega(n^2)$ and at most $O(n^3)$~\cite{bhattacharyya2013convergence, mohajer2013convergence, wedin2015quadratic}. For multidimensional opinions, Chazelle~\cite{chazelle2011total} initially establishes an upper bound of \( n^{O(n)} \) on the stabilization time. This bound was later improved to a polynomial upper bound~\cite{etesami2013termination}. Subsequently, Bhattacharyya et al.~\cite{bhattacharyya2013convergence} prove that for \( n \) nodes in any dimension \( d \geq 2 \), the stabilization time is $O(n^{10}d^2)$. This bound was subsequently improved to $O(n^8)$ by Etesami and Başar~\cite{etesami2015game}. Finally, Martinsson~\cite{martinsson2016improved} shows that the maximal stabilization time for the \( n \)-node HK model in any dimension \( d \geq 2 \) can be improved to $O(n^4)$, and this is the best bound known.

\subsubsection{Abelson Model}
Considering Abelson's opinion dynamic (\ref{abelson_sec}), the system can be proven to converge, i.e., 
\[
\lim_{t \to \infty} \textbf{s}(t) = \text{average}(s(0)) \textbf{1}_n,
\]
in the case that the graph is connected~\cite{bullo2018lectures}. A more general convergence to a consensus result for the continuous-time Laplacian flow has been proved in Theorem 7.4 in \cite{bullo2018lectures}. 

In the case of time-varying coefficients, suppose that the gains $a_{ij}(t)$ satisfy the following type-symmetry condition  
\[
K^{-1} a_{ji}(t) \leq a_{ij}(t) \leq K a_{ji}(t), \quad \forall t \geq 0,
\]
for some $K \geq 1$ constant. 
Then, the limit  $\textbf{s}' = \lim_{t \to \infty} \textbf{s}(t)$ exists~\cite{hendrickx2012convergence}. Moreover, if nodes $i$ and $j$ interact persistently in the sense that  
\[
\int_{0}^{\infty} a_{ij} (t) \, dt = \infty,
\]
then their final opinions coincide, i.e.,  $s'_i = s'_j$~\cite{hendrickx2012convergence}.

\begin{table}[htbp!]\scriptsize
\centering
\renewcommand{\arraystretch}{1.3}
\begin{tabular}{|c|p{2.5cm}|p{5cm}|p{4cm}|}
\hline
\textbf{Model} & \textbf{Assumptions} & \textbf{Stabilization Time} & \textbf{Parameters Explanation} \\
\hline
\multirow{5}{*}{Voter Model (p=1)} 
& General graph & $O(n^3 \log n)$~\cite{hassin2001distributed} & under synchronous update \\
\cline{2-4}
& General graph & $O(n^3)$~\cite{kanade2023coalescence} & under synchronous update \\
\cline{2-4}
& General graph  & $O\left(\frac{1}{1 - \lambda_2} \left( \log^4 n + \frac{n}{\rho} \right) \right)$~\cite{cooper2013coalescing} & $\lambda_2$ is the second eigenvalue of the transition matrix of a random walk on $G$, and $\rho = \frac{(\sum_{u \in V} d(u))^2}{\sum_{u \in V} d(u)^2}$, and the model unfold synchronously \\
\cline{2-4}
& Graphs with rewired edges & $O\left(\frac{|E|}{d_{\text{min}} \cdot \phi}\right)$~\cite{berenbrink2016bounds} & $\phi$ is the conductance and $d_{\text{min}}$ is the minimum degree, and the model unfold synchronously \\
\cline{2-4}
& Scale-free graph & $O(n)$ for $\alpha>3$, $O(n/\ln n)$ for $\alpha=3$, $O(n^{2(\alpha-2)/(\alpha-1)})$ for $2<\alpha<3$, $O(\ln(n)^n)$ for $\alpha=2$, $O(1)$ for $\alpha<2$~\cite{sood2008Voter} & Degree distribution is $n_k \sim k^{-\alpha}$, and the model updates asynchronously \\
\hline
\multirow{2}{*}{Galam Model (p=1)} & - & $O(\log(\log(n)))$~\cite{gartner2020threshold} & If all the rooms are of size greater than $3$ \\
\cline{2-4}
& - & $O(\log(n))$~\cite{gartner2020threshold} & If all the rooms are of size $2$ \\
\hline
\multirow{4}{*}{Majority Model (p=2)} & Unweighted graph & $O(n^2)$~\cite{fogelman1983transient} & $-$ \\
\cline{2-4}
& Unweighted graph & $\Omega\left(\frac{n^2}{\log^2 n}\right)$~\cite{frischknecht2013convergence} & $-$ \\
\cline{2-4}
& random $d$-regular graph & $O (\log_d n)$~\cite{zehmakan2020opinion} & $-$ \\
\cline{2-4}
& Weighted graph & $\min\{2\omega(G), 2^n\}$~\cite{keller2014even} & $\omega(G)=\sum_{\{u,v\}\in E} w_{uv}$ \\
\hline
\multirow{1}{*}{Random Majority (p=exp)} & General graphs & Exponential in $n$~\cite{zehmakan2023random} & $-$ \\
\hline
Ising (p=2) & General graphs &  $ O(n \log(n))$~\cite{chen2021optimal}, \cite{mossel2013exact} & if $(d_{\max}-1)\tanh(\beta)<1$, where $d_{\max}$ denotes maximal degree and $\beta$ is bounded. \\
\hline
Sznajd (p=2) & Cycle &  $T(n) \propto n^{-3/2}$~\cite{sznajd2000opinion} & The result is based on Monte Carlo simulations \\
\hline
\multirow{2}{*}{PUSH (p=1)} 
& Complete graph & $(1 + o(1))(\log_2 n + \log n)$~\cite{frieze1985shortest} & $-$ \\
\cline{2-4}
& Hypercubes and Erd\H{o}s--R\'enyi graphs & $O(\log n)$~\cite{feige1990randomized} & $p = \frac{c\log n}{n}$ where $c > 1$ \\
\cline{2-4}
\hline
\multirow{1}{*}{PULL (p=1)} 
& General graph & $50(\beta+2)\, \log n ({\phi}^{-1} +  \frac{d_{\max}}{\lceil \phi\, \mathrm{vol}(S) \rceil})$ with probability at least $1 - O(n^{-\beta})$~\cite{giakkoupis2011tight} &   $S \subseteq V$ is an (arbitrary) initial set of $+1$ vertices,  $\beta > 0$ is a fixed parameter, and $\mathrm{vol}(S)$ denotes the sum of degrees of vertices in $S$. \\
\hline
\multirow{4}{*}{PUSH-PULL Protocol (p=1)} 
& Regular graphs (Lower bound) & $\log_2 n + \log n - o(\log n)$~\cite{elsasser2009runtime} & $-$ \\
\cline{2-4}
& Dynamic graphs & $O(\log n)$~\cite{panagiotou2015randomized} & Edges appear and disappear over time \\
\cline{2-4}
& General graphs (Asynchronous update) & $O(T(G)+\log n)$~\cite{giakkoupis2016asynchrony} & $T(G)$ is the stabilization time of the synchronous PUSH-PULL protocol \\
\cline{2-4}
& Preferential attachment graphs & $O(\log^2 n)$~\cite{chierichetti2011rumor} & $-$ \\
\hline
\multirow{1}{*}{\textit{r}-Bootstrap Percolation (p=1)} 
& General graph & \(O(|E|)\)~\cite{fogelman1983transient} & $-$ \\
\cline{2-4}
\hline
\multirow{1}{*}{Linear Threshold (p=1)} 
& General graph & \(O(n)\) & $-$ \\
\cline{2-4}\hline
\multirow{1}{*}{Independent Cascade (p=1)} 
& General graph & \(O(n)\) & Simple graphs \\
\cline{2-4}
\hline
\multirow{2}{*}{French-DeGroot model (p=1)} 
& Synchronous & $O(\frac{-1}{\log(|\lambda_2|)})$~\cite{ding2019consensus} & $\lambda_2$ is the second largest eigenvalue of the transition matrix \\
\cline{2-4}
& Asynchronous & $O(4d_{\max}n^2 \lceil \log(1/\epsilon\rceil))$~\cite{elboim2024asynchronous} & $\varepsilon$-stabilization time $\tau_\varepsilon = \min\{ t : \max_{v} s_t(v) - \min_{v} s_t(v) \leq \varepsilon \}$ \\
\cline{2-4}
\hline
\multirow{3}{*}{The DW (p=2)} 
& Under Consensus &  $Cn^2 \exp(-C' t)$~\cite{zhang2015convergence} & $C$ and $C'$ are constants and $t$ is time. \\
\cline{2-4}
& Under Polarization & $Cn^2 t^{-1}$~\cite{zhang2015convergence} & $C$ is a constant. \\
\cline{2-4}
\hline
\multirow{5}{*}{The HK (p=2)} 
& Scalar opinion & at least $\Omega(n^2)$ and at most $O(n^3)$ \cite{bhattacharyya2013convergence, mohajer2013convergence, wedin2015quadratic} & $-$ \\
\cline{2-4}
& Multidimensional opinions & \( n^{O(n)} \)~\cite{chazelle2011total} & $-$ \\
\cline{2-4}
& \( d \geq 2 \) & $O(n^{10}d^2)$~\cite{bhattacharyya2013convergence},  $O(n^8)$~\cite{etesami2015game},  $O(n^4)$~\cite{martinsson2016improved}, $O(n^4)$~\cite{martinsson2016improved}   & $-$ \\
\cline{2-4}
\hline
\end{tabular}
\caption{Summary of convergence properties of opinion dynamic models. The letter p stands for periodicity.}
\label{tab:model_stabilization}
\end{table}

\section{Viral Marketing}\label{Viral_Marketing}

\textit{Viral marketing} is a strategy that aims to maximize the adoption of a product or idea by strategically selecting a small set of seed users to trigger large cascades in social networks. This approach has gained significant attention in both theoretical and applied research. At its core, viral marketing seeks to understand how information, behaviors, or innovations spread through networks. In the context of opinion dynamics and information spreading models, while the term ``viral marketing'' is commonly associated with the influence maximization problem in the IC and LT models popularized by \cite{kempe2003maximizing}, it encompasses a broader spectrum of research questions and optimization problems. These questions can be categorized into four distinct groups, each addressing a unique aspect of information diffusion and activation within networks.

\begin{itemize}
\item \textbf{Influence Maximization.} The first group focuses on the classic influence maximization (IM) problem, introduced initially in~\cite{domingos2001mining,richardson2002mining}, where the goal is to identify a set of initial nodes (called seed nodes, which are the initially $+1$ nodes), constrained by a budget (typically representing the maximum number of seed nodes), that maximizes the expected final number of $+1$ nodes under a given diffusion model. This problem has been extensively studied for models such as IC and LT, e.g.,~\cite{kempe2003maximizing}.

\item \textbf{Minimum Conversion Set.} The second group shifts the focus to optimization from a different perspective, where the goal is to determine the minimum number of initial $+1$ nodes (also called \textit{conversion sets} by \cite{dreyer2009irreversible}) required for the entire network to become $+1$ at the end of the diffusion process.

\item \textbf{Structural Diffusion Bounds.} 
The third line of research focuses on theoretical bounds rather than algorithmic solutions. Again, the aim is to study the minimum number of $+1$ nodes required at the beginning of the process for the entire graph to become $+1$ at the end. However, in this setup, the objective is to establish bounds on this value in terms of various graph and model parameters. This is usually referred to as bounds on the minimum size of \textit{target sets}~\cite{abebe2018opinion}, \textit{contagious sets}~\cite{feige2017contagious}, \textit{winning sets}~\cite{zehmakan2023random}, \textit{dynamic monopolies}~\cite{jeger2019dynamic} (or \textit{dynamos}), or \textit{percolating sets}~\cite{janson2012bootstrap} in the literature. Still, we will continue to use the conversion set for consistency.

\item \textbf{Threshold Analysis.} Finally, the fourth group investigates a random initial setup and is interested in threshold behaviour. Suppose that nodes initially adopt $+1$ with some probability $p$ and $-1$ with probability $1-p$, and the question is to find the minimum $p$ for which the process results in a full $+1$ at the end with high probability. This phenomenon, known as \textit{phase transition} or \textit{threshold behaviour}, has been studied in models such as bootstrap percolation~\cite{balogh2006bootstrap}, epidemiological processes~\cite{newman2002spread}, and the Galam model~\cite{gartner2020threshold}, providing a probabilistic lens on network activation.
\end{itemize}

Together, these four groups provide a comprehensive framework for understanding the multifaceted nature of viral marketing and information diffusion in networks. 

Viral marketing finds compelling interpretations across domains: in opinion dynamics, it identifies small groups whose early adoption of a belief can shift entire populations; in epidemiology, it models critical \textit{superspreaders} whose infection guarantees an epidemic’s dominance; in economics, it optimizes product launches by seeding influencers to maximize peer-driven adoption; and in cybersecurity, it reveals vulnerable nodes whose compromise would accelerate malware or misinformation. Defence strategies thus aim to either eliminate small conversion sets, making networks resistant to cascades, or ensure these sets are large enough to deter malicious exploitation.

It should be noted, however, that not all four problem groups have been studied for every diffusion model, highlighting opportunities for further research in this area. Furthermore, for some models, such as the PUSH-PULL protocol, these questions may be trivial, since a single node suffices to reach consensus.

\subsection{Preliminaries}

In this section, we review key concepts and tools essential to our analysis of viral marketing results. 

\begin{definition*}[\textbf{Monotonically Increasing Function}]
A function $f: 2^V \to \mathbb{R}$ is \emph{monotonically increasing} if for all subsets $A \subseteq B \subseteq V$, the inequality  $f(A) \leq f(B)$ holds. 
\end{definition*}

\begin{definition*}[\textbf{Submodular Function}]
A function $f: 2^V \to \mathbb{R}$ is \emph{submodular} if for all subsets $A \subseteq B \subseteq V$ and any element $e \in V \setminus B$, it satisfies $ f(A \cup \{e\}) - f(A) \geq f(B \cup \{e\}) - f(B)$. 
\end{definition*}

\begin{definition*}[\textbf{$\alpha$-Approximation Algorithm}]
For a maximization problem, an algorithm $\mathrm{ALG}$ achieves an $\alpha$-approximation ($0 < \alpha \leq 1$) if for all inputs $x$, $\mathrm{ALG}(x) \geq \alpha \cdot \mathrm{OPT}(x)$. Similarly, for minimization case, an algorithm $\mathrm{ALG}$ achieves an $\alpha$-approximation ($\alpha \geq 1$) if for all inputs $x$, $\mathrm{ALG}(x) \leq \alpha \cdot \mathrm{OPT}(x)$ where $\mathrm{OPT}(x)$ denotes the optimal value for input $x$.
\end{definition*}

\paragraph{Greedy Algorithm.} The greedy algorithm for an optimization problem constructs a solution by solving the problem iteratively. At each iteration, it selects the best possible option based on a locally optimal choice. For maximization problems involving monotone and submodular functions, the greedy algorithm provides a $(1 - 1/e)$-approximation guarantee, where $e$ is the base of the natural logarithm~\cite{nemhauser1978analysis}. This result is foundational in the analysis of approximation algorithms for submodular optimization.

\begin{definition*}[\textbf{Max-$k$-Coverage Problem}]
Given a collection of sets and an integer~$k$, the goal is to select $k$ sets that maximize the number of covered elements. Formally, let $U$ be a universe of elements and $\mathcal{S} = \{A_1, A_2, \dots, A_m\}$ be a collection of subsets of $U$. The objective is to find a subset $\mathcal{S}' \subseteq \mathcal{S}$ with $|\mathcal{S}'| \leq k$ such that $\left| \bigcup_{A \in \mathcal{S}'} A \right|$ is maximized.
\end{definition*}
This problem is NP-hard, but the greedy algorithm provides a $(1 - 1/e)$-approximation, see e.g., \cite{hochbaum1998analysis}. 

\begin{definition*}[\textbf{Vertex Cover Problem}]
Given an undirected graph $G = (V, E)$, a vertex cover is a subset $C \subseteq V$ such that every edge in $E$ is incident to at least one node in $C$. The goal is to find a minimum-size vertex cover.
\end{definition*}
This problem is also NP-hard, but it admits a $2$-approximation algorithm based on greedy strategies or linear programming relaxations~\cite{vazirani2001approximation}. 

\paragraph{Monte Carlo Method.} The Monte Carlo method is a class of computational algorithms that rely on repeated random sampling to obtain numerical results. It is beneficial for estimating quantities that are difficult to compute deterministically, such as expected values or probabilities, and is often used in simulation-based evaluations or randomized algorithms.

\subsection{Discrete Models}

\subsubsection{Voter Model}

Hassin and Peleg~\cite{hassin2001distributed} prove lower and upper bounds for the size of the conversion set that can bring the Voter model to consensus with high probability. It is shown that for probability $1/2$, a constant-size conversion set suffices. However, for probabilities slightly higher than $1/2$ (specifically, $1 - \varepsilon$ for $\frac{1}{n} \leq \varepsilon < \frac{1}{2}$), the smallest conversion sets are of size $\Theta\left(\sqrt{\frac{n}{\varepsilon}}\right)$. Considering the Voter model under competing products, however, the IM problem has been proven to be NP-hard by Zehmakan
et al.~\cite{zehmakan2023viral} through a reduction from the maximum coverage problem. Zehmakan
et al.~\cite{zehmakan2023viral} also develop a polynomial-time approximation algorithm with the best possible approximation guarantee by leveraging the monotonicity and submodularity of the objective function. Even-Dar and Shapira~\cite{even2007note} show that for the IM problem, the most natural heuristic solution, which selects the nodes in the network with the highest degree, is indeed the optimal solution. In the same setup, Zhou et al.~\cite{zhou2014maximizing} consider the IM problem, but instead of focusing on the expected $+1$ nodes at the end of the process, they focus on the expected value of the total $+1$ nodes \textit{from start to end}. They provided the exact solution for this setup.

A growing body of research focuses on maximizing the fixation probability in the Moran process, that is, the likelihood that the \( +1 \) state successfully takes over the entire network (referred to as \textit{invasion}), quantified by the probability of consensus on state \( +1 \). In this context, given a budget \( k \), the problem of selecting a set of \( k \) agents to initially adopt opinion \( +1 \) to maximize the overall fixation probability is NP-hard. As a result, greedy algorithms are commonly used for seed selection as an efficient approach; see, for example,~\cite{petsinis2023maximizing, brendborg2022fixation, durocher2022invasion, petsinis2024seed}.

\subsubsection{Majority Models}

\paragraph{Galam Majority Model. }
Gärtner and Zehmakan~\cite{gartner2020threshold} study this model from a mathematical perspective and investigate its threshold behavior. They demonstrate that if all groups are of the same size, with each group consisting of at least three individuals, there exists a threshold value $\alpha$ (which depends on the rooms size) such that for any $\epsilon > 0$, if the initial density $P^+(0) \geq \alpha + \epsilon$, opinions will converge to $+1$. Conversely, if $P^+(0) \leq \alpha - \epsilon$, opinions will converge to $-1$.

Through experiments, Li and Zehmakan~\cite{li2023graph} investigate the IM problem on graphs using a graph variant of the Galam model, where seeds are selected based on re, such as centrality measures.

\paragraph{Graph-based Majority Model. } 
It is proven that there are arbitrarily large graphs for which dynamos can be surprisingly small; Berger~\cite{berger2001dynamic} proves that for every $n$, there exists a graph with at least $n$ nodes containing a conversion set of size $18$. Additionally, Peleg~\cite{peleg1998size,peleg2002local} shows that for all graphs, the size of a conversion set that converts all nodes to state $+1$ in a single step must be at least $\Omega(\sqrt{n})$. The same bound is proven for two-round conversion sets~\cite{peleg1998size}, namely, those that lead to all $+1$ states in exactly two rounds. 

The problem of finding the minimum size of a conversion set for a given graph $G$ is known to be NP-hard for various majority-based opinion dynamics models~\cite{karia2023hard}. Mishra et al.~\cite{mishra2002hardness} prove that this problem cannot be approximated within a factor of $\left(\frac{1}{3} - \epsilon\right) \ln n$ unless $\text{NP} \subseteq \text{DTIME}(n^{O(\log \log n)})$. However, for general graphs, Auletta et al.~\cite{auletta2018reasoning} establish that every graph has a conversion set of size at most $n/2$ under the asynchronous variant of the majority model. Additionally, Out and
Zehmakan~\cite{out2021majority} explores the minimum size of a conversion set in real-world social networks for a modified majority model, where high-degree nodes (referred to as ``elites'') exert a disproportionately larger influence than other nodes. Considering the minimum size of a conversion set, Fazli et al.~\cite{fazli2014non} consider modeling the diffusion of two competing technologies over a social network.

Considering random $d$-regular graphs,  Gärtner and Zehmakan~\cite{gartner2018majority} demonstrate that, if the initial probability of being in state $+1$ is less than $\frac{1}{2} - \epsilon$, for a given $\epsilon>0$, then the final states of all nodes will be $-1$ in $O(\log_d \log n)$ rounds with high probability, provided that $d \geq \frac{c}{\epsilon^2}$ for some constant $c$. Zehmakan~\cite{zehmakan2020opinion} proves that, if $\lambda_2 = o(d)$ where $\lambda_2$ denotes the second-largest absolute eigenvalue of its adjacency matrix, the majority model, starting from $(\frac{1}{2} - \delta)n$ $-1$ nodes (for an arbitrarily small constant $\delta > 0$), results in a fully $+1$ configuration in sub-logarithmically many rounds.

Faliszewski et al.~\cite{faliszewski2022opinion} study the problem of selecting $k$ nodes to change their states (opinions) under majority-based preference diffusion (both synchronous and asynchronous updates), where nodes have preferences over candidates and update their states based on the majority influence in their neighborhood, to reach consensus on the preferred candidate. The problem is proven to be NP-hard, and an integer linear programming-based algorithm is proposed to solve it.

Considering the majority model with asynchronous update, Bredereck and Elkind~\cite{bredereck2017manipulating} show that two extreme update sequences, one maximizing the spread of opinion $+1$ (best-case for manipulation) and the other maximizing $-1$ (worst-case), and show that both of these opposing extremes ultimately lead to the very same final distribution of opinions.

\subsubsection{Ising Model}
Galam et al.~\cite{galam1982sociophysics} introduce a dissatisfaction function that the system minimizes to reach stability, leading to two possible states: a \textit{work state} and a \textit{strike state}. These states are separated by a critical point where minor disturbances can cause sudden, irreversible shifts in group behavior.

\subsubsection{Snajad Model}

Among the four groups of viral marketing problems, the phase transition problem is very well-studied under this model. In an empirical study, Stauffer et al.~\cite{stauffer2000generalization} generalize the Sznajd model on a square lattice, introducing six variants with different dynamical rules. In one such variant, a \(2\times2 \) plaquette of four neighbouring nodes influences its surrounding eight neighbours only if all nodes within the plaquette are aligned. In this variant, a phase transition is observed: when the initial density of \(+1\) nodes exceeds 0.5, the system eventually reaches a consensus configuration with all nodes taking the value \(+1\); conversely, if the initial density is below 0.5, the system converges to all nodes taking the value \(-1\). This behaviour emerges only in sufficiently large systems (specifically, lattices of size \(101 \times 101\) or larger). Notably, such a transition does not occur in the one-dimensional Snajad model, as reviewed by Sznajd-Weron et al.~\cite{sznajd2021review}. 

Ochrombel~\cite{ochrombel2001simulation} introduces a significant simplification of the model, fundamentally altering its mechanism. Instead of requiring a pair of nodes to influence their neighbours, a single node could convince its nearest neighbours. However, Schulze~\cite{schulze2003advertising} demonstrates, through experiments, that this modification eliminated the previously observed transition, a result later confirmed through analytical calculations on a complete graph by Slanina and Lavicka~\cite{slanina2003analytical}. Even introducing long-range interactions in place of nearest-neighbour interactions failed to restore the phase transition, as shown by Schulze~\cite{schulze2003long}. 

Extensions of the Sznajd model have introduced additional complexity. For example, Sabatelli and Richmond~\cite{sabatelli2003phase} propose a consensus model incorporating memory effects. Building on the Sznajd model, this paper introduces a consensus model in which individuals retain memory of past states. Using synchronous updating, they examined the conditions under which the system transitions from a configuration without consensus to one with complete consensus. Their findings indicate that consensus is reached only if the initial asymmetry (measured by the absolute difference between the densities of $+1$ and $-1$ nodes) exceeds a certain critical threshold. This threshold depends on the length of memory retained by the individuals. 

\subsubsection{Bootstrap Percolation Model}
For $r$-BP (where $r \geq 3$), it is NP-complete to determine if a graph has a conversion set of size at most (a positive integer) $k$ as proven by  Dreyer Jr and Roberts~\cite{dreyer2009irreversible}. On the other hand, for the undirected path and cycle on \( n \) nodes, the size of the smallest conversion set for $r=2$, denoted as $C_2$, is given by $\lceil \frac{n+1}{2} \rceil$ and $\lceil \frac{n}{2} \rceil$, respectively~\cite{dreyer2009irreversible}. For the complete bipartite graph \( K_{n_1,n_2} \), this number is given as \cite{dreyer2009irreversible} 
\[
C_r(K_{n_1,n_2}) =
\begin{cases} 
r, & \text{if } n_1, n_2 \geq r \\ 
n_1, & \text{if } n_2 < r \leq n_1 \\ 
n_2+n_1, & \text{if } n_1,n_2 < r 
\end{cases}
\]
If \( G \) is a tree in which every internal node (any node that is not a leaf) has degree greater than \( r \) for \( r > 1 \), then, $C_r(G)=|L_G|$ where $|L_G|$ denotes the set of leaves of $G$~\cite{dreyer2009irreversible}. 

Janson et al.~\cite{janson2012bootstrap} demonstrate that for Erd\H{o}s--R\'enyi graphs with the probability parameter $p=p(n)$ of the edges in the graph, the final number of $+1$ nodes exhibits a threshold behavior: with high probability, it is either small (at most twice the initial number of $+1$ nodes) or large, sometimes reaching $n$. However, if $p$ is sufficiently small such that the graph contains many nodes of degree less than $r$, then, with high probability, these nodes cannot become $+1$ (except if they are initially), and the final number of $+1$ nodes is at most $n - o(n)$. For these graphs with $p=\frac{d}{n}$ and $1 \ll d \ll (\frac{n\log (\log n)}{\log^2 n})^{\frac{r-1}{r}}$, Feige et al.~\cite{feige2017contagious} prove that the minimal size of a conversion set is $\Theta(\frac{n}{d^{\frac{r}{r-1}}\log d})$ with high probability.

Balogh and Bollobás~\cite{balogh2006bootstrap} study the $d$-dimensional hypercube, where in the initial configuration each of the $2^d$ nodes is $+1$ with probability $p$ and $-1$ with probability $1 - p$, independently of the others. They prove that there exist constants $c_1, c_2 > 0$ such that for $p(d) \geq \frac{c_1}{d^2} 2^{-2\sqrt{d}}$, the probability that the entire graph becomes $+1$ tends to $1$ as $d \to \infty$, while for $p(d) \leq \frac{c_2}{d^2} 2^{-2\sqrt{d}}$, this probability tends to $0$.

Considering the case when $G$ is the $d$-dimensional grid $[L]^d = \{1, \cdots, L\}^d$, Balogh et al.~\cite{balogh2012sharp} prove that the phase transition for every $d$ and $L$, where $d \gg r \gg 2$, is given by 
\[
\left(\frac{C(d,r)+o(1)}{\log_{(r-1)}(L)}\right)^{d-r+1},
\]
where $C(d,r) > 0$ is a constant and $\log(r)$ denotes an $r$-times iterated logarithm. In the same setup, but considering two-way $r$-BP on a $d$-dimensional grid, Zehmakan~\cite{zehmakan2021threshold} identifies the threshold values. Several papers are devoted to identifying this threshold for specific values of $d$ and $r$. We refer the interested reader to the references therein~\cite{balogh2012sharp,zehmakan2021threshold} for more on grids and $r$-BP. 

The existence of the phase transition has also been established in random $d$-regular graphs~\cite{balogh2007bootstrap} for $d \geq 3$,  random graphs with a power-law degree distribution~\cite{amini2014bootstrap},  preferential attachment graphs~\cite{amin2018phase},  homogeneous trees~\cite{fontes2008bootstrap}, stochastic block models~\cite{torrisi2023bootstrap}, and infinite trees~\cite{balogh2006bootstrap}.

\subsubsection{Linear Threshold and Independent Cascade Models}

Considering the IM problem in the context of the IC and IM model, Kempe et al.~\cite{kempe2003maximizing} demonstrate that the objective function exhibits submodularity, which allows the greedy algorithm to achieve a $(1 - (1 - 1/k)^k)$-approximation where $k$ is the seed size (this finding was later extended to all diffusion models that are locally submodular by Kempe et al.~\cite{kempe2005influential} and by Mossel and Roch~\cite{mossel2007submodularity}). 
Khanna and Lucier show that the $(1 - (1 - 1/k)^k)$ barrier can be overcome in the IC model when graphs are restricted to be undirected. They prove that the greedy algorithm for IM under the IC model for undirected graphs achieves a $(1 - (1 - 1/k)^k + c)$-approximation for some constant $c > 0$ that does not even depend on $k$. Considering the LT model with directed graphs, this ratio is proven to be  $\left(1 - \left(1 - \frac{1}{k} \right)^k + \Omega\left(\frac{1}{k^3} \right)\right)$-approximation \cite{schoenebeck2020limitations}.
Furthermore, Schoenebeck et al.~\cite{schoenebeck2020limitations} prove that the $(1 - (1 - 1/k)^k)$-approximation ratio is tight in two scenarios: 1) when the graph is directed, and 2) when the nodes are weighted. In both cases, the greedy algorithm cannot achieve a $(1 - (1 - 1/k)^k + f(k))$-approximation for any positive function $f(k)$. 

Since computing the objective function for the IM problem is computationally hard, for example \#P-hard for the IC and LT models~\cite{kempe2003maximizing}, Monte Carlo simulations are typically used. However, their high computational cost has motivated various optimization efforts. In this line, Leskovec et al.~\cite{leskovec2007cost} reduce redundant evaluations using submodularity, while Goyal et al.~\cite{goyal2011celf++} speed up computations by processing multiple nodes simultaneously. Despite these improvements, memory usage remains an issue. To address this, Rostamnia and Kianian~\cite{rostamnia2019vertex} propose filtering inefficient nodes before applying the greedy algorithm, while Li and Liu~\cite{li2019modeling} leverage clique structures to identify top \( k \) influential nodes more efficiently. These methods enhance both speed and scalability.

IM has also been studied in the adaptive setting, where seeds are selected iteratively. In this setting, the seed-picker can observe the cascade resulting from previously chosen seeds before deciding the next one, as shown by Golovin and Krause~\cite{golovin2011adaptive}, Han et al.~\cite{han2018efficient}, Peng and Chen~\cite{peng2019adaptive,chen2019adaptivity}. 

Building on the classic IM problem, researchers have explored the competitive IM problem, where multiple entities compete to maximize their influence in a social network. Given an influence propagation model and the probability that each node is pre-seeded by a competitor, the goal is to select a set of $k$ seed nodes to maximize the expected influence cascade in the presence of competing influences; see, for example,~\cite{borodin2010threshold,pathak2010generalized,bharathi2007competitive,wu2015maximizing,carnes2007maximizing,datta2010viral,lin2015analyzing,zhu2016minimum},

It should be noted that, in addition to greedy-based algorithms, other approximation algorithms have been applied to the IM under these two models. For instance, Goldberg and Liu~\cite{goldberg2013diffusion} propose an approximation algorithm based on linear programming, while Angell and Schoenebeck~\cite{angell2017don} and Schoenebeck and Tao~\cite{schoenebeck2019beyond} develop a dynamic programming approach.

In addition to the IM problem, Chen~\cite {chen2009approximability} studies conversion set selection for the LT model. Unlike prior work, here threshold values are deterministic and are given by $t(v) \in \mathbb{N}$, where $1 \leq t(v) \leq d(v)$. The conversion set selection problem (finding the minimum size of a conversion set) in this setup cannot be approximated within a ratio of \( O(2^{\log^{1-\epsilon} n}) \) for any fixed constant \( \epsilon > 0 \), unless \( \text{NP} \subseteq \text{DTIME}(n^{\text{polylog}(n)}) \). The conversion set selection problem has also been analyzed under the IC model, where different influences compete with one another in a social network~\cite {zhu2016minimum}.

\subsubsection{Epidemiological Processes}

A fundamental question in epidemiological studies is whether a disease (or rumor) will spread widely or fade out, a topic that falls under the category of threshold analysis. This problem has been extensively explored using methods from statistical physics, see e.g.,~\cite{wang2017unification}. The outcome depends on key factors such as the infection rate, recovery rate, and the underlying structure of connections between individuals. 

\paragraph{The SIS Model.}
Boguná et al.~\cite{boguna2003absence} show that for scale-free networks with parameter (\(2 < \alpha \leq 3\)), there's no epidemic threshold, meaning epidemics can always happen, regardless of how the graph is structured. In a similar setup (scale-free graphs), and by analyzing real data from computer virus infections to determine the average lifetime and prevalence of viral strains on the Internet, Pastor-Satorras and Vespignani~\cite{pastor2001epidemic} find no epidemic threshold, allowing viruses to spread continuously. The epidemic threshold, in the presence of explicit correlations among the degrees of connected nodes, was shown by Boguná and Pastor-Satorras~\cite{boguna2002epidemic} to be inversely proportional to the largest eigenvalue of the adjacency matrix. The same result has been derived for interconnected networks by  Wang et al.~\cite{wang2013effect}. Dykman et al.~\cite{dykman2008disease} reveal a fundamental connection between extinction dynamics and epidemic thresholds in this model.  The system exhibits critical behavior near the epidemic threshold ($R_0 = 1$), indicating a continuous phase transition, with disease persistence becoming increasingly fragile as $R_0 \to 1^+$. We refer the interested reader to~\cite{cai2016solving} and~\cite{valdano2018epidemic} for the closed-form expression of the epidemic threshold, derived from the analytical solution of the SIS differential equations model.

\paragraph{The SIR Model.}
Newman~\cite{newman2002spread} develops a theoretical framework for analyzing the SIR model on networks using percolation theory and generating function methods. The author derives exact solutions (for the governing differential equations) for networks where transmission probabilities between nodes may or may not be correlated, using generating functions to compute key properties such as outbreak sizes and critical thresholds. For power-law networks (where degree distribution is given by $\sim k^{-\alpha}$), the epidemic threshold vanishes when $\alpha \leq 3$, enabling epidemics at any transmission rate. When $\alpha > 3$, a finite threshold emerges, requiring transmission to exceed this critical value. Dickison et al.~\cite{dickison2012epidemics} consider interconnected graphs (graphs composed of multiple distinct subgraphs) and show that in strongly coupled graphs, epidemics always spread globally and are accelerated by interconnections. In contrast, epidemics may remain localized in weakly coupled graphs, with interconnections only amplifying the spread within vulnerable subgraphs. Wang et al.~\cite{wang2016predicting} compare some mean-field-based methods for predicting epidemic thresholds.

\subsection{Continuous Models}

\subsubsection{French-DeGroot model}

Zhou et al.~\cite{zhou2023sublinear} investigate the problem of maximizing the overall opinion in equilibrium by adding \(k\) edges, each connecting a 1-leader (a node with a fixed opinion value of 1) to a follower (a node that updates its opinion through weighted averaging of in-neighbors’ opinions). The objective function is proven to be monotonically increasing and submodular; hence, the greedy algorithm yields an approximation ratio of \(1 - 1/e\). In a similar setup, Zhou and Zhang~\cite{zhou2023opinion} investigate the leader selection problem in the French-DeGroot model where, given \(n_0 = \Theta(1)\) leaders with opinion $0$, the objective is to maximize the average equilibrium opinion by selecting \(k = \Theta(1)\) leaders with opinion $+1$ from the remaining \(n - n_0\) nodes. As the objective function is monotonically increasing and submodular, the greedy algorithm again yields an approximation factor of \(1 - 1/e\).

By incorporating leadership into this model, Dong et al.~\cite{dong2017managing} show that the final consensus opinion is a linear combination of the leaders' initial opinions. They also examine the problem of adding the minimum number of edges required to achieve consensus in the presence of leaders.

\subsubsection{Friedkin-Johnsen Model}
Gionis et al.~\cite{gionis2013opinion} employ the FJ model to identify a set of conversion nodes whose expressed opinions can be modified to $+1$ to optimize the sum of expressed opinions (at equilibrium). By reducing from vertex cover on regular graphs, this problem has been proven to be NP-hard. Also, they showed that the objective function is monotonically increasing and submodular, and thus, it can be approximated within a factor of \((1-1/e)\) using a greedy algorithm. A similar optimization problem has been studied for unweighted directed graphs~\cite{sun2023opinion}.
In a similar work, Xu et al.~\cite{xu2020opinion} consider this optimization problem: Given a graph \(G\), innate opinion vectors, and a budget amount \(k\), determine the modification of innate opinions such that the overall opinion is maximized. They also study this problem: Given a graph \(G\), the innate opinion vector, and an integer \(k\), seek a set \(S\) of \(k\) nodes and fix \(s_{i}\) to \(u_{i}\) for \(v_i \in S\), such that the overall opinion is maximized. 

Additionally, Abebe et al.~\cite{abebe2018opinion} study this problem under the FJ model: given a graph with an internal opinion reflecting their internal opinion and their susceptibility-to-persuasion parameter, how should the nodes' susceptibilities be modified to maximize (or minimize) the total sum of opinions at equilibrium? The optimization problem is NP-hard by a reduction from the vertex cover problem on $d$-regular graphs, and the objective function is neither sub- nor super-modular. The problem of recommending $k$ new edges between a $1$-leader and follower nodes to maximize the equilibrium overall opinion is studied by Zhu and Zhang~\cite{zhu2025opinion}.

There is also a line of research that focuses on selecting a set \(S\) of \(k\) missing edges to maximize the expected final number of \(+1\) nodes in the LT and IC models. The approach integrates the LT and IC models (which capture influence diffusion) and the FJ model (which governs opinion updates through averaging the opinions of in-neighborhoods). This combination simulates how opinions evolve dynamically after activation. See \cite{tu2022viral} and \cite{he2021positive} for relevant references.

\subsubsection{Hegselmann-Krause and  Deffuant-Weisbuch Models}
Hegselmann et al.~\cite{hegselmann2014optimal} examine a scenario where normal nodes update their opinions based on the HK model, while strategic nodes treat opinions as strategies to influence the dynamics. The research focuses on a benchmark problem where a strategic node aims to control opinion dynamics over a campaign period by strategically placing opinions to steer as many normal nodes as possible toward a desired opinion interval by the campaign's end. The authors find that even in simple cases, such as those involving a small number of normal nodes, a single strategic node, and a short campaign period, the problem is complex and challenging to solve.

He et al.~\cite{he2022dynamic} study the influence maximization problem of selecting seed nodes using a combination of the LT model (for activation) and the HK model (for opinion evolution). 

Pineda and Buendía~\cite{pineda2015mass} study the effect of external mass media on the DW and the HK models. They consider two scenarios: (1) two distinct confidence levels, and (2) each individual having a unique confidence level. The findings show that without mass media, confidence diversity can help the system reach consensus. When mass media is present, it is most effective at convincing people if the group has a moderate mix of different confidence levels, neither too similar nor too diverse.

\subsection{Alternative Objective Functions: Polarization and Disagreement}

So far, we have focused on objective functions related to overall opinion or the expected number of activated nodes in a network. However, recent research has explored alternative optimization goals beyond these conventional measures, particularly in continuous models. In this section, we review some of these approaches, focusing on various optimization objectives and their implications for network analysis.

Within the FJ model framework, where polarization is defined as the norm of expressed opinions (divided by $n$), Matakos et al.~\cite{matakos2017measuring} investigate the mitigation of polarization through opinion moderation. Their approach encourages more neutral stances via education, exposure to diverse viewpoints, or incentives. This problem was proven to be NP-hard through a reduction from the $m$-subset sum problem~\footnote{Defined as: given a set of $n$ positive integers $v_1,...,v_n$, a value $m$, and target $b$, determine whether there exists a subset $B$ of size $m$ such that $\sum_{v_i \in B} v_i = b$}. Musco et al.~\cite{musco2018minimizing} subsequently redefine polarization as the variance of opinions at equilibrium, demonstrating that the polarization index alone is non-convex as an objective function. The combined polarization-disagreement index (disagreement is defined as the sum of squared differences between nodes' expressed opinions at equilibrium) is convex, and then minimizing this combined index by modifying the innate opinions of $k$ nodes admits a polynomial-time solution. To mitigate this index, Liwang and Zhongzhi~\cite{zhu2022nearly} propose a greedy-based algorithm.
Building on this foundation, Zhu et al.~\cite{zhu2021minimizing} minimize the index by adding $k$ edges to the network. In contrast, Wang and Kleinberg~\cite{wang2024relationship} prove that link recommendation algorithms do not increase the polarization-disagreement index, though this doesn't hold for polarization alone. Chitra and Musco~\cite{chitra2020analyzing} demonstrate that small edge weight changes (modeling filter bubbles) can significantly increase polarization. Further developments by Bhalla et al.~\cite{bhalla2023local} show that friend-of-friend connections and confirmation bias amplify polarization, while their removal or randomization mitigates it. Indeed, Rácz and Rigobon~\cite{racz2023towards} examine how network structure affects polarization evolution. Notably, beyond polarization and disagreement, other adversarial network phenomena have recently been formulated as optimization problems within standard opinion dynamics models, as noted by Gaitonde et al.~\cite{gaitonde2020adversarial}.

In the FJ model, Bindel et al.~\cite{bindel2015bad} investigate network design problems: Given an unweighted graph and a vector \(\mathbf{u}\) of internal opinions, determine a set of edges (from missing edges) to add to the graph such that the augmented graph minimizes the social cost which combines two key components: (1) the internal disagreement between each node's expressed opinion and its innate opinion, and (2) the interpersonal disagreement between connected nodes with differing opinions. Finding the best set of edges to add from a specific node \(u\) is NP-hard by reducing the subset sum problem to this problem. 

\subsection{Experiments}

While theoretical results (such as approximation guarantees, hardness results, and lower/upper bounds), provide foundational insights into viral marketing strategies, researchers often validate their results using real-world datasets to ensure practical applicability. Testing on empirical data helps verify scalability, efficiency, and real-world performance, particularly for viral marketing problems. We emphasize that such results typically aim to complement theoretical findings or illuminate aspects that are difficult to analyze theoretically. Experiments can be conducted on both real-world datasets (such as Facebook or Twitter sub-networks) and synthetic graphs (such as Erd\H{o}s--R\'enyi, Barab\'{a}si-Albert, or stochastic block models).

Several publicly available real-world datasets serve as benchmarks for evaluating viral marketing algorithms. Among the most widely used is the Stanford Large Network Dataset Collection (\href{https://snap.stanford.edu}{SNAP}), which includes social networks like Twitter, which represents Twitter interactions among members of the 117\textsuperscript{th} U.S. Congress, where nodes are members and edges indicate retweets, mentions, or replies~\cite{fink2023centrality}; Facebook, which is a page-page graph of verified Facebook sites where nodes represent official Facebook pages and links are mutual likes between sites~\cite{rozemberczki2019gemsec,rozemberczki2019multiscale}; GitHub Social Network, where nodes are developers who have starred at least 10 repositories and edges are mutual follower relationships between them~\cite{rozemberczki2019multiscale}; Twitch Gamers Social Network, where nodes are Twitch users and edges are mutual follower relationships between them~\cite{rozemberczki2021twitch}; Collaboration Networks, which are from the e-print arXiv and cover scientific collaborations between authors' papers in a specific category, if an author $i$ co-authored a paper with author $j$, the graph contains an undirected edge from $i$ to $j$~\cite{leskovec2007graph}; and Citation Graphs, sourced from the e-print arXiv and encompassing all citations within a dataset in the high-energy physics theory field, if a paper $i$ cites paper $j$, the graph contains a directed edge from $i$ to $j$~\cite{leskovec2005graphs}. Other notable dataset collections are the \href{https://networkrepository.com/}{Network Repository}~\cite{rossi2015network}, and the Koblenz Network Collection~\cite{kunegis2013konect}, among others. These datasets enable researchers to compare algorithmic performance across diverse real-world scenarios, ensuring robustness and generalizability beyond synthetic or idealized theoretical settings.

As a notable example, for the LT and the IC model, the IM has been tested on a collaboration graph by  Kempe et al.~\cite{kempe2003maximizing}. Particularly, the performance of the proposed greedy algorithm was tested against various centrality-based baselines. Conducting experiments on Twitter, (where the network depicts the debate over the Delhi legislative assembly elections of 2013), and Reddit (where nodes are users who posted in the politics subreddit and there is an edge between two users if there exist two subreddits, other than politics, that both users posted in during the given period) De et al.~\cite{de2014learning}, and Chitra and Musco~\cite{chitra2020analyzing} confirm that link recommendations reinforce filter bubbles in social networks. These datasets have been widely used in the context of the FJ model to study polarization under different setups, see e.g.,~\cite{musco2018minimizing,bhalla2023local,racz2023towards}. Another good example of using datasets, but under the French-DeGroot opinion dynamic, is~\cite{zhou2023opinion}, which considers a wide range of small networks, like the WikiVote network where nodes are users and edges represent votes for adminship, with fewer than a thousand nodes, and large datasets like YouTube, where YouTube users are nodes and subscriptions represent edges, with more than a million nodes. Under the majority-based models, Avin et al.~\cite{avin2019majority} is a good example of using a wide variety of datasets from these repositories. Under the Voter model and the IM, we refer the reader to the work by~\cite{zehmakan2023viral}.

\section{Users Characteristics}\label{Users_Characteristics}
Most opinion dynamics models assume homogeneous user behavior, which is a significant simplification that fails to capture real-world heterogeneity. In reality, users exhibit distinct behavioral patterns, which can be characterized as follows: (1) \textbf{Biased} who are inclined toward one specific opinion, modeled as a fixed bias term influencing their opinion updates; (2) \textbf{Stubborn} who are reluctant to change their opinions, often represented by a low (or zero) susceptibility to external influence in update rules; (3) \textbf{Contrarian} nodes who may take opposing views to disagree, which can be formalized as adopting the inverse of the weighted neighborhood opinion; (4) \textbf{Agnostic} nodes who do not form an opinion on a topic, for example because they are uninterested, typically modeled as having a static neutral opinion; and (5) \textbf{Activeness} which makes some nodes more likely to engage in discussions than others, captured by node-specific interaction probabilities. These distinctions require precise mathematical formulations to ensure accurate modeling and analysis of opinion dynamics.

In this section, we review core results in this area, focusing on how these user characteristics shape outcomes such as stabilization time, polarization, and group influence.

\subsection{Discrete Models}

\subsubsection{Voter Model}

In the presence of biased agents (who choose $+1$ with probability $\alpha$ and with probability $1-\alpha$ revise their opinion based on those held by their neighbors) in the asynchronous Voter model, the stabilization time is proven to be $O(\frac{n \log(n)}{\alpha})$ time steps with high probability regardless of the underlying network~\cite{anagnostopoulos2022biased}.

In contrast, for the synchronous Voter model on a graph with minimum degree $\omega(\log(n))$, consider the setup where nodes are initially in state $+1$ (with probability $q > 1/2$) or $-1$. An external agent attempts to subvert the $+1$ majority by introducing a bias toward $-1$ in two ways: (1) altering communication links, or (2) directly corrupting nodes, with success probability $p$ per round. In this setting, no phase transition occurs; even a slight bias ($p > 0$) is sufficient to overturn the $+1$ majority within $O(1)$ rounds.

In the presence of stubborn nodes with opposing views, the system cannot reach consensus; instead, opinions converge to a distribution with disagreement~\cite{yildiz2013binary}. Mobilia et al.~\cite{mobilia2007role} analyze the Voter model with a stubborn node density $Z$ in directed graphs. When stubborn nodes are equally divided between opposing views, opinions follow a Gaussian distribution in the mean-field limit. The distribution's standard deviation scales as $\sigma \propto 1/\sqrt{Z}$, independent of system size. This demonstrates that even a few stubborn nodes can prevent both consensus and the formation of a stable majority. 

Romero Moreno et al.~\cite{romero2020zealotry} explore IM in the Voter model with stubborn agents on scale-free networks. The external controller aims to have a limited budget of unidirectional connections to maximize the expected vote share at equilibrium. The results show that when the budget is large or stubbornness levels are low, optimal strategies primarily target stubborn users. For small budgets or highly stubborn users, the optimal strategy focuses more on normal agents. Additionally, Masuda~\cite{masuda2015opinion} examines opinion maximization in the Voter model with stubborn users.

Considering agnostic nodes in the model, Gauy et al.~\cite{gauy2025Voter} introduce a variant of the consensus problem by combining the voter model and rumour spreading processes. The authors define a martingale describing the probability of consensus 
and show that estimating consensus probabilities via a Markov chain Monte Carlo method has complexity \(O(n^2 \log n)\) for general graphs and \(O(n \log n)\) for Erd\H{o}s-R\'enyi graphs (with parameter $p \gg \frac{\log(n)}{n}$) in the case of asynchronous update. For the synchronous update, these bounds are given as \(O(n \log n)\) and $O( \log(n))$ for general graphs and Erd\H{o}s--R\'enyi graph.

\subsubsection{Majority Models}

\paragraph{Galam Majority Model.} 
This model incorporates a range of psychological traits to reflect diverse user behaviors in opinion dynamics. Galam~\cite{galam2004contrarian} conducts an analytical study on the impact of introducing contrarian individuals into the model. In this case, and with a few contrarians, both opinions can coexist, but there’s still a clear majority and minority. However, with many contrarians present, no opinion dominates, and individuals continue to switch opinions without any clear majority formation.  

Galam~\cite{galam2016stubbornness} shows that by adding stubborn individuals, the usual threshold can be removed. This ensures that the outcome of the debate does not depend on how it starts. Galam and Jacobs~\cite{galam2007role} also show that when these nodes are added, they force the dynamics to shift in favor of their opinion (while trying to convince open-minded people may not always work~\cite{galam2010public}). If there are more stubborn nodes on one side, their opinion becomes dominant, leaving the minority around them stuck. If both sides have an equal number of stubborn nodes, the system remains balanced with two possible outcomes. Martins and Galam~\cite{martins2013building} show that a small but suborn minority can reliably win public debates by reinforcing their convictions through prolonged discussion.

Cheon and Morimoto~\cite{cheon2016balancer} introduce \textit{balancer} nodes into the Glam opinion dynamics model, representing skeptics who oppose stubborn opinions. Their inclusion creates a threshold separating three distinct opinion phases with sharp transitions, revealing how skepticism disrupts the dominance of the majority.

By incorporating activeness into the Galam model and limiting interactions to smaller groups, the model exhibits a minority counteroffensive effect, in which the minority opinion can prevail even if the majority threshold is $50\%$~\cite{qian2015activeness}. This effect still happens when everyone has the same level of activity~\cite{qian2015activeness}.

This model also examines phenomena such as polarization, as discussed in~\cite{galam2023unanimity}. In a community where most individuals share similar views, opinions naturally converge toward consensus. However, when opposing groups with strong, divergent opinions (contrarian or stubborn nodes) are present, polarization begins to emerge. According to the findings of this paper, when the proportion of contrarian nodes exceeds a certain threshold (approximately $16\%$), the community splits into two polarized groups. Similarly, when $22\%$ or more of a group consists of stubborn nodes, polarization occurs.

\paragraph{Graph-based Majority Model. }
Avin et al.~\cite{avin2019majority} reveal how small, well-connected elite groups can disproportionately influence majority outcomes in social networks. They measured this effect through an \textit{influence factor}, a multiplier representing how much more weight each elite member's vote carries compared to ordinary members (e.g., an influence factor of $2$ means each elite vote counts as 2). Analyzing real and synthetic networks, they found that an elite coalition as small as the square root of all network connections ($\sqrt{m}$) can reliably dominate voting outcomes, requiring only minimal influence factors (for example 2) when elite opinions are fixed (``irreversible'' case) and up to $8$ when all members can change opinions (``reversible'' case). However, Out and Zehmakan~\cite{out2021majority} propose two countermeasures that individual nodes can adopt relatively easily and ensure that elites will not have disproportionate power to engineer the dominant output. The first countermeasure essentially requires each node to make random new connections, while the second demands that nodes be stubborn.

For the majority rule model under the complete graph with biased agents (agents who choose one opinion, called the preferred opinion, with higher probability than the other opinion, Mukhopadhyay et al.~\cite{mukhopadhyay2020Voter} show that the network reaches consensus on the preferred opinion with high probability only if the initial fraction of agents with the preferred opinion is above a certain threshold determined by the biases of the agents.

Incorporating biased agents into the asynchronous majority models, Anagnostopoulos et al.~\cite{anagnostopoulos2022biased} show that the expected stabilization time grows exponentially with the minimum degree under the majority update rule. In particular, the stabilization time is super-polynomial in expectation whenever the minimum degree is $\omega(\log(n))$. However, in this setup (biased majority model), there are some graph structures for which the stabilization time is exponential~\cite{lesfari2022biased}. Indeed, a sharp phase transition has been observed in stabilization times, with a threshold that depends on the bias and the graph's degree distribution~\cite{anagnostopoulos2022biased}. In the same setup (asynchronous majority models with bias agents favoring state $+1$), Cruciani et al.~\cite{cruciani2023phase} demonstrate a sharp phase transition governed by a critical bias threshold. When the bias influence remains below this threshold, an initial $-1$ majority persists for super-polynomial time ($n^{\omega(1)}$ rounds) with high probability. However, when the bias strength exceeds the threshold, the system rapidly converges to the $+1$ consensus in constant time ($O(1)$ rounds). 

\subsubsection{Ising Model}
Galam~\cite{galam2024spontaneous,galam2025spontaneous} uses an Ising-inspired model to demonstrate that a group's ultimate consensus is not predetermined but is path-dependent, shaped by its unique history of interactions. It reveals that echo chambers are the dynamic end-product of local opinion exchange, with polarization being a byproduct of this process. This implies that social media algorithms play a key role in shaping these processes. Galam~\cite{galam1997rational} models group decision-making using the Ising framework, showing that while social interactions naturally push a group toward unanimous consensus, the outcome is highly sensitive to small external pressures. The introduction of individual biases (representing personal experiences and values) weakens overall group polarization.

\subsubsection{Sznajd Model}
He et al.~\cite{he2004sznajd}, through an empirical study, show that the opinion supported by a larger number of stubborn users ultimately dominates, leading the majority of the population to adopt it.

Karan et al.~\cite{karan2017modeling} investigate the Sznajd model with stubborn users where initial opinions of the population are randomly distributed. The results show that for populations larger than $100$, the system reaches a stable steady state (with low variance) and becomes independent of initial conditions.

Crokidakis and De Oliveira~\cite{crokidakis2011sznajd} and Crokidakis and Forgerini~\cite{crokidakis2012competition} demonstrate the robustness of the Sznajd model when extended with a probabilistic setup. Each node successfully convinces its neighbors with probability $p \in (0,1]$ and fails with probability $1-p$ (the persuasion power). The system reaches consensus, and the expected consensus times follow a log-normal distribution for all values $p$. 

Considering age, Sun et al.~\cite{sun2005opinion} employ the Sznajd model to study how opinions change, but with added factors like the influence of parents, social impact, and age differences between individuals. The results show that while age does not affect the final consensus of the Sznajd model, it does influence the opinion distribution, particularly when the initial state is balanced at $50\%$ for each opinion.

Considering memory, Sabatelli and Richmond~\cite{sabatelli2003phase} introduce a memory feature in the Sznajd consensus model, where nodes remember their past $T$ opinions. This memory helps nodes make decisions when they are frustrated by current interactions. The model suggests that as the memory length $T$ increases, the phase transition to complete consensus becomes easier, allowing consensus to be achieved with larger memory and lattice sizes. Also, by incorporating memory features into the model, the stabilization time decreases as the graph's connectivity increases~\cite{guzman2006small}.

\subsubsection{Epidemiological Processes}
In epidemiological models, stubbornness is arguably the most extensively studied property of nodes. This stubbornness can be acquired, for instance, through vaccination, which is highly effective based on historical data~\cite{reluga2007resistance}. From a network-based perspective, Chatterjee and Zehmakan~\cite{chatterjee2023effective} propose a hybrid vaccination strategy that combines network centrality and disease dynamics to minimize the impact of epidemics across networks.
From a modeling perspective, a prominent approach to representing this property is to incorporate stubbornness into the governing differential equations (see, for example, \cite{khaloufi2023continuous, arino2003global}).

\subsection{Continuous Models}

\subsubsection{French-DeGroot model}

Wai et al.~\cite{wai2016active} examine French-DeGroot opinion dynamics with stubborn nodes (who influence their out-neighbors but always stick to their innate opinion) and prove that consensus can still be reached in this model if every ordinary node has a nonzero weight placed on their neighbors' information, represented by the stochastic matrix $\textbf{P}$, in the stubborn nodes. Later, Abrahamsson et al. ~\cite{abrahamsson2019opinion} show that under a weaker assumption, where instead of requiring every ordinary node to have a nonzero weight in the stubborn nodes, it suffices for at least one ordinary node to have a nonzero weight for the consensus still to be achieved.

Dandekar et al.~\cite{dandekar2013biased} introduce a generalization of the DeGroot model that incorporates biased assimilation, where individuals weight confirming evidence more heavily than disconfirming evidence based on a bias parameter $b_i \ge 0$. Each node updates its opinion according to its neighbors' opinions, with $b_i = 0$ recovering the standard DeGroot model. The authors define the network disagreement index (the weighted sum of squared opinion differences between connected nodes) to measure opinion divergence and classify a process as polarizing if this index increases over time. They demonstrate that classical DeGroot averaging cannot induce polarization, even in homophilous networks. In contrast, the biased model can lead to polarization when $b_i > 1$ and to either persistent disagreement or consensus when $b_i < 1$, depending on the degree of homophily.

Xia et al.~\cite{xia2020analysis} introduce a nonlinear generalization of the French-DeGroot opinion dynamics model, in which each individual has a bias parameter that determines how they weight confirming versus disconfirming opinions. When bias is zero, the classic DeGroot consensus emerges, but with bias, richer dynamics occur: an extreme consensus configuration (all opinions at $-1$ or $+1$) is locally stable in strongly connected networks, while the neutral consensus ($1/2$) is unstable. 

\subsubsection{Friedkin-Johnsen Model}
Xu et al.~\cite{xu2022effects} show that stubbornness strongly affects almost all aspects of opinion dynamics, such as the expressed opinion, convergence velocity to equilibrium (increasing a node's stubbornness can speed up convergence, while decreasing it may slow it down), and overall opinion. Shirzadi and Zehmakan~\cite{shirzadi2024stubborn} show that when all nodes in the graph share the same stubbornness factor, any increase in this factor increases polarization--disagreement index, while in the case that nodes have different stubbornness factors, increasing the stubbornness of a neutral node can reduce this index. 

\subsubsection{Deffuant-Weisbuch Model}
Sîrbu et al.~\cite{sirbu2019algorithmic} modify the DW model to incorporate the bias parameter into the model (by increasing the likelihood of pairing individuals with similar opinions, mimicking online platforms that suggest interactions with like-minded peers). The results show that this modification leads to: (a) increased opinion fragmentation, even in scenarios where the original model would predict consensus, (b) heightened polarization, and (c) a significant slowdown in the convergence to the asymptotic state, making the system unstable.

Carletti et al.~\cite{carletti2006make} investigate the effectiveness of propaganda, modeled as an external opinion $s_p$ influencing all individuals. Individuals whose opinions differ from $s_p$ by less than $r$ adjust toward it with strength $\alpha_p$. The propaganda efficiency $E_p$ is defined as the fraction of individuals adopting $s_p$ at equilibrium. The stabilization time shown to be as $\frac{n}{2r \ln(2)} \ln(2rn)$. 

\subsubsection{Hegselmann-Krause Model}

When nodes are classified as open-minded, moderate-minded, or closed-minded based on confidence levels, simulations Fu et al.~\cite{fu2015opinion} show that closed-minded nodes (functionally analogous to stubborn agents) dominate final opinion clusters by anchoring extreme positions. Open-minded nodes hinder consensus unless they reach a critical mass, whereas moderate-minded nodes have negligible effects. 

Chen et al.~\cite{chen2017opinion} modify the HK model to incorporate the \textit{social similarity}, representing the social relationship between a node and its in-neighbors. Based on individual characteristics, social similarity affects how individuals interact with their out-neighbors during the opinion formation process. By considering both confidence bounds and social similarity, the model changes how out-neighbors are selected. The results show that the model captures essential features such as fragmentation, polarization, and consensus, and that consensus is more easily achieved with the right similarity threshold.

\section{Conclusion and Future Works}\label{Conclusion}

\subsection{Conclusion}
Opinions play a crucial role in shaping societies, and understanding how they spread through social networks is more critical than ever. This survey provides a unified overview of opinion diffusion models across disciplines, using consistent notation to bridge gaps between approaches. We have highlighted key theoretical results on convergence and explored essential applications, such as viral marketing and optimization problems related to opinion spread. Additionally, we emphasized the growing importance of nodes' characteristics in influencing opinion dynamics. This work aims to foster cross-disciplinary collaboration in opinion dynamics.

\subsection{Future Works}
\paragraph{Convergence Properties.}
Although the convergence properties of many opinion dynamics models have been well studied, there are still important aspects that require further investigation. For example, the expected stabilization time of the IC and LT models remained unproven (although for both models, the upper bound $O(n)$ is trivial). As another case, the convergence properties of epidemiological processes have mainly been analyzed using differential equation frameworks. However, from a graph-based perspective, which offers a more realistic representation, the convergence behavior of these models has received limited attention and warrants further study. Moreover, most of the proven bounds apply to general graphs or to special graphs of theoretical interest. It would be valuable to provide bounds that are more relevant to real-world networks. Two possible approaches include: (1) providing bounds in terms of parameters that are well-understood for real-world networks, such as diameter or degree distribution, and (2) providing bounds for synthetic graph models, such as Barab\'{a}si-Albert graphs. 

\paragraph{Viral Marketing.} Several essential aspects of viral marketing remain unexplored. For instance, in the Sznajd model on general networks, only threshold analysis has been conducted, while other forms of viral marketing have not been studied. A similar gap exists for $r$-BP diffusion and epidemiological processes, where different classes of viral marketing strategies have yet to be investigated. Recently, for the IC and LT models, there has been considerable work focusing on (1) dynamic networks, where the network structure changes over time~\cite{kim2017influence}, and (2) settings where seeds can be selected in multiple rounds rather than only at the initial time~\cite{soltani2023minimum}. Similar setups could be studied for other opinion dynamics models to advance understanding of viral marketing strategies across different frameworks.

\paragraph{User Characteristics.} The impact of user properties, such as stubbornness or agnosticism, has been largely overlooked in popular opinion dynamics models such as the LT, IC, $r$-BP, and PUSH-PULL protocols. It remains unclear how the presence of such users affects fundamental aspects such as stabilization time and the effectiveness of viral marketing strategies. This gap highlights a missing dimension in the existing literature on opinion dynamics.

\paragraph{Neural Message Passing and Opinion Dynamics. }

Neural message passing is the foundational framework of graph neural networks (see \cite{hamilton2020graph} for a comprehensive review), providing a mathematically intuitive mechanism for propagating and aggregating information across interconnected nodes. In this process, node representations (also known as embeddings) are dynamically updated by integrating both their intrinsic features and the states (embeddings) of their neighbors. Although originating from different disciplines, neural message passing and opinion dynamics share core principles: both explore how local interactions among nodes (or individuals) give rise to emergent global behaviors in networks (or social systems), see e.g.,~\cite{li2025unigo,hevapathige2025graph}.

This conceptual overlap motivated a recent line of work that investigated whether opinion dynamics models can serve as efficient and expressive alternatives to traditional neural message passing, see, e.g., \cite{lv2023unified}, \cite{wang2025understanding}. Specifically, it raises the following key question: How can opinion dynamics models be effectively integrated into the neural message-passing framework? What properties of opinion dynamics models could enhance the expressive power of GNNs, that is, their ability to distinguish between different graph structures and capture complex relational patterns? There is a growing body of research aimed at addressing these questions. For example, Lv et al.~\cite{lv2023unified} design a neural message-passing scheme based on a modification of the HK opinion dynamics model to mitigate oversmoothing (a phenomenon in which node representations become increasingly similar as the number of GNN layers increases). Graph neural diffusion~\cite{chamberlain2021grand}, which is closely related to the Abelson opinion dynamics model, exemplifies the application of continuous message passing. The interested reader is referred to the survey~\cite{han2023continuous} for more information on the continuous message passing mechanism. In another study, Wang et al.~\cite{wang2025understanding} propose a modified time-varying opinion dynamics model that mitigates oversmoothing while maintaining satisfactory performance. This line of research is likely to continue, as the field of opinion dynamics is rich and offers many techniques that can be adapted to design more effective neural message-passing methods.

 \paragraph{Acknowledgments.} 
We would like to acknowledge the insightful comments and suggestions of \textbf{Serge Galam}, which have contributed significantly to the development of this survey.


\bibliographystyle{unsrt}

\end{document}